%% file: Impacts.tex
\documentclass{vldb}
\usepackage[cp1251]{inputenc}
\usepackage{graphics,epstopdf,epsfig,subfigure,graphicx,url}
\usepackage{multirow,amsmath}
\usepackage{amssymb}
\pdfoutput=1
\newtheorem{myDef}{Definition}
\newtheorem{myEg}{Example}
\title{Impacts of Dirty Data: an Experimental Evaluation}
\begin{document}
\numberofauthors{1}
\author{
\alignauthor
Zhixin Qi\hspace{0.8cm}Hongzhi Wang\hspace{0.8cm}Jianzhong Li\hspace{0.8cm}Hong Gao\\
\affaddr{Harbin Institute of Technology}
\email{zhixin.qi@foxmail.com\hspace{0.8cm}\{wangzh,lijzh,honggao\}@hit.edu.cn}
}
\date{}
\maketitle
\begin{abstract}
Data quality issues have attracted widespread attention due to the negative impacts of dirty data on data mining and machine learning results. The relationship between data quality and the accuracy of results could be applied on the selection of the appropriate algorithm with the consideration of data quality and the determination of the data share to clean. However, rare research has focused on exploring such relationship. Motivated by this, this paper conducts an experimental comparison for the effects of missing, inconsistent and conflicting data on classification and clustering algorithms. Based on the experimental findings, we provide guidelines for algorithm selection and data cleaning.
\end{abstract}

\input{Introduction}
\input{Preliminaries}
\input{Classification}
\input{Clustering}

\input{Experiments}

\vspace{-2mm}
\section{GUIDELINES AND FUTURE WORK}
\label{sec:guidelines}
Based on the discussions, we give guidelines for algorithm selection and data cleaning.

\underline{\emph{Classification Guidelines.}} We suggest users select classification algorithm and clean dirty data according to the following steps.

First, users are suggested to detect error rates (e.g., missing rate, inconsistent rate, conflicting rate) from the given data.

Second, according to the given task requirements (e.g., well performance on Precision/Recall/F-measure), we suggest users select candidate algorithms of which Precision/Recall/F-measure on the given data is better than 70\%.

Third, if the given data size is small, we recommend Logistic Regression.

Fourth, according to task requirements and the error type which takes the largest proportion, we suggest users find the corresponding $sensibility$ order and choose the least sensitive classification algorithm.

Finally, according to the selected algorithm, task requirements, and error rates of the given data, we suggest users find the corresponding $keeping$ $point$ orders and clean each type of dirty data to its $keeping$ $point$.

\underline{\emph{Clustering Guidelines.}} We suggest users select clustering algorithm and clean dirty data according to the following steps.

First, users are suggested to detect error rates (e.g., missing rate, inconsistent rate, conflicting rate) from the given data.

Second, according to the given task requirements (e.g., well performance on Precision/Recall/F-measure), we suggest users select candidate algorithms of which Precision/Recall/F-measure on the given data is better than 70\%.

Third, if the given data size is large, we recommend DBSCAN.

Fourth, according to task requirements and the error type which takes the largest proportion, we suggest users find the corresponding $sensibility$ order and choose the least sensitive clustering algorithm.

Finally, according to the selected algorithm, task requirements, and error rates of the given data, we suggest users find the corresponding $keeping$ $point$ orders and clean each type of dirty data to its $keeping$ $point$.

In addition, this work opens many noteworthy avenues for future work, which are listed as follows.

For researchers and practitioners in the fields related to data analytics and data mining. ($i$) Since dirty-data impacts on classification and clustering are valuable, their effects on other kinds of algorithms (e.g., association rules mining) need to be tested. ($ii$) Dirty-data impacts are related to error type, error rate, data size, and algorithm performance on original datasets. Hence, constructing a model with these parameters to predict dirty-data impacts is in demand.

For researchers and practitioners in data-quality and data-cleaning related fields. ($i$) Since the error-tolerance ability of different algorithms on different error types are different, we are unnecessary to clean all dirty data before data mining and machine learning tasks. Instead, data cleaning to an appropriate rate (e.g., $keeping$ $point$) is suggested. However, which part of dirty data has priority to be repaired first is a challenging problem. ($ii$) Since different users have different task requirements, how to clean data on demand needs a solution.

\bibliographystyle{unsrt}
\bibliography{Impacts.bbl}

\input{Appendix}
\end{document}

%% file: Introduction.tex
\section{INTRODUCTION}
Data quality has become a serious issue which cannot be overlooked in both data mining and machine learning communities. We call the data with data quality problems as \emph{dirty data}. Since dirty data affect the accuracy of a data mining or machine learning (e.g., classification or clustering) task, we have to know the relationship between the quality of input data set and accuracy of the results. Based on such relationship, we could select an appropriate algorithm with the consideration of data quality issues and determine the share of data to clean.

Due to the large collection of classification and clustering algorithms, it is difficult for users to decide which algorithm should be adopted. The effects of data quality on algorithms are helpful for algorithm selection. Therefore, impacts of dirty data are in urgent demand.

Before a classification or clustering task, data cleaning is necessary to guarantee data quality. Various data cleaning approaches have been proposed, e.g., data repairing with integrity constraints~\cite{beskales2013relative,chu2013holistic}, knowledge-based cleaning
systems~\cite{chu2015katara,hao2017cleaning}, and crowdsourced data cleaning~\cite{chu2015katara,wang2012crowder}, etc. These methods improve data quality dramatically, but the costs of data cleaning are still expensive~\cite{dallachiesa2013nadeef}. If we know how dirty data affect accuracy of the results, we could clean data selectively according to the accuracy requirements instead of cleaning the entire dirty data. As a result, the data cleaning costs are reduced. Therefore, the study of the relationship between data quality and accuracy of results is in demand.

Unfortunately, rare research has been conducted to explore the specific impacts of dirty data on different algorithms. Thus, this paper aims to fill this gap. This brings following challenges.

\begin{enumerate}
\item Due to the great number of classification and clustering algorithms, the first challenge is how to choose algorithms for experiments.

\item Since existing measures (e.g., Precision, Recall, F-measure) are unable to quantify the fluctuation degrees of results, they are insufficient to evaluate the impacts of dirty data on algorithms. Thus, how to define new metrics for evaluation is the second challenge.

\item Since there is no well-planned dirty data benchmark, we have to generate data sets with the consideration of error type, error rate, data size, and etc. Therefore, the third challenge is how to design data sets to test the impacts of dirty data.
\end{enumerate}

In the challenge of these problems, this paper selects twelve classical algorithms in data mining and machine learning communities. We make comprehensive analyses of possible dirty-data impacts on these algorithms. Then, we evaluate the specific effects of different types of dirty data on different algorithms. Based on the experimental results, we provide suggestions for algorithm selection and data cleaning. In the research field, dirty data are classified into a variety of types~\cite{fan2012foundations}, such as missing data, inconsistent data, and conflicting data. Most existing researches focus on improving data quality for these three kinds of dirty data~\cite{fan2010capturing,cong2007improving,getoor2012entity}. Thus, this paper focuses on the three main types.

In summary, our contributions in this paper are listed as follows.
\begin{enumerate}
\item We conduct an experimental comparison for the effects of missing, inconsistent, and conflicting data on classification and clustering algorithms, respectively. To the best of our knowledge, this is the first paper that studies this issue.

\item We introduce two novel metrics, $sensibility$ and $keeping$ $point$, to evaluate dirty-data impacts on algorithms.

\item Based on the evaluation results, we provide guidelines of algorithm selection and data cleaning for users. We also give suggestions for future work to researchers and practitioners.
\end{enumerate}

The rest paper is organized as follows. Dimensions of data quality are reviewed in Section~\ref{sec:dimension}. Section~\ref{sec:classification} analyzes dirty-data impacts on six classical classification algorithms. We discuss impacts of dirty data on six clustering methods in Section~\ref{sec:clustering}. Our experiments are described in Section~\ref{sec:experiments}, and our guidelines and suggestions are presented in Section~\ref{sec:guidelines}.

%% file: Preliminaries.tex
\section{Dimensions of Data Quality}
\label{sec:dimension}
Data quality has many dimensions~\cite{sidi2012data}. In this paper, we focus on three basic dimensions, completeness, consistency, and entity identity~\cite{fan2012foundations}. For these dimensions, the corresponding dirty data types are missing data, inconsistent data, and conflicting data. In this section, we introduce these three kinds of dirty data.

Missing data refer to values that are missing from databases. For example, on Table~\ref{table:student}, the values of $t_1$[Country] and $t_2$[City] are missing data.

\begin{table}[!htb]
\small
\centering
\vspace{-5mm}
\caption{Student Information}
\label{table:student}
\begin{tabular}{|c|c|c|c|c|}
\hline
        & Student No. & Name &  City  &  Country  \\
\hline
  $t_1$ & 170302 & Alice  & NYC &   \\
\hline
  $t_2$ & 170302 & Steven  &   & FR  \\
\hline
  $t_3$ & 170304 & Bob &   NYC  & U.S.A \\
\hline
  $t_4$ & 170304 & Bob &   LA  & U.S.A  \\
\hline
\end{tabular}
\end{table}
\vspace{-2mm}

Inconsistent data are identified as violations of consistency rules which describe the semantic constraint of data. For example, a consistency rule ``[Student No.] $\rightarrow$ [Name]'' on Table~\ref{table:student} means that Student No. determines Name. As the table shows, $t_1$[Student No.] = $t_2$[Student No.], but $t_1$[Name] $\neq$ $t_2$[Name]. Thus, the values of $t_1$[Student No.], $t_1$[Name], and $t_2$[Name] are inconsistent.

Conflicting data refer to different values which describe an attribute of the same entity. For example, on Table~\ref{table:student}, both $t_3$ and $t_4$ describe $Bob$'s information, but $t_3$[City] and $t_4$[City] are different. Thus, $t_3$[City] and $t_4$[City] are conflicting data.

%% file: Classification.tex
\section{CLASSIFICATION ALGORITHMS}
\label{sec:classification}
In this section, we analyze possible dirty-data impacts on six classical classification algorithms, Decision Tree, K-Nearest Neighbor Classifier, Naive Bayes, Bayesian Network, Logistic Regression, and Random Forests. We choose these algorithms since they are always used as competitive classification algorithms~\cite{caruana2006empirical,caruana2008empirical,elish2008predicting,ghotra2015revisiting}.

For simplicity, we define the notations used through this paper in Table~\ref{table:notations}.

\begin{table}[!htb]
\small
\centering
\vspace{-5mm}
\caption{Notation Definition}
\label{table:notations}
\begin{tabular}{|c|c|}
\hline
 \textbf{notation} & \textbf{definition}  \\
\hline
  $X_i$ & the $i$th attribute \\
\hline
  $Y$ &  class/target attribute  \\
\hline
  $Y_j$ & the $j$th class label  \\
\hline
  $n$ &  the number of attributes  \\
\hline
  $n_c$ &  the number of class labels  \\
\hline
  $D_{train}$ & the training set of given data \\
\hline
  $D_{test}$ &  the test set of given data \\
\hline
\end{tabular}
\end{table}
\vspace{-2mm}

\subsection{Decision Tree}
The decision tree~\cite{quinlan1986induction} splits $D_{train}$ according to an attribute set that optimizes certain criterion in the training process. To determine the best split, a measure of node impurity is needed. Popular measures include Gini index~\cite{prabhu2014fastxml}, information gain~\cite{thornton2013auto}, and misclassification error~\cite{thornton2013auto}.

Given a node $t$, the Gini index for $t$ is defined as follows.
\begin{equation}
Gini(t)=\sum_{i=1}^{n_c}\sum_{i\neq j}P(C_i|t)P(C_j|t)=1-\sum_{i=1}^{n_c}P(C_i|t)^2
\end{equation}
where $P(C_i|t)$ is the relative frequency of class $C_i$ at $t$.

Entropy at $t$ is defined as follows.
\begin{equation}
Entropy(t)=-\sum_{i=1}^{n_c}P(C_i|t)logP(C_i|t)
\end{equation}
where $P(C_i|t)$ is the relative frequency of class $C_i$ at $t$.

Suppose a parent node $p$ is split into $k$ partitions. Based on entropy, information gain for $p$ is defined as follows.
\begin{equation}
Gain(p)=Entropy(p)-\sum_{i=1}^{k}\frac{n_i}{n}Entropy(N_i)
\end{equation}
where $N_i$ is the $i$th partition, and $n_i$ is the number of records in $N_i$.

Misclassification error at $t$ is defined as follows.
\begin{equation}
	Error(t)=1-max_{i=1}^{n_c}P(C_i|t)
\end{equation}
where $P(C_i|t)$ is the relative frequency of class $C_i$ at $t$.

In the decision tree induction, the attribute with minimum Gini index/maximum information gain/minimum misclassification error is chosen as the split node first. With the induced decision tree, the records in $D_{test}$ are classified.

When some dirty values exist in $D_{train}$, incorrect data could affect the value of $P(C_i|t)$. Since $P(C_i|t)$ determines the measure of node purity (e.g., $Gini(t)$, $Gain(p)$, $Error(t)$), the split attribute might be poorly chosen. Thus, the poor induced decision tree could cause an inaccurate classification result. When some values of ($X_1$, $X_2$,..., $X_n$) in $D_{test}$ are dirty, these data might lead to a wrong branch of the decision tree, which results in a wrong class label.

\subsection{K-Nearest Neighbor Classifier}
K-nearest neighbor classifier~\cite{hastie1996discriminant} (KNN for brief) requires three things, $D_{train}$, distance metric (e.g., Euclidean distance~\cite{begum2015accelerating}) to compute the distance between records, and the value of $k$, which is the number of nearest neighbors to retrieve. To classify a record in $D_{test}$, we first compute its distance to other training records. Then, $k$ nearest neighbors are identified. Finally, we use class labels of the nearest neighbors to determine the class label of the unknown record (e.g., by taking majority vote).

Given two records $p$ and $q$, the Euclidean distance between them is defined as follows.
\begin{equation}
	d(p,q)=\sqrt{\sum_{i=1}^n(p_i-q_i)^2}
\end{equation}
where $p_i$ is the $i$th attribute of $p$, and $q_i$ is the $i$th attribute of $q$.

When some dirty data exist in ($X_1$, $X_2$,..., $X_n$) of $D_{train}$ or $D_{test}$, the value of $d(p,q)$ may change. Accordingly, the $k$ nearest neighbors list would be affected, which leads to a wrong class label. When some values of $Y$ in $D_{train}$ are dirty, these incorrect data might affect the vote of class labels, which causes a wrong classification result.

\subsection{Bayes Classifiers}
Bayes classifier is a probabilistic framework for solving classification problems. It computes the posterior probability $P(Y|X_1X_2...X_n)$ for all values of $Y$ in $D_{test}$ according to the Bayes theorem, which is extended as follows.
\begin{equation}
    P(Y|X_1X_2...X_n)=\frac{P(X_1X_2...X_n|Y)P(Y)}{P(X_1X_2...X_n)}
\end{equation}

The goal of Bayes classifiers is to choose the value of $Y$ that maximizes $P(Y|X_1X_2...X_n)$, which is equivalent to maximize $P(X_1X_2...X_n|Y)P(Y)$. Naive Bayes and Bayesian network are classical Bayes classifiers.

\subsubsection{Naive Bayes}
Naive Bayes classifier~\cite{bayes1763letter} assumes the independence among attributes $X_i$ when the class is given, i.e. $P(X_1X_2...X_n|Y_j)$ = $P(X_1|Y_j)P(X_2|Y_j)...P(X_n|Y_j)$. Since  $P(X_i|Y_j)$ for all $X_i$ and $Y_j$ in $D_{train}$ can be estimated, records in $D_{test}$ are classified to $Y_j$ to maximize $P(Y_j)\prod P(X_i|Y_j)$.

When some dirty values exist in ($X_1$, $X_2$,..., $X_n$) of $D_{train}$ or $D_{test}$, incorrect data could affect the value of $P(X_i|Y_j)$, which leads to the wrong value of $P(Y_j)\prod P(X_i|Y_j)$. Since the maximal $P(Y_j)\prod P(X_i|Y_j)$ determines the final $Y_j$, classification result would be impacted. When some values of $Y$ in $D_{train}$ are dirty, the values of both $P(Y_j)$ and $P(X_i|Y_j)$ may change. Accordingly, $P(Y_j)\prod P(X_i|Y_j)$ could be affected, which causes an incorrect class label.

\subsubsection{Bayesian Network}
Bayesian network~\cite{pearl1986fusion} is a directed acyclic diagram (DAG for brief) based on conditional probability tables (CPT for brief). In DAG, a node is conditional independent of its non-descendants, if its parents are known.

When Bayesian network structure is fixed, we estimate conditional probabilities based on $D_{train}$ and learn CPT. When Bayesian network structure is unknown, we first estimate Bayesian network using minimum description length~\cite{wu2017decomposed}, and then learn Bayesian network and CPT. Based on the learned model, we make Bayeaian network inference and compute the maximal posterior probability to determine class labels of $D_{test}$.

Given $D_{train}$, the description length of Bayesian network $B$ is defined as follows.
\begin{equation}
    cost(B|D_{train})=b|B|-\sum\limits_{i=1}^{|D_{train}|}logP(x_i|B)
\end{equation}
where $|B|$ is the number of parameters of $B$, $b$ is the number of bits for describing a parameter, and $x_i$ is the $i$th instance in $D_{train}$.

When some dirty values exist in $D_{train}$, dirty data may affect the value of $cost(B|D_{train})$. Accordingly, the learned Bayesian network and CPT would be incorrect, which leads to inaccurate inference based on the network. Since probabilities are computed with inference, the maximal posterior probability could be impacted, which results in a wrong class label. When some values of ($X_1$, $X_2$,..., $X_n$) in $D_{test}$ are dirty, wrong values might affect the estimation of the maximal posterior probability, which leads to an incorrect classification result.

\subsection{Logistic Regression}
Logistic regression~\cite{mcfadden1972conditional} is a binary classifier. In order to establish a regression function (Sigmoid function~\cite{jain2016extreme}) of classification boundary line, we use optimization methods (e.g., gradient ascent method~\cite{chang2015heterogeneous}) to determine the best regression coefficient of the function based on $D_{train}$. Once the regression function is constructed, we use it to classify records in $D_{test}$.

Given input data $X_1$, $X_2$,..., $X_n$, the input of Sigmoid function is computed as follows.
\begin{equation}
    z=w_1X_1+w_2X_2+...+w_nX_n
\end{equation}

The Sigmoid function is defined as follows.
\begin{equation}
    f(z)=\frac{1}{1+e^{-z}}
\end{equation}

When some dirty values exist in ($X_1$, $X_2$,..., $X_n$) of $D_{train}$ or $D_{test}$, wrong values could affect the value of $z$. Accordingly, the learned Sigmoid function would change, which leads to inaccurate class labels of $D_{test}$. When some values of $Y$ in $D_{train}$ are dirty, incorrect class labels may mislead the establishment of Sigmoid function. Based on the imprecise function, the classification of $D_{test}$ might be affected.

\subsection{Random Forests}
In the training process, the random forests algorithm~\cite{cui2015optimal} constructs a set of base classifiers of decision trees. In the testing process, class labels of $D_{test}$ are predicted by aggregating predictions made by multiple classifiers.

When some dirty values exist in $D_{train}$, incorrect data might affect splitting attribute selection in decision trees, which causes inaccurate decision tree induction and wrong predictions of $D_{test}$. When some values of ($X_1$, $X_2$,..., $X_n$) in $D_{test}$ are dirty, these data might mislead class labels of corresponding records, which causes incorrect classification.

%% file: Clustering.tex
\section{CLUSTERING ALGORITHMS}
\label{sec:clustering}
In this section, we discuss possible dirty-data impacts on six classical clustering algorithms, K-Means, LVQ, CLARANS, DBSCAN, BIRCH, and CURE. We choose these algorithms since they are always used as competitive clustering algorithms~\cite{khanmohammadi2017improved,kirchner2016facilitating,wu2013clustering,gulati2015clustering}.

\subsection{Prototype-Based Clustering}
Prototype-based clustering assumes that clustering structure is portrayed by a group of prototypes. This kind of algorithms initialize prototypes at first, and then update them in the iterative process. K-Means~\cite{macqueen1967some}, learning vector quantization~\cite{kohonen1995learning} (LVQ for brief), and CLARANS~\cite{ng1994cient} are three classical clustering methods.

\subsubsection{K-Means}
K-Means clustering approach selects $K$ points as the initial centroids firstly. Then, we form $K$ clusters by assigning all points to the closest centroid, and recompute the centroid of each cluster. The iterative process ends until the centroids do not chagne.

When some dirty values exist, incorrect data might affect computation of centroids. Accordingly, some points would be clustered to wrong class labels.

\subsubsection{LVQ}
LVQ clustering method assumes that there are class labels in data samples, and these marked labels can assist clustering. Given sample set $D_s$ = \{($\emph{\textbf{X}}_1$, $Y_1$), ($\emph{\textbf{X}}_2$, $Y_2$),...,($\emph{\textbf{X}}_m$, $Y_m$)\}, each $\emph{\textbf{X}}_j$ (1 $\leq$ $j$ $\leq$ $m$) is a feature vector ($X_{j1}$; $X_{j2}$;...;$X_{jn}$), which is expressed with $n$ attributes. $Y_j$ $\in$ $Y$ is the class label of $\emph{\textbf{X}}_j$. The goal of LVQ is to learn a group of $n$-dimensional prototype vector \{$\emph{\textbf{p}}_1$, $\emph{\textbf{p}}_2$,..., $\emph{\textbf{p}}_q$\}, each vector denotes a cluster, and each cluster label $t_i$ $\in$ $Y$.

First, LVQ algorithm initializes prototype vectors. Then, vectors are optimized in the iterative process. In each iteration, algorithm selects a marked training sample randomly, and finds a prototype vector with the shortest distance from the selected sample. If their labels are not the same, the prototype vector is updated.

When some values in $\emph{\textbf{X}}_j$ ($j$ = 1, 2,..., m) of $D_s$ are dirty, wrong values could mislead the label updating of prototype vector \{$\emph{\textbf{p}}_1$, $\emph{\textbf{p}}_2$,..., $\emph{\textbf{p}}_q$\}. When some dirty values exist in $Y_j$ ($j$ = 1, 2,..., m) of $D_s$, incorrect class labels would directly affect the labels of \{$\emph{\textbf{p}}_1$, $\emph{\textbf{p}}_2$,..., $\emph{\textbf{p}}_q$\}. When some values in $\emph{\textbf{X}}_j$ ($j$ = 1, 2,..., q) of \{$\emph{\textbf{p}}_1$, $\emph{\textbf{p}}_2$,..., $\emph{\textbf{p}}_q$\} are dirty, wrong values might impact the distance computing, which leads to an incorrect class label.

\subsubsection{CLARANS}
CLARANS algorithm selects $K$ points as centroids at first. Then, we randomly choose a centroid $k_1$ as the current point, and a neighbor point of $k_1$ $k_2$. We compute the cost difference between $k_1$ and $k_2$. If the cost of $k_2$ is less, we set it as the current point, and select a neighbor of $k_2$. If not, we find another neighbor of $k_1$. The iteration ends until the number of sampling is achieved.

When some dirty values exist, incorrect data might affect computation of the cost difference. Accordingly, some points could be clustered to wrong class labels.

\subsection{Density-Based Clustering}
Density-based clustering locates regions of high density that are separated from one another by regions of low density. DBSCAN~\cite{ester1996density} is a basic density-based clustering algorithm. All noise points are discarded in DBSCAN, and we perform clustering on the remaining points. At first, we put an edge between all core points that are within $Eps$ (a specified radius) of each other. Then, each connected component is taken as a separate cluster, and each border point are assigned to one of the clusters of its associated core points.

When some values are dirty, wrong values would impact computation of density and point distances since density is the number of points within $Eps$ of a point. Accordingly, some points might be assigned to incorrect clusters.
\subsection{Hierarchical Clustering}
Hierarchical clustering produces a set of nested clusters organized as a hierarchical tree. BIRCH~\cite{zhang1996birch} and CURE~\cite{guha1998cure} are classical hierarchical algorithms.

\subsubsection{BIRCH}
BIRCH algorithm introduces a clustering feature tree (CF tree for brief) to summarize the inherent clustering structure of data. Firstly, we scan the given data to build an initial in-memory CF tree. Then, we use an arbitrary clustering algorithm (e.g., an agglomerative hierarchical clustering algorithm) to cluster the leaf nodes of the CF tree. Finally, we scan the data again and assign the data points using the cluster centroids found in the previous step as seeds.

When some dirty data exist, incorrect values could affect construction of CF tree and computation of cluster centroids, which makes data points assigned to wrong clusters.
\subsubsection{CURE}
Instead of representing clusters by their centroids, CURE algorithm uses a collection of representative points. There are three steps in this method. The first step is initialization. At first, we take a small sample of the given data and cluster it in main memory using a hierarchical method in which clusters are merged when they have a close pair of points (e.g., MIN clustering method). Then, we select a small set of points from each cluster to be representative points. These points should be chosen to be as far from one another as possible. Lastly, we move each of the representative points a fixed fraction of the distance between its location and the centroid of its cluster.

The second step is merging clusters. We merge two clusters if they have a pair of representative points, one from each cluster, that are sufficiently close.

The third step is point assignment. Each point $p$ is brought from secondary storage and compared with the representative points. We assign $p$ to the cluster of the representative point that is closest to $p$.

When some values in the given data are dirty, wrong values would effect the location of representative points and the computation of distance between representative point and the centroid of its cluster. Accordingly, data points might be clustered into incorrect clusters.

%% file: Experiments.tex
\section{EXPERIMENTAL STUDY}
\label{sec:experiments}
We evaluated dirty-data impacts on twelve classical algorithms discussed in Section~\ref{sec:classification} and~\ref{sec:clustering}.
\subsection{Experimental Setting}
\underline{\textbf{Datasets}} We selected 9 typical data sets from UCI public datasets\footnote{http://archive.ics.uci.edu/ml/datasets.html} with various types and sizes. Their basic information is shown in Table~\ref{table:datasets}. Due to the completeness and correctness of these original datasets, we injected errors into them, and then evaluated their performance on different algorithms. Thus, the original datasets were used as the baseline, and the accuracy of algorithms was measured based on the results on original datasets.

\begin{table}[!htb]
\small
\centering
\vspace{-5mm}
\caption{Datasets Information}
\label{table:datasets}
\begin{tabular}{|c|c|c|c|}
\hline
 \textbf{Name} & \textbf{Number of} & \textbf{Number of} & \textbf{Algorithm} \\
 & \textbf{Attributes} & \textbf{Records} &  \\
\hline
   &  &  & Classification\\
  Iris  & 4 & 150 & Clustering \\
     & & & Regression \\
\hline
  Ecoli & 8 & 336 & Classification \\
\hline
  Car & 6 & 1728 & Classification  \\
\hline
  Chess & 36 & 3196 & Classification  \\
\hline
  Adult & 14 & 48842 & Classification  \\
\hline
  Seeds & 7 & 210 & Clustering \\
\hline
  Abalone & 8 & 4177 & Clustering \\
\hline
  HTRU & 9 & 17898 & Clustering \\
\hline
  Activity & 3 & 67651 & Clustering \\
\hline
\end{tabular}
\end{table}

\underline{\textbf{Setup}} All experiments were conducted on a machine powered by two Intel(R) Xeon(R) E5-2609 v3@1.90GHz CPUs and 32GB memory, under CentOS7. All the algorithms were implemented in C++ and compiled with g++ 4.8.5.

\underline{\textbf{Metrics}} First, we used standard Precision, Recall, and F-measure to evaluate the effectiveness of classification and clustering algorithms. These measures were computed as follows.
\begin{equation}
    Precision=\frac{\sum_{i=1}^{n_c}\frac{rc_i}{rn_i}}{n_c}
\end{equation}
where $rc_i$ is \#-of records which are correctly classified to class $i$, and $rn_i$ is \#-of records which are classified to class $i$.

\begin{equation}
    Recall=\frac{\sum_{i=1}^{n_c}\frac{rc_i}{r_i}}{n_c}
\end{equation}
where $rc_i$ is \#-of records which are correctly classified to class $i$, and $r_i$ is \#-of records of class $i$.

\begin{equation}
    F-measure=\frac{2\times Precision\times Recall}{Precision+Recall}
\end{equation}

However, these metrics only showed us the variations of accuracy. They were not possible to measure the fluctuation degrees quantitatively. Therefore, novel metrics were introduced to evaluate dirty-data impacts on algorithms. We defined the first metric $sensibility$ as follows.

\begin{myDef}
Given the values of a measure $y$ of an algorithm with a\%, (a+x)\%, (a+2x)\%,..., (a+bx)\% (a$\geq$0, x$>$0, b$>$0) error rate. $Sensibility$ of an algorithm on dirty data is computed as $|y_{a}$ - $y_{a+x}|$ + $|y_{a+x}$ - $y_{a+2x}|$ + ... + $|y_{a+(b-1)x}$ - $y_{a+bx}|$.
\end{myDef}

$Sensibility$ is defined to measure the sensibility of an algorithm to the data quality. The larger the value of $sensibility$ is, the more sensitive an algorithm is to the data quality. Therefore, $sensibility$ shows the fluctuation degrees of algorithms quantitatively. Here, we explain the computation of $sensibility$ with Figure~\ref{fig:dt-miss-p} as an example.

\begin{myEg}
Since the values of $Precision$ ($P$) of the decision tree algorithm with 0\%, 10\%,..., 50\% missing rate are given, $sensibility$ is computed as $|P_{0\%}$ - $P_{10\%}|$ + $|P_{10\%}$ - $P_{20\%}|$ +...+ $|P_{40\%}$ - $P_{50\%}|$. Thus, in Iris dataset, $sensibility$ is $|$78.37\%-84.16\%$|$ + $|$84.16\%-78.08\%$|$ + $|$78.08\%-74.36\%$|$ + $|$74.36\%-64.99\%$|$ + $|$64.99\%-58.71\%$|$ = 31.24\%. In Ecoli dataset, $sensibility$ is $|$63.47\%-62.93\%$|$ + $|$62.93\%-53.97\%$|$ + $|$53.97\%-50.93\%$|$ + $|$50.93\%-48.07\%$|$ + $|$48.07\%-34.5\%$|$ = 28.97\%. In Car dataset, $sensibility$ is $|$81.33\%-60.93\%$|$ + $|$60.93\%-43.7\%$|$ + $|$43.7\%-42.87\%$|$ + $|$42.87\%-40.47\%$|$ + $|$40.47\%-35.47\%$|$ = 45.86\%. In Chess dataset, $sensibility$ is $|$82.17\%-78.17\%$|$ + $|$78.17\%-76.53\%$|$ + $|$76.53\%-75.77\%$|$ + $|$75.77\%-75.9\%$|$ + $|$75.9\%-75.57\%$|$ = 6.86\%. And in Adult dataset, $sensibility$ is $|$80.5\%-75.27\%$|$ + $|$75.27\%-71.3\%$|$ + $|$71.3\%-72.93\%$|$ + $|$72.93\%-71.53\%$|$ + $|$71.53\%-67.23\%$|$ = 16.53\%. Thus, the average of $sensibility$ is 25.89\%.
\end{myEg}

Though $sensibility$ tells the fluctuation degrees of algorithms, we could not determine the error rate at which an algorithm is unacceptable. Motivated by this, we defined the second novel metric $keeping$ $point$ as follows.

\begin{myDef}
Given the values of a measure $y$ of an algorithm with a\%, (a+x)\%, (a+2x)\%,..., (a+bx)\% (a$\geq$0, x$>$0, b$>$0) error rate, and a number $k$ ($k$ $>$ 0). If the larger value of $y$ causes the better accuracy, and $y_{a\%}$-$y_{(a+ix)\%}$ $>$ $k$ (0 $<$ i $\leq$ b), we take min\{(a+(i-1)x)\%\} as the $keeping$ $point$. If $y_{a\%}$-$y_{(a+bx)\%}$ $\leq$ $k$, we take min\{(a+bx)\%\} as the $keeping$ $point$. If the smaller value of $y$ causes the better accuracy, and $y_{(a+ix)\%}$-$y_{a\%}$ $>$ $k$ (0 $<$ i $\leq$ b), we take min\{(a+(i-1)x)\%\} as $keeping$ $point$. If $y_{(a+bx)\%}$-$y_{a\%}$ $\leq$ $k$, we take min\{(a+bx)\%\} as the $keeping$ $point$.
\end{myDef}

$Keeping$ $point$ is defined to measure the dirty-data tolerability of an algorithm. The larger the value of $keeping$ $point$ is, the higher error-tolerability of an algorithm is. Therefore, $keeping$ $point$ is useful to show the error rate at which an algorithm is acceptable. Here, we take Figure~\ref{fig:dt-miss-p} as an example to explain $keeping$ $point$ of an algorithm.

\begin{myEg}
We know the values of $Precision$ of the decision tree algorithm with 0\%, 10\%,..., 50\% missing rate, and set 10\% as the value of $k$. In Iris dataset, when the missing rate is 40\%, $y_{0\%}$-$y_{40\%}$ = 78.37\%-64.99\% = 13.38\% $>$ 10\%, we take 30\% as the $keeping$ $point$. In Ecoli dataset, when the missing rate is 30\%, $y_{0\%}$-$y_{30\%}$ = 63.47\%-50.93\% = 12.54\% $>$ 10\%, we take 20\% as the $keeping$ $point$. In Car dataset, when the missing rate is 10\%, $y_{0\%}$-$y_{10\%}$ = 81.33\%-60.93\% = 20.4\% $>$ 10\%, we take 0\% as the $keeping$ $point$. In Chess dataset, when the missing rate is 50\%, $y_{0\%}$-$y_{50\%}$ = 82.17\%-75.57\% = 6.6\% $\leq$ 10\%, we take 50\% as the $keeping$ $point$. In Adult dataset, when the missing rate is 50\%, $y_{0\%}$-$y_{50\%}$ = 80.5\%-67.23\% = 13.27\% $>$ 10\%, we take 40\% as the $keeping$ $point$. Thus, the average of $keeping$ $point$ is 28\%.
\end{myEg}

In addition, we used running time to evaluate the efficiency of all algorithms. We ran each test 5 times and reported logarithms of average time.

\subsection{Evaluation on Classification Algorithms}
Since various kinds of dirty data could affect the performance of classification algorithms, we varied error rates, including missing rate, inconsistent rate, and conflicting rate to evaluate the classification methods in Section~\ref{sec:classification}.

\begin{table*}[!htb]
\small
\centering
\caption{$Sensibility$ Results of Classification and Clustering Algorithms (Unit: \%)}
\label{table:sensibility}
\begin{tabular}{|c|c|c|c|c|c|c|c|c|c|}
\hline
       & \multicolumn{3}{|c|}{\textbf{Missing}} &  \multicolumn{3}{|c|}{\textbf{Inconsistent}} & \multicolumn{3}{|c|}{\textbf{Conflicting}}\\
\hline
  \textbf{Algorithm} & \textbf{P} & \textbf{R} & \textbf{F} & \textbf{P} & \textbf{R} & \textbf{F} & \textbf{P} & \textbf{R} & \textbf{F} \\
\hline
  Decision Tree & 25.89 & 31.11 & 26.64 & 35.41 & 40.94 & 38.33 & 16.09 & 21.56 & 16.45 \\
\hline
  KNN &  18.09 & 13.18 & 17.45 & 21.84 & 19.21 & 20.93 & 11.39  & 6.70 & 9.32 \\
\hline
  Naive Bayes & 27.04 & 23.37 & 26.40 & 29.48 & 37.18 & 35.49 & 15.10 & 21.85 & 20.33 \\
\hline
  Bayesian Network & 46.40 & 34.04 & 35.37 & 33.29 & 21.53 & 23.15 & 17.26 & 15.18 & 16.01 \\
\hline
  Logistic Regression & 38.26 & 18.73 & 30.69 & 37.84 & 28.10 & 38.83 & 31.74 & 18.51 & 25.60 \\
\hline
  Random Forests & 25.77 & 24.57 & 29.39 & 39.21 & 34.86 & 40.74 & 27.93 & 15.85 & 27.53 \\
\hline
  K-Means & 31.06 & 27.80 & 32.08 & 31.83 & 32.21 & 35.63 & 23.79  & 21.86 & 25.17 \\
\hline
  LVQ & 11.94 & 21.14 & 19.61 & 20.55 & 18.83 & 21.41 & 9.20  & 19.57 & 20.13 \\
\hline
  CLARANS & 34.26 & 40.16 & 39.48 & 31.11 & 29.45 & 31.56 & 20.67  & 22.64 & 24.04 \\
\hline
  DBSCAN & 15.89 & 22.88 & 17.16 & 20.40 & 10.39 & 12.34 & 18.64  & 9.55 & 16.10 \\
\hline
  BIRCH & 32.58 & 44.56 & 32.90 & 24.32 & 22.48 & 19.40 & 15.16  & 22.44 & 16.52 \\
\hline
  CURE & 38.68 & 32.71 & 39.23 & 28.81 & 32.90 & 32.67 & 32.74  & 29.11 & 32.62 \\
\hline
\end{tabular}
\end{table*}

\subsubsection{Classification - Varying Missing Rate}
\label{subsec:class-miss}
To evaluate the impacts of missing data on classification algorithms, we deleted values from original datasets randomly varying missing rate from 10\% to 50\%. We used 10-fold cross validation, and generated training data and testing data randomly. In the testing process, we imputed numerical missing values with the average values and captured categorical ones with the maximum values. Experimental results were depicted in Figure~\ref{fig:dt-miss},~\ref{fig:knn-miss},~\ref{fig:nb-miss},~\ref{fig:bn-miss},~\ref{fig:log-miss}, and~\ref{fig:ad-miss}.

Based on the results, we had the following observations. First, for well-performed algorithms whose Precision/Recall/F-measure is larger than 80\% on original datasets, as the data size increases, Precision, Recall, or F-measure of algorithms becomes stable, except \textbf{Logistic Regression}. The reason is that the amount of clean data is larger for larger data size. Accordingly, the impacts of missing data on algorithms reduce. However, Logistic regression establishes a regression function as the model. For regression functions, the parameter computation is more sensitive to missing data. Thus, when data size rises, the amount of missing data becomes larger, which has larger impacts on Logistic Regression.

Second, as shown in Table~\ref{table:sensibility}, for Precision, the $sensibility$ order is ``Bayesian Network $>$ Logistic Regression $>$ Naive Bayes $>$ Decision Tree $>$ Random Forests $>$ KNN''. For Recall, the $sensibility$ order is ``Bayesian Network $>$ Decision Tree $>$ Random Forests $>$ Naive Bayes $>$ Logistic Regression $>$ KNN''. For F-measure, the $sensibility$ order is ``Bayesian Network $>$ Logistic Regression $>$ Random Forests $>$ Decision Tree $>$ Naive Bayes $>$ KNN''. Thus, the least sensitive algorithm is \textbf{KNN}. This is because that as missing rate rises, the increasing missing values may not affect $k$ nearest neighbors. Even if $k$ nearest neighbors are affected, they are not necessarily voted for the final class label. In addition, the most sensitive algorithm is \textbf{Bayesian Network}. The reason is that the increasing missing data could affect the computation of posterior probabilities, which would directly impact classification results.

Third, as shown in Table~\ref{table:keeping}, for Precision, the $keeping$ $point$ order is ``Decision Tree $>$ Naive Bayes $=$ Random Forests $>$ KNN $>$ Logistic Regression $>$ Bayesian Network''. For Recall, the $keeping$ $point$ order is ``Random Forests $>$ KNN $>$ Naive Bayes $=$ Logistic Regression $>$ Decision Tree $=$ Bayesian Network''. For F-measure, the $keeping$ $point$ order is ``Decision Tree $>$ Naive Bayes $=$ Bayesian Network $>$ KNN $>$ Logistic Regression $>$ Random Forests''. Therefore, for Precision and F-measure, the most incompleteness-tolerant algorithm is \textbf{Decision Tree}. This is because that decision tree models only use splitting features for classification. As the missing rate rises, the increasing missing data may not affect splitting features. For Recall, the most incompleteness-tolerant algorithm is \textbf{Random Forests}. This is because the increasing missing values may not affect splitting attributes. Even if impacted, there is little chance to cause inaccurate classification since the final result is made by multiple base classifiers. For Precision and Recall, the least incompleteness-tolerant algorithm is \textbf{Bayesian Network}. This is because the increasing missing data would change the posterior probabilities, which could affect classification results directly. For F-measure, the least incompleteness-tolerant algorithm is \textbf{Random Forests}. The reason is that F-measure on original datasets (error rate is 0\%) is high. When few missing values exist in the datasets, F-measure drops a lot.

Fourth, as data size increases, the running time of algorithms fluctuates more. This is because that as data size rises, the amount of missing data becomes larger, which introduces more uncertainty to algorithms. Accordingly, the uncertainty of running time increases.

\begin{table}[!htb]
\small
\centering
\caption{$Keeping$ $Point$ Results of Classification and Clustering Algorithms ($k$=10\%, Unit: \%)}
\label{table:keeping}
\begin{tabular}{|c|c|c|c|c|c|c|c|c|c|}
\hline
       & \multicolumn{3}{|c|}{\textbf{Missing}} &  \multicolumn{3}{|c|}{\textbf{Inconsistent}} & \multicolumn{3}{|c|}{\textbf{Conflicting}}\\
\hline
  \textbf{Algorithm} & \textbf{P} & \textbf{R} & \textbf{F} & \textbf{P} & \textbf{R} & \textbf{F} & \textbf{P} & \textbf{R} & \textbf{F} \\
\hline
  Decision & 28 & 26 & 28 & 18 & 16 & 16 & 50 & 50 & 50 \\
  Tree  & & & & & & & & & \\
\hline
  KNN & 24 & 32 & 20 & 22 & 22 & 22 & 40  & 50 & 40 \\
\hline
  Naive & 26 & 28 & 24 & 22 & 12 & 12 & 50 & 40 & 40 \\
  Bayes  & & & & & & & & &  \\
\hline
  Bayesian & 20 & 26 & 24 & 16 & 26 & 26 & 46 & 50 & 50 \\
  Network  & & & & & & & & & \\
\hline
  Logistic   & 22 & 28 & 16 & 16 & 14 & 16 & 32 & 34 & 32 \\
  Regression & & & & & & & & & \\
\hline
  Random & 26 & 50 & 10 & 26 & 14 & 8 & 42 & 38 & 34 \\
  Forests  & & & & & & & & & \\
\hline
  K-Means & 38 & 32 & 32 & 28 & 22 & 22 & 44  & 38 & 38 \\
\hline
  LVQ & 44 & 40 & 48 & 28 & 14 & 20 & 44  & 44 & 40 \\
\hline
  CLARANS & 2 & 2 & 0 & 22 & 18 & 18 & 34  & 34 & 28 \\
\hline
  DBSCAN & 30 & 40 & 30 & 32 & 44 & 34 & 36  & 50 & 36 \\
\hline
  BIRCH & 24 & 20 & 24 & 20 & 26 & 26 & 50  & 34 & 38 \\
\hline
  CURE & 18 & 18 & 16 & 20 & 18 & 16 & 32  & 34 & 24 \\
\hline
\end{tabular}
\end{table}

\subsubsection{Classification - Varying Inconsistent Rate}
To evaluate the impacts of inconsistency on classification algorithms, we injected inconsistent values to original datasets randomly according to consistency rules on the given data. The inconsistent rate was varied from 10\% to 50\%. We used 10-fold cross validation, and generated training data and testing data randomly. Experimental results were depicted in Figure~\ref{fig:dt-incons}, \ref{fig:knn-incons}, \ref{fig:nb-incons}, \ref{fig:bn-incons}, \ref{fig:log-incons}, and \ref{fig:ad-incons}.

Based on the results, we had the following observations. First, as shown in Table~\ref{table:sensibility}, for Precision, the $sensibility$ order is ``Random Forests $>$ Logistic Regression $>$ Decision Tree $>$ Bayesian Network $>$ Naive Bayes $>$ KNN''. For Recall, the $sensibility$ order is ``Decision Tree $>$ Naive Bayes $>$ Random Forests $>$ Logistic Regression $>$ Bayesian Network $>$ KNN''. For F-measure, the $sensibility$ order is ``Random Forests $>$ Logistic Regression $>$ Decision Tree $>$ Naive Bayes $>$ Bayesian Network $>$ KNN''. Thus, the least sensitive algorithm is \textbf{KNN}. The reason is similar as that of the least sensitive algorithm varying missing rate. For Precision and F-measure, the most sensitive algorithm is \textbf{Random Forests}. And for Recall, the most sensitive algorithm is \textbf{Decision Tree}. These are due to the fact that as the inconsistent rate increases, more and more incorrect values cover the correct ones in decision tree training models, which leads to inaccurate classification results. Since base classifiers in Random Forests are decision trees, the reason for Random Forests is the same as that for Decision Tree.

Second, as shown in Table~\ref{table:keeping}, for Precision, the $keeping$ $point$ order is ``Random Forests $>$ KNN $=$ Naive Bayes $>$ Decision Tree $>$ Bayesian Network $=$ Logistic Regression''. For Recall, the $keeping$ $point$ order is ``Bayesian Network $>$ KNN $>$ Decision Tree $>$ Logistic Regression $=$ Random Forests $>$ Naive Bayes''. For F-measure, the $keeping$ $point$ order is ``Bayesian Network $>$ KNN $>$ Decision Tree $=$ Logistic Regression $>$ Naive Bayes $>$ Random Forests''. Therefore, for Precision, the most inconsistency-tolerant algorithm is \textbf{Random Forests}. The reason has been discussed in Section~\ref{subsec:class-miss}. For Recall and F-measure, the most inconsistency-tolerant algorithm is \textbf{Bayesian Network}. This is because inconsistent values contain incorrect ones and correct ones. Hence, incorrect values have little effect on the computation of posterior probabilities. Accordingly, classification results may not be affected. For Precision, the least inconsistency-tolerant algorithms are \textbf{Bayesian Network} and \textbf{Logistic Regression}. For Recall, the least inconsistency-tolerant algorithm is \textbf{Naive Bayes}. And for F-measure, the least inconsistency-tolerant algorithm is \textbf{Random Forests}. These are because that Precision/Recall/F-measure of these algorithms on original datasets (error rate is 0\%) is high. When few inconsistent values are injected, Precision/Recall/F-measure drops dramatically.

Third, the observation of running time on experiments varying missing rate was still true when the inconsistent rate was varied.

\subsubsection{Classification - Varying Conflicting Rate}
To evaluate the impacts of conflicting data on classification algorithms, we injected conflicting values to original datasets randomly varying conflicting rate from 10\% to 50\%. We used 10-fold cross validation, and generated training and testing data randomly. Experimental results were depicted in Figure~\ref{fig:dt-conf}, \ref{fig:knn-conf}, \ref{fig:nb-conf}, \ref{fig:bn-conf}, \ref{fig:log-conf}, and \ref{fig:ad-conf}.

First, the observation of the relationship between the data size and algorithm stability on experiments varying missing rate was still true when the conflicting rate was varied.

Second, as shown in Table~\ref{table:sensibility}, for Precision, the $sensibility$ order is ``Logistic Regression $>$ Random Forests $>$ Bayesian Network $>$ Decision Tree $>$ Naive Bayes $>$ KNN''. For Recall, the $sensibility$ order is ``Naive Bayes $>$ Decision Tree $>$ Logistic Regression $>$ Random Forests $>$ Bayesian Network $>$ KNN''. For F-measure, the $sensibility$ order is ``Random Forests $>$ Logistic Regression $>$ Naive Bayes $>$ Decision Tree $>$ Bayesian Network $>$ KNN''. Thus, the least sensitive algorithm is \textbf{KNN}. The reason is similar as that of the least sensitive algorithm varying missing rate. For Precision, the most sensitive algorithm is \textbf{Logistic Regression}. This is because parameter computation of the regression function is easily affected by the increasing conflicting values, which causes an inaccurate logistic regression model. For Recall, the most sensitive algorithm is \textbf{Naive Bayes}. This is because that incorrect values in the increasing conflicting values affect the computation of posterior probabilities in Bayes theorem. For F-measure, the most sensitive algorithm is \textbf{Random Forests}. The reason is the same as that of the most sensitive algorithm varying inconsistent rate.

Third, as shown in Table~\ref{table:keeping}, for Precision, the $keeping$ $point$ order is ``Decision Tree $=$ Naive Bayes $>$ Bayesian Network $>$ Random Forests $>$ KNN $>$ Logistic Regression''. For Recall, the $keeping$ $point$ order is ``Decision Tree $=$ KNN $=$ Bayesian Network $>$ Naive Bayes $>$ Random Forests $>$ Logistic Regression''. For F-measure, the $keeping$ $point$ order is ``Decision Tree $=$ Bayesian Network $>$ KNN $=$ Naive Bayes $>$ Random Forests $>$ Logistic Regression''. Therefore, the most conflict-tolerant algorithm is \textbf{Decision Tree}. The reason is similar as that of the most incompleteness-tolerant algorithm. The least conflict-tolerant algorithm is \textbf{Logistic Regression}. This is due to the fact that conflicting data have much effect on parameter computation of logistic regression models.

Fourth, the observation of running time on experiments varying missing rate was still true when the conflicting rate was varied.

\subsubsection{Discussion}
In classification experiments, we first found that dirty-data impacts are related to error type and error rate. Thus, the error rate of each error type in the given data is necessary to be detected. Second, we observed that for algorithms whose Precision/Recall/F-measure is larger than 80\% on original datasets, Precision, Recall, or F-measure of algorithms become stable as data size rises, except Logistic Regression. Since the parameter $k$ in $keeping$ $point$ was set as 10\%, candidate algorithms of which Precision/Recall/F-measure is larger than 70\% are acceptable. Third, we prefer to choose stable algorithms. Hence, Logistic Regression is suitable for smaller data size. Fourth, we compared the fluctuation degrees of classification algorithms in our experiments. When dirty data exist, the algorithm with the least degree is the most stable. Fifth, beyond $keeping$ $point$, the accuracy of the selected algorithm becomes unacceptable. Thus, the error rate of each type needs to be controlled within its $keeping$ $point$.

\subsection{Evaluation on Clustering Algorithms}
Since various types of dirty data could affect the performance of clustering algorithms, we varied error rates, involving missing rate, inconsistent rate, and conflicting rate to evaluate clustering approaches in Section~\ref{sec:clustering}.
\subsubsection{Clustering - Varying Missing Rate}
\label{subsec:clus-miss}
To evaluate missing-data impacts on clustering algorithms, we deleted values from original datasets randomly varying missing rate from 10\% to 50\%. In clustering process, we imputed numerical missing values with the average values, and captured categorical ones with the maximum values. Experimental results were depicted in Figure~\ref{fig:km-miss}, \ref{fig:lvq-miss}, \ref{fig:clr-miss}, \ref{fig:db-miss}, \ref{fig:bir-miss}, and \ref{fig:cure-miss}.

Based on the results, we had the following observations. First, as shown in Table~\ref{table:sensibility}, for Precision, the $sensibility$ order is ``CURE $>$ CLARANS $>$ BIRCH $>$ K-Means $>$ DBSCAN $>$ LVQ''. For Recall, the $sensibility$ order is ``BIRCH $>$ CLARANS $>$ CURE $>$ K-Means $>$ DBSCAN $>$ LVQ''. For F-measure, the $sensibility$ order is ``CLARANS $>$ CURE $>$ BIRCH $>$ K-Means $>$ LVQ $>$ DBSCAN''. Thus, for Precision and Recall, the least sensitive algorithm is \textbf{LVQ}. This is because that LVQ is a supervised clustering algorithm on the basis of marked labels. Hence, there is little chance to be affected by missing values. For F-measure, the least sensitive algorithm is \textbf{DBSCAN}. This is due to the fact that DBSCAN eliminates all noise points at the beginning of the algorithm, which makes it more resistant to missing values. For Precision, the most sensitive algorithm is \textbf{CURE}. This is because that the location of representative points in CURE is easily effected by missing values, which causes inaccurate clustering results. For Recall, the most sensitive algorithm is \textbf{BIRCH}. This is due to the fact that missing data could impact the construction of CF tree in BIRCH, which directly leads to wrong clustering results. For F-measure, the most sensitive algorithm is \textbf{CLARANS}. This is because that the computation of cost difference in CLARANS is susceptible to missing values, which makes some points clustered incorrectly.

Second, as shown in Table~\ref{table:keeping}, for Precision, the $keeping$ $point$ order is ``LVQ $>$ K-Means $>$ DBSCAN $>$ BIRCH $>$ CURE $>$ CLARANS''. For Recall, the $keeping$ $point$ order is ``LVQ $=$ DBSCAN $>$ K-Means $>$ BIRCH $>$ CURE $>$ CLARANS''. For F-measure, the $keeping$ $point$ order is ``LVQ $>$ K-Means $>$ DBSCAN $>$ BIRCH $>$ CURE $>$ CLARANS''. Therefore, the most incompleteness-tolerant algorithm is \textbf{LVQ}. This is because that LVQ is a supervised clustering algorithm based on marked labels. Hence, there is little chance for it to be affected by missing values. The least incompleteness-tolerant algorithm is \textbf{CLARANS}. This is due to the fact that the computation of cost difference in CLARANS is susceptible to missing data, which causes inaccurate clustering results.

Third, as data size increases, running time of algorithms fluctuates more. This is because that as data size rises, the amount of missing data becomes larger, which introduces more uncertainty to algorithms. Accordingly, the uncertainty of running time increases.

\subsubsection{Clustering - Varying Inconsistent Rate}
To evaluate inconsistent-data impacts on clustering algorithms, we injected inconsistent values to original datasets randomly according to consistency rules on the given data. Inconsistent rate was varied from 10\% to 50\%. Experimental results were depicted in Figure~\ref{fig:km-incons}, \ref{fig:lvq-incons}, \ref{fig:clr-incons}, \ref{fig:db-incons}, \ref{fig:bir-incons}, and \ref{fig:cure-incons}.

Based on the results, we had the following observations. First, for well-performed algorithms (Precision/Recall/F-measure is larger than 80\% on original datasets), as data size increases, Precision, Recall, or F-measure of algorithms fluctuates more widely, except \textbf{DBSCAN}. This is because that the amount of inconsistent values becomes larger as data size rises. The increasing incorrect data have more effect on clustering process. However, DBSCAN discards noise points at the beginning of the algorithm. When data size rises, the number of clean data becomes larger. Accordingly, the proportion of eliminated points reduces, which has less impact on DBSCAN.

Second, as shown in Table~\ref{table:sensibility}, for Precision, the $sensibility$ order is ``K-Means $>$ CLARANS $>$ CURE $>$ BIRCH $>$ LVQ $>$ DBSCAN''. For Recall, the $sensibility$ order is ``CURE $>$ K-Means $>$ CLARANS $>$ BIRCH $>$ LVQ $>$ DBSCAN''. For F-measure, the $sensibility$ order is ``K-Means $>$ CURE $>$ CLARANS $>$ LVQ $>$ BIRCH $>$ DBSCAN''. Thus, the least sensitive algorithm is \textbf{DBSCAN}. The reason is similar as that of the least sensitive algorithm varying missing rate. For Precision and F-measure, the most sensitive algorithm is \textbf{K-Means}. This is due to the fact that the computation of centroids are susceptible to incorrect values, which causes wrong clustering results. For Recall, the most sensitive algorithm is \textbf{CURE}. The reason is similar as that of the most sensitive algorithm varying missing rate.

Third, as shown in Table~\ref{table:keeping}, for Precision, the $keeping$ $point$ order is ``DBSCAN $>$ K-Means $=$ LVQ $>$ CLARANS $>$ BIRCH $=$ CURE''. For Recall, the $keeping$ $point$ order is ``DBSCAN $>$ BIRCH $>$ K-Means $>$ CLARANS $=$ CURE $>$ LVQ''. For F-measure, the $keeping$ $point$ order is ``DBSCAN $>$ BIRCH $>$ K-Means $>$ LVQ $>$ CLARANS $>$ CURE''. Therefore, the most inconsistency-tolerant algorithm is \textbf{DBSCAN}. This is because that DBSCAN eliminates all noise points at the beginning of the algorithm, which makes it more resistant to inconsistent data. For Precision, the least inconsistency-tolerant algorithms are \textbf{BIRCH} and \textbf{CURE}. For Recall, the least inconsistency-tolerant algorithm is \textbf{LVQ}. For F-measure, the least inconsistency-tolerant algorithm is \textbf{CURE}. These are due to the fact that the distance computation of these algorithms is susceptible to incorrect values, which causes inaccurate clustering results.

Fourth, the observation of running time on experiments varying missing rate was still true when the inconsistent rate was varied.

\subsubsection{Clustering - Varying Conflicting Rate}
To evaluate impacts of conflicting data on clustering algorithms, we injected conflicting values to original datasets randomly varying conflicting rate from 10\% to 50\%. Experimental results were depicted in Figure~\ref{fig:km-conf}, \ref{fig:lvq-conf}, \ref{fig:clr-conf}, \ref{fig:db-conf}, \ref{fig:bir-conf}, and \ref{fig:cure-conf}.

Based on the results, we had the following observations. First, as shown in Table~\ref{table:sensibility}, for Precision, the $sensibility$ order is ``CURE $>$ K-Means $>$ CLARANS $>$ DBSCAN $>$ BIRCH $>$ LVQ''. For Recall, the $sensibility$ order is ``CURE $>$ CLARANS $>$ BIRCH $>$ K-Means $>$ LVQ $>$ DBSCAN''. For F-measure, the $sensibility$ order is ``CURE $>$ K-Means $>$ CLARANS $>$ LVQ $>$ BIRCH $>$ DBSCAN''. Thus, for Precision, the least sensitive algorithm is \textbf{LVQ}. The reason is similar as that of the least sensitive algorithm varying missing rate. For Recall and F-measure, the least sensitive algorithm is \textbf{DBSCAN}. The reason has been discussed in Section~\ref{subsec:clus-miss}. The most sensitive algorithm is \textbf{CURE}. The reason is similar as that of the most sensitive algorithm varying missing rate.

Second, as shown in Table~\ref{table:keeping}, for Precision, the $keeping$ $point$ order is ``BIRCH $>$ K-Means $=$ LVQ $>$ DBSCAN $>$ CLARANS $>$ CURE''. For Recall, the $keeping$ $point$ order is ``DBSCAN $>$ LVQ $>$ K-Means $>$ CLARANS $=$ BIRCH $=$ CURE''. For F-measure, the $keeping$ $point$ order is ``LVQ $>$ K-Means $=$ BIRCH $>$ DBSCAN $>$ CLARANS $>$ CURE''. Therefore, for Precision, the most conflict-tolerant algorithm is \textbf{BIRCH}. This is because that conflicting data contain correct ones and incorrect ones, which makes the construction of CF tree insusceptible to incorrect values. For Recall, the most conflict-tolerant algorithm is \textbf{DBSCAN}. The reason is similar as that of the most inconsistency-tolerant algorithm varying inconsistent rate. For F-measure, the most conflict-tolerant algorithm is \textbf{LVQ}. The reason is similar as that of the most incompleteness-tolerant algorithm varying missing rate. The least conflict-tolerant algorithm is \textbf{CURE}. This is due to the fact that the location of representative points in CURE could be easily affected by conflicting values, which makes data points clustered inaccurately.

Third, the observation of running time on experiments varying missing rate was still true when the conflicting rate was varied.

\subsubsection{Discussion}
In clustering experiments, we first found that dirty-data impacts are related to error type and error rate. Thus, the error rate of each error type in the given data is necessary to be detected. Second, we observed that for algorithms whose Precision/Recall/F-measure is larger than 80\% on original datasets, Precision, Recall, or F-measure of algorithms becomes unstable as data size rises, except DBSCAN. Since the parameter $k$ in $keeping$ $point$ was set as 10\%, candidate algorithms of which Precision/Recall/F-measure is larger than 70\% are acceptable. Third, we prefer to choose stable algorithms. Hence, DBSCAN is suitable for larger data size. Fourth, we compared the fluctuation degrees of clustering algorithms in our experiments. When dirty data exist, the algorithm with the least degree is the most stable. Fifth, beyond $keeping$ $point$, the accuracy of the selected algorithm becomes unacceptable. Thus, the error rate of each type needs to be controlled within its $keeping$ $point$.

%% file: Appendix.tex
\begin{figure*}[!htb]
\centering
\subfigure{
\includegraphics[width=1.4in,height=1.0in]{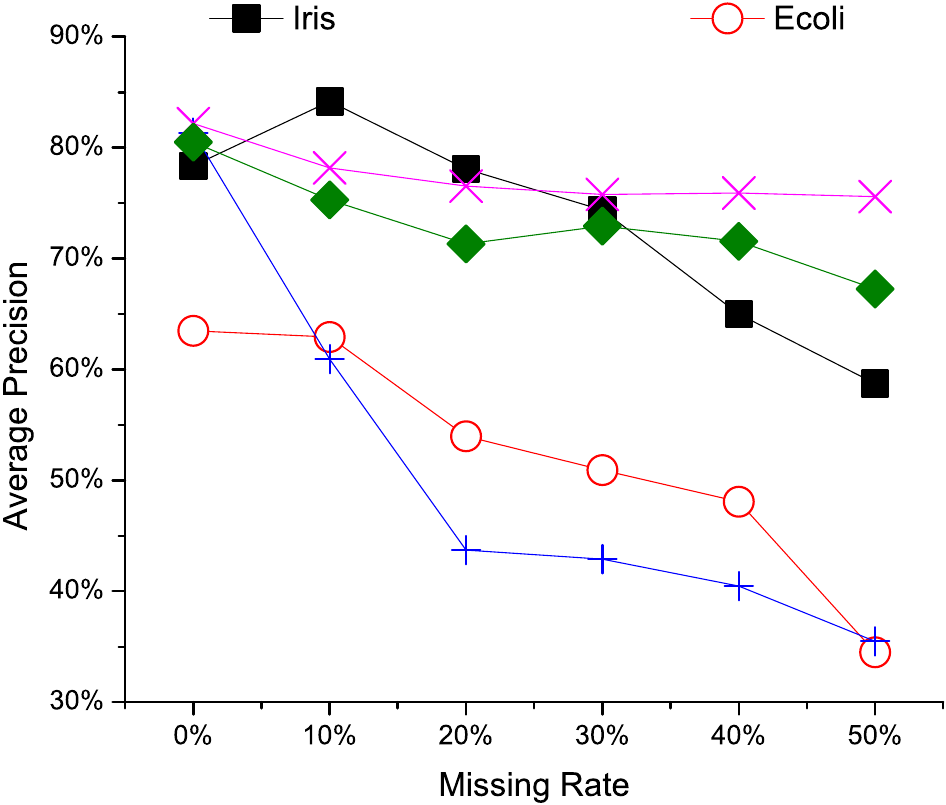}
\label{fig:dt-miss-p}
}
\subfigure{
\includegraphics[width=1.4in,height=1.0in]{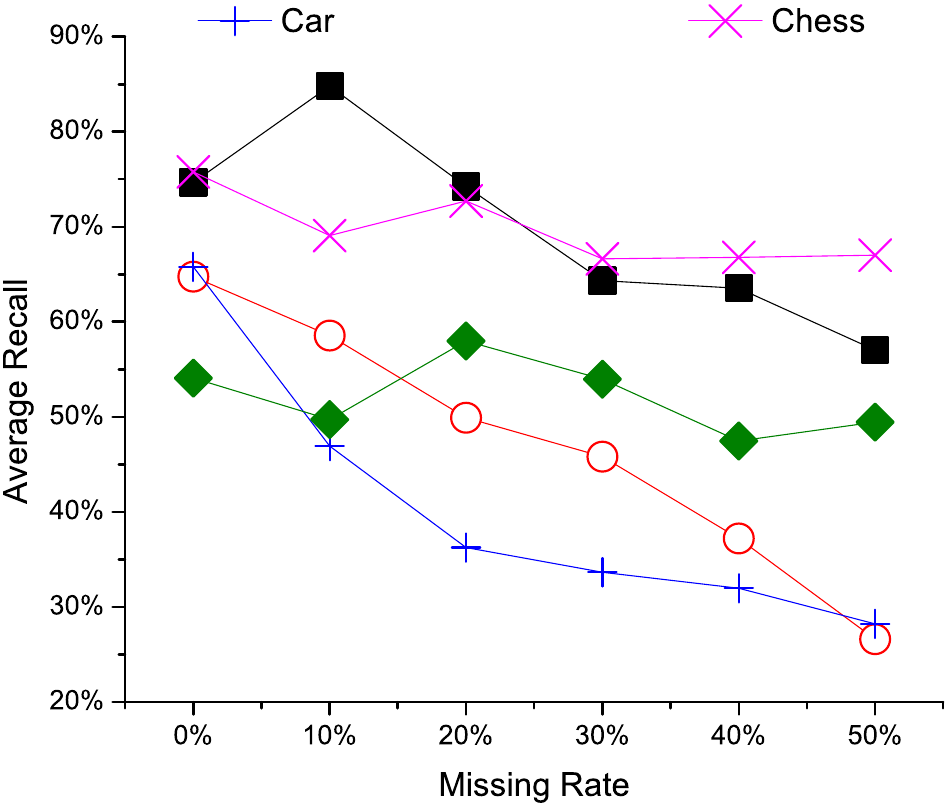}
\label{fig:dt-miss-r}
}
\subfigure{
\includegraphics[width=1.4in,height=1.0in]{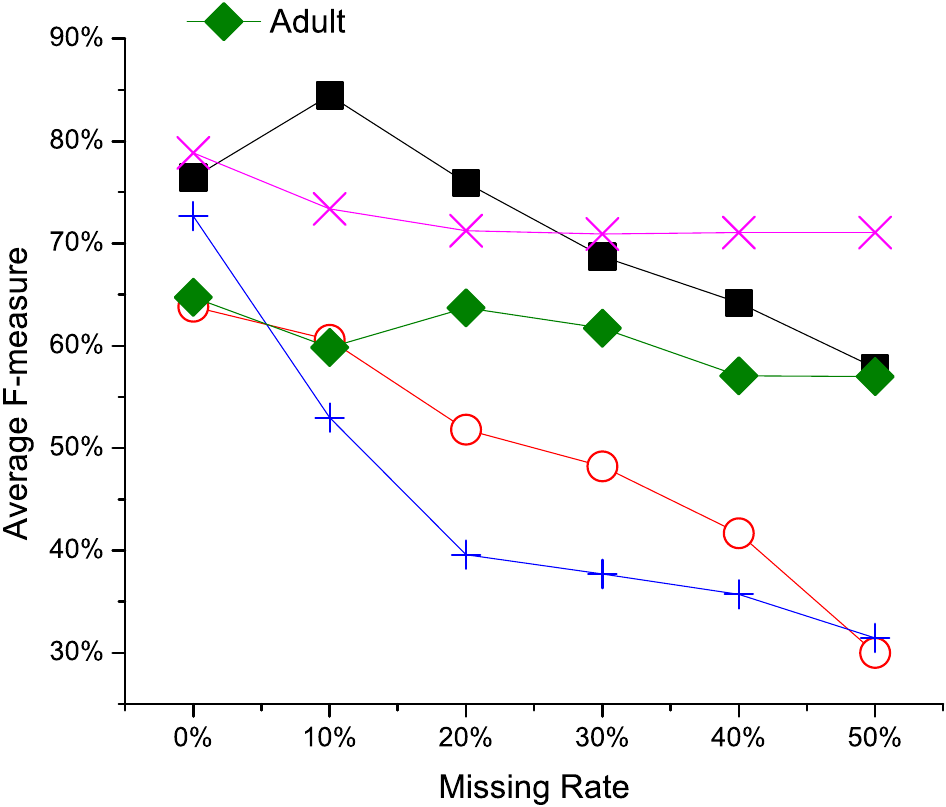}
\label{fig:dt-miss-f}
}
\subfigure{
\includegraphics[width=1.4in,height=1.0in]{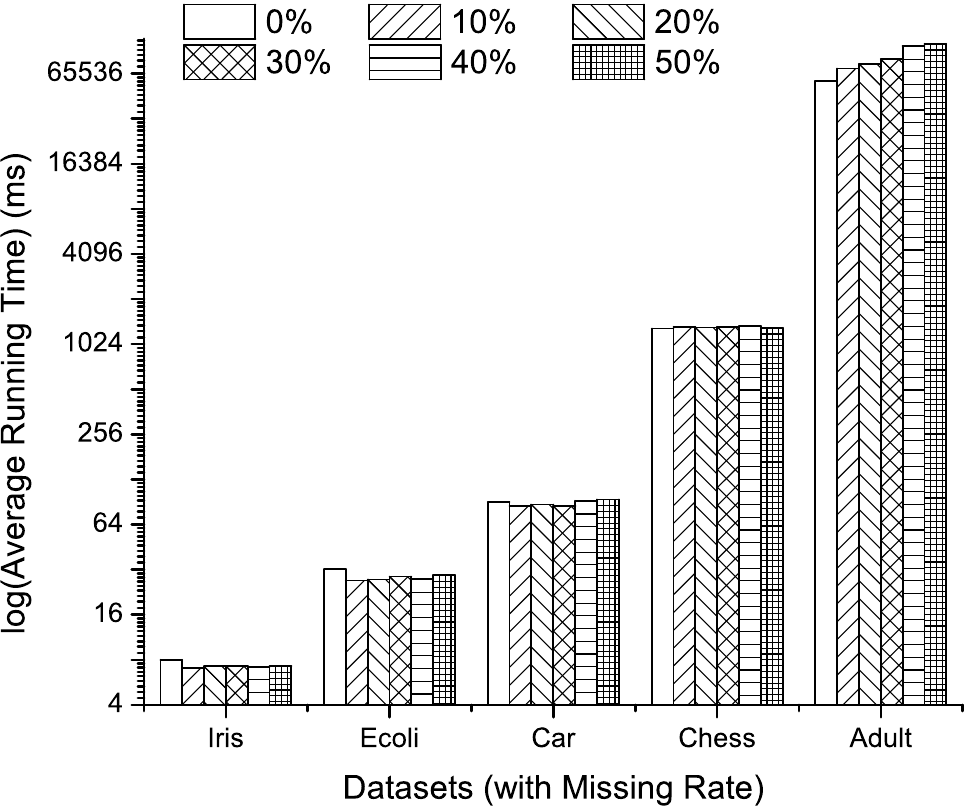}
\label{fig:dt-miss-t}
}
\vspace{-2mm}
\caption{Results on Classification for Decision Tree Algorithm: Varying Missing Rate.}
\vspace{-2mm}
\label{fig:dt-miss}
\end{figure*}

\begin{figure*}[!htb]
\centering
\subfigure{
\includegraphics[width=1.4in,height=1.0in]{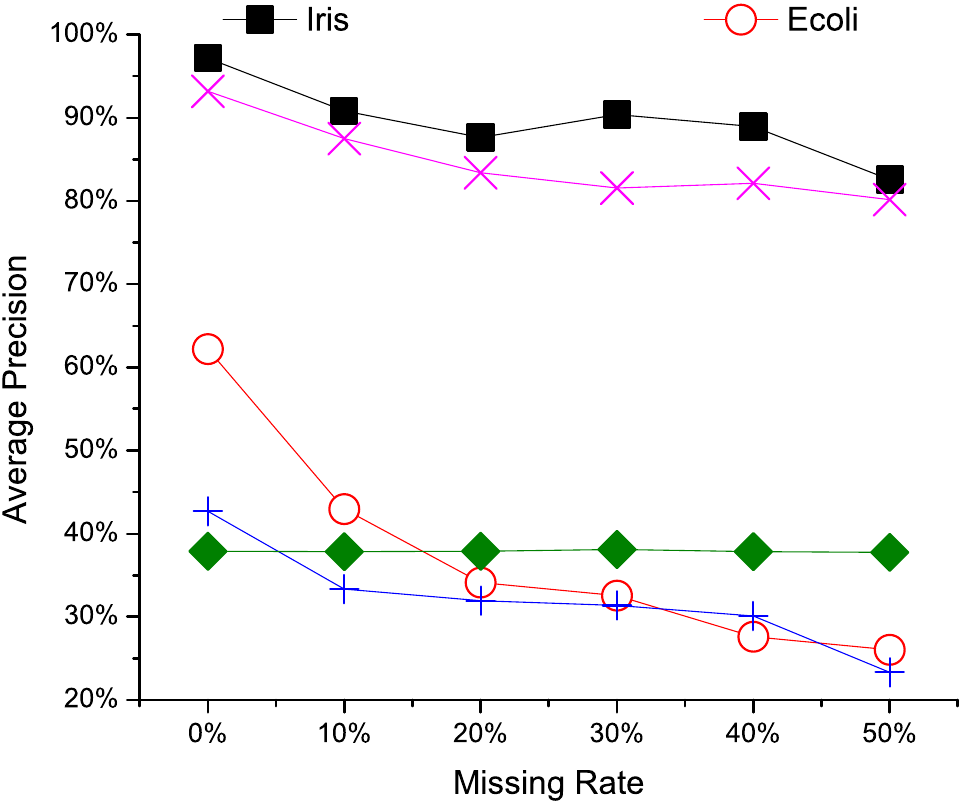}
\label{fig:knn-miss-p}
}
\subfigure{
\includegraphics[width=1.4in,height=1.0in]{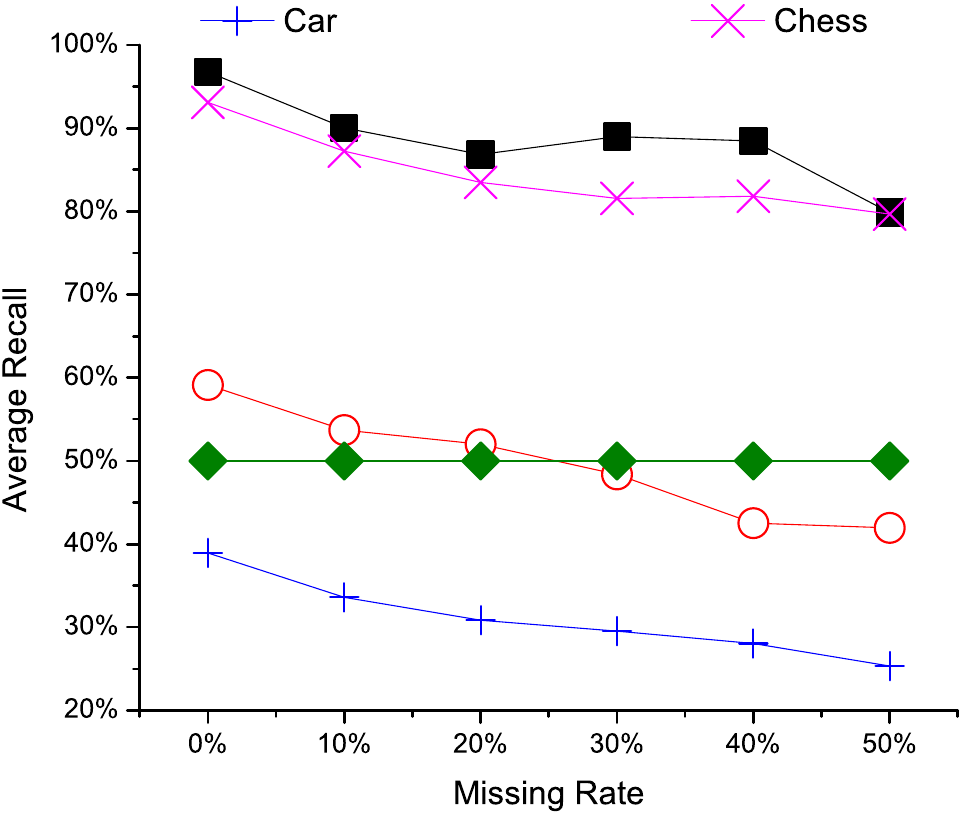}
\label{fig:knn-miss-r}
}
\subfigure{
\includegraphics[width=1.4in,height=1.0in]{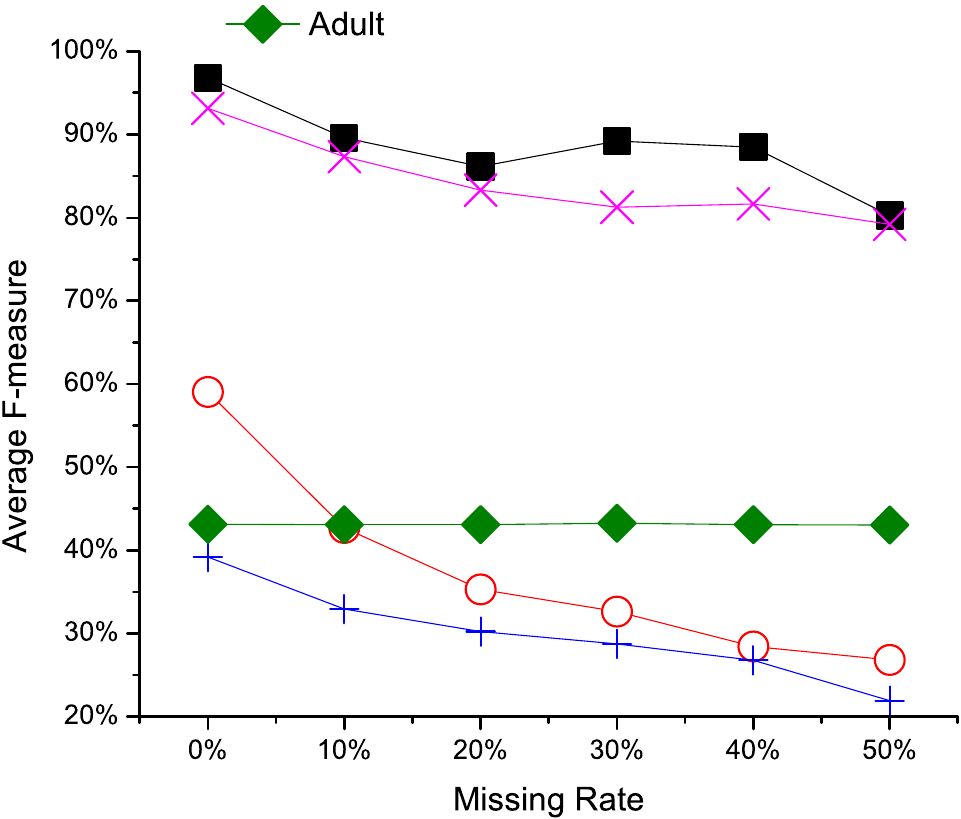}
\label{fig:knn-miss-f}
}
\subfigure{
\includegraphics[width=1.4in,height=1.0in]{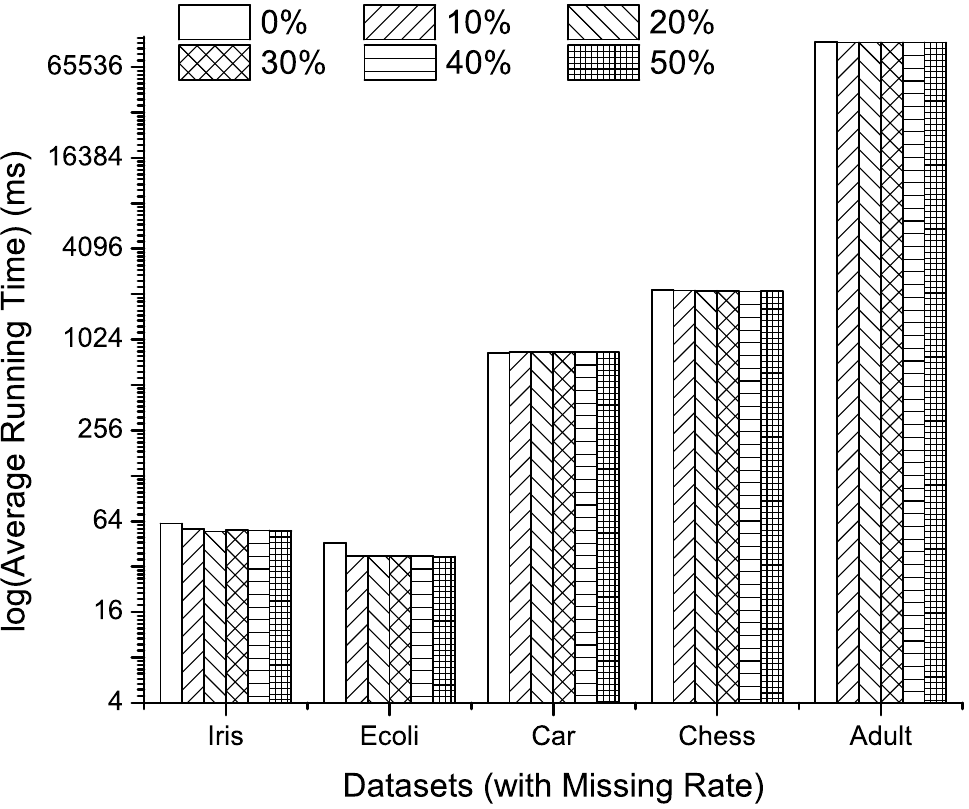}
\label{fig:knn-miss-t}
}
\vspace{-2mm}
\caption{Results on Classification for KNN Algorithm: Varying Missing Rate.}
\vspace{-2mm}
\label{fig:knn-miss}
\end{figure*}

\begin{figure*}[!htb]
\centering
\subfigure{
\includegraphics[width=1.4in,height=1.0in]{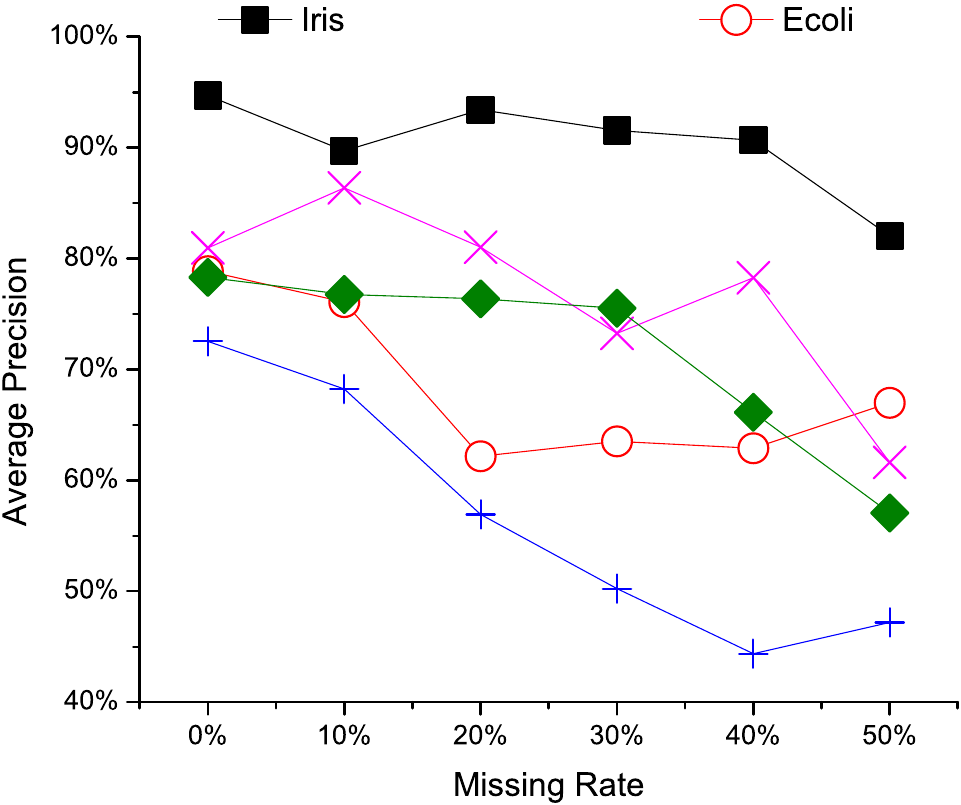}
\label{fig:nb-miss-p}
}
\subfigure{
\includegraphics[width=1.4in,height=1.0in]{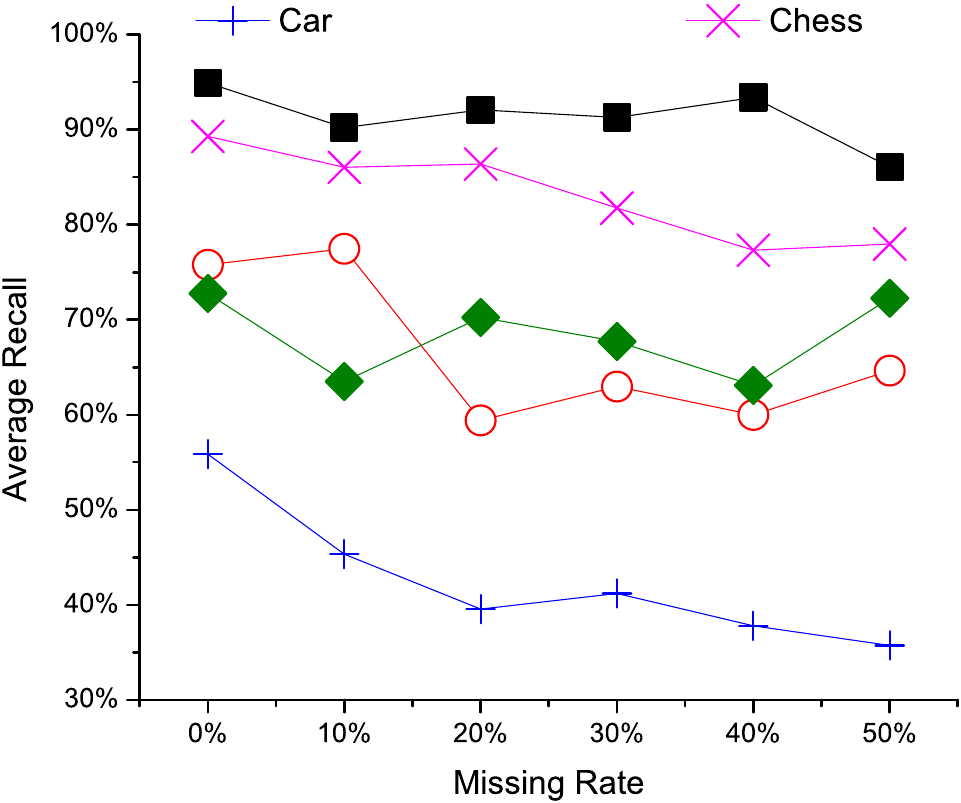}
\label{fig:nb-miss-r}
}
\subfigure{
\includegraphics[width=1.4in,height=1.0in]{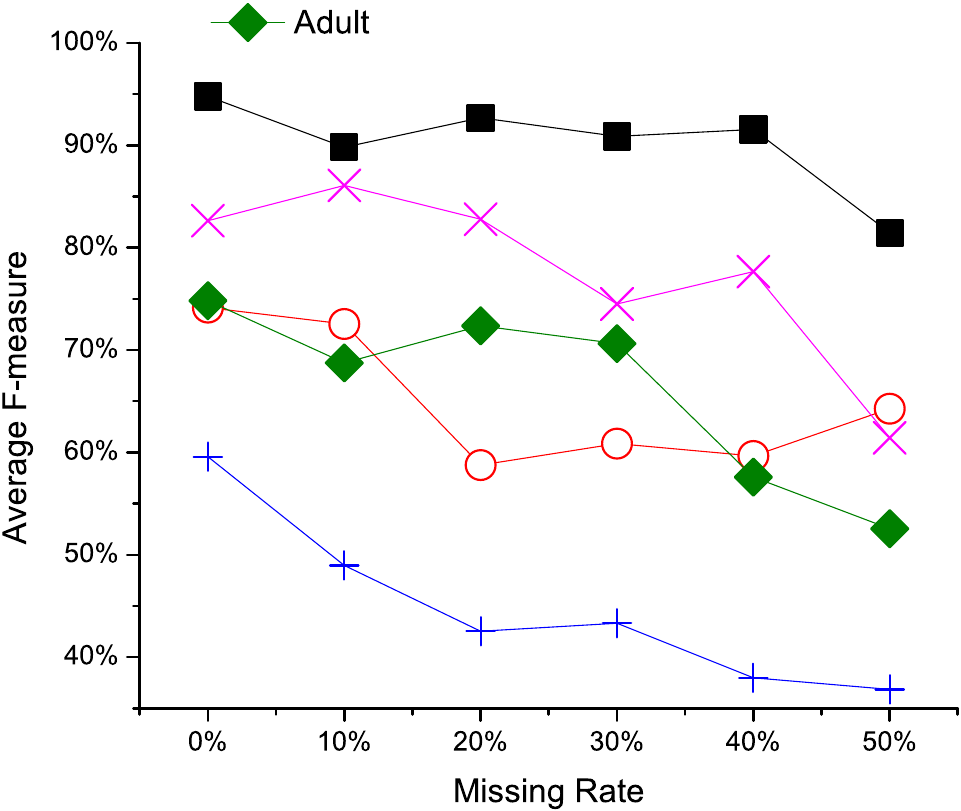}
\label{fig:nb-miss-f}
}
\subfigure{
\includegraphics[width=1.4in,height=1.0in]{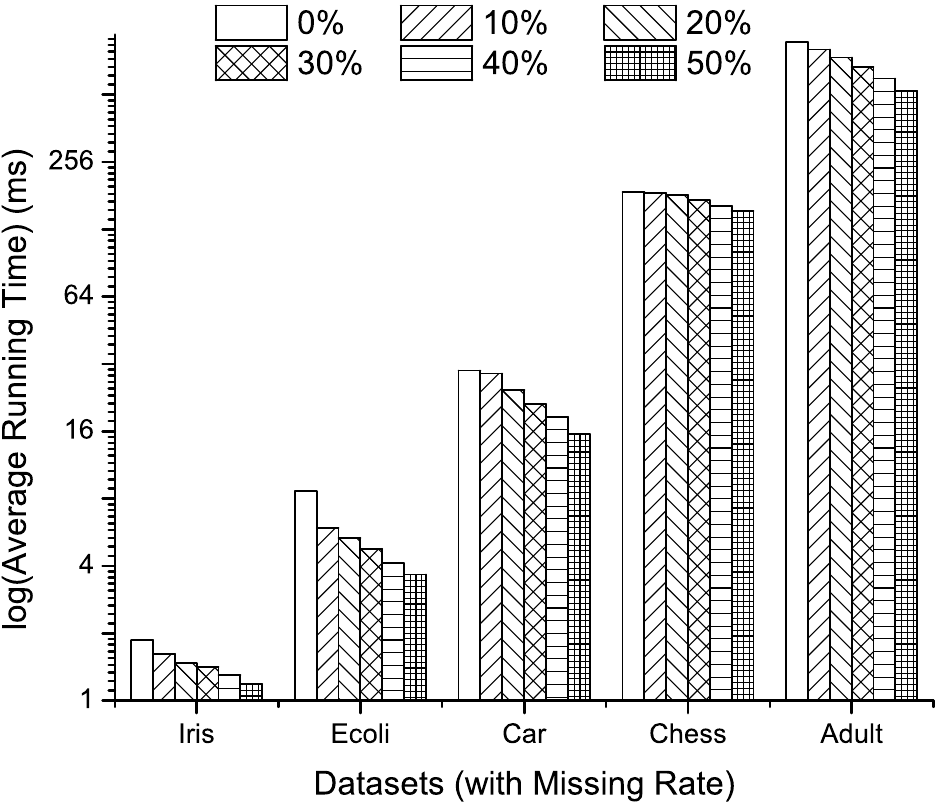}
\label{fig:nb-miss-t}
}
\vspace{-2mm}
\caption{Results on Classification for Naive Bayes Algorithm: Varying Missing Rate.}
\vspace{-2mm}
\label{fig:nb-miss}
\end{figure*}

\begin{figure*}[!htb]
\centering
\subfigure{
\includegraphics[width=1.4in,height=1.0in]{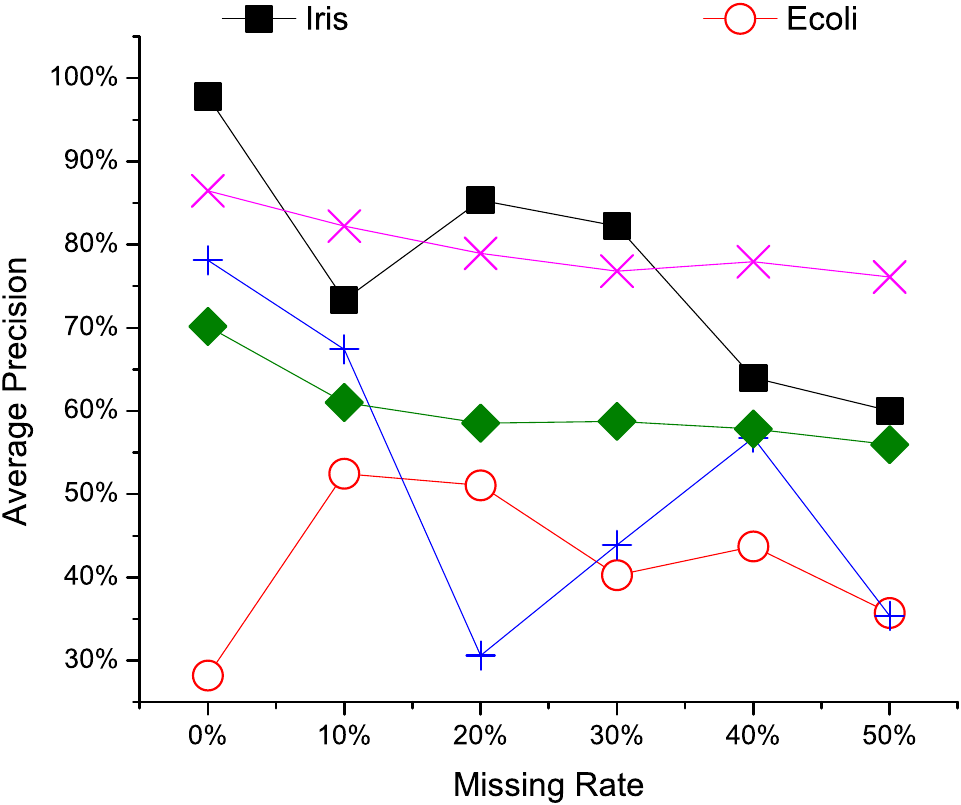}
\label{fig:bn-miss-p}
}
\subfigure{
\includegraphics[width=1.4in,height=1.0in]{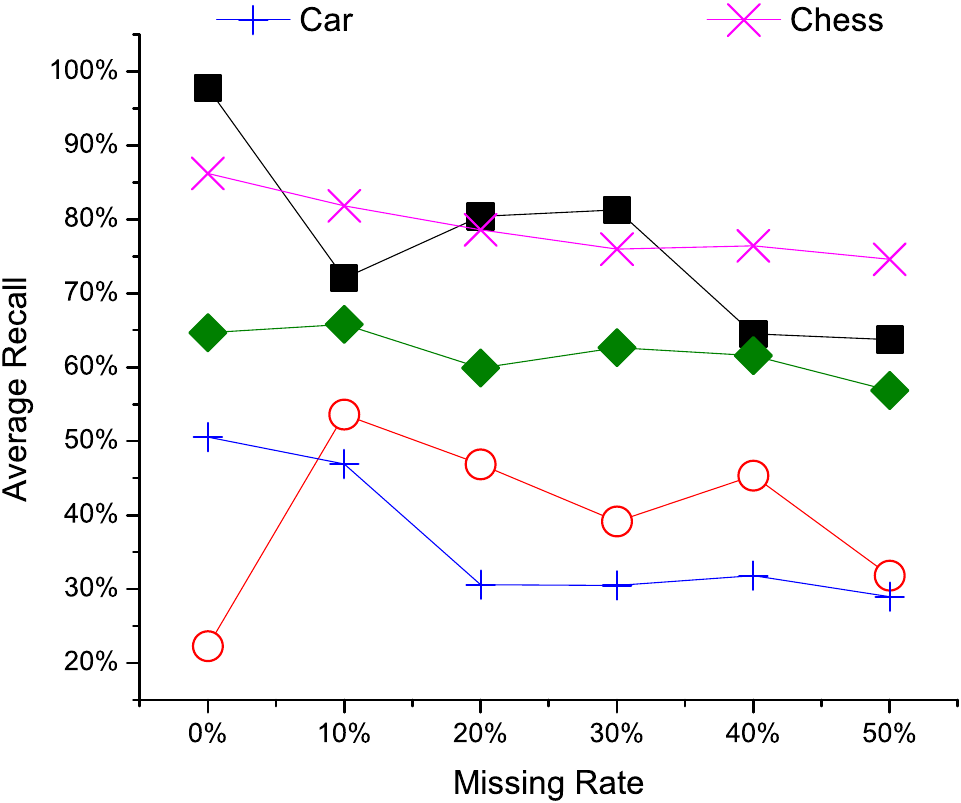}
\label{fig:bn-miss-r}
}
\subfigure{
\includegraphics[width=1.4in,height=1.0in]{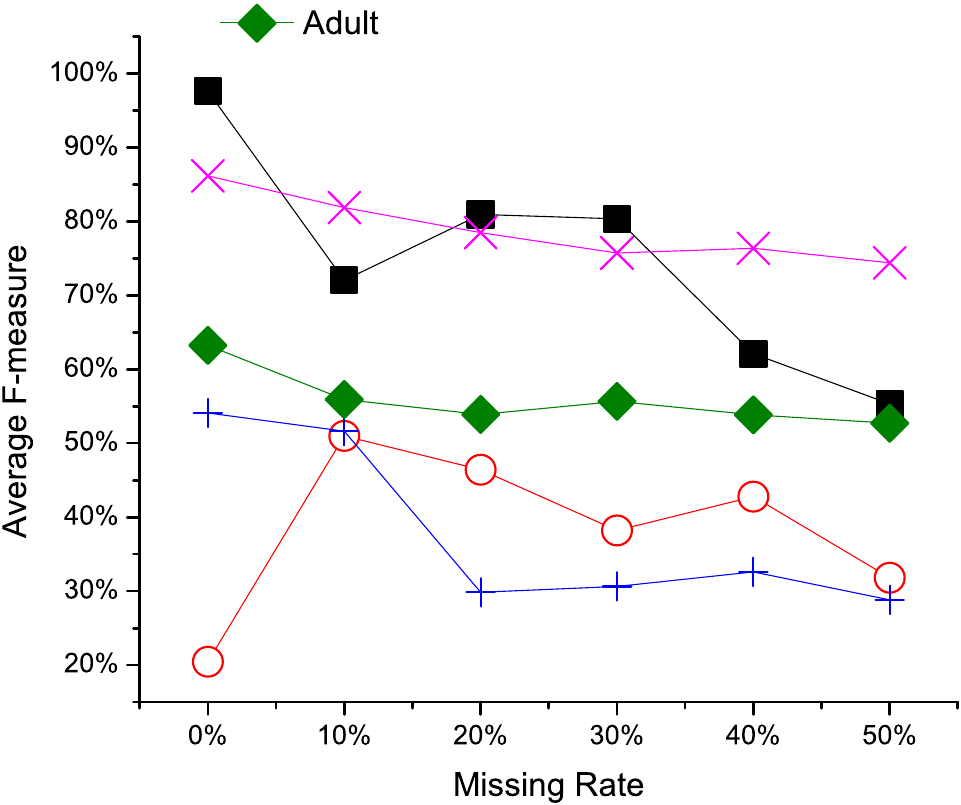}
\label{fig:bn-miss-f}
}
\subfigure{
\includegraphics[width=1.4in,height=1.0in]{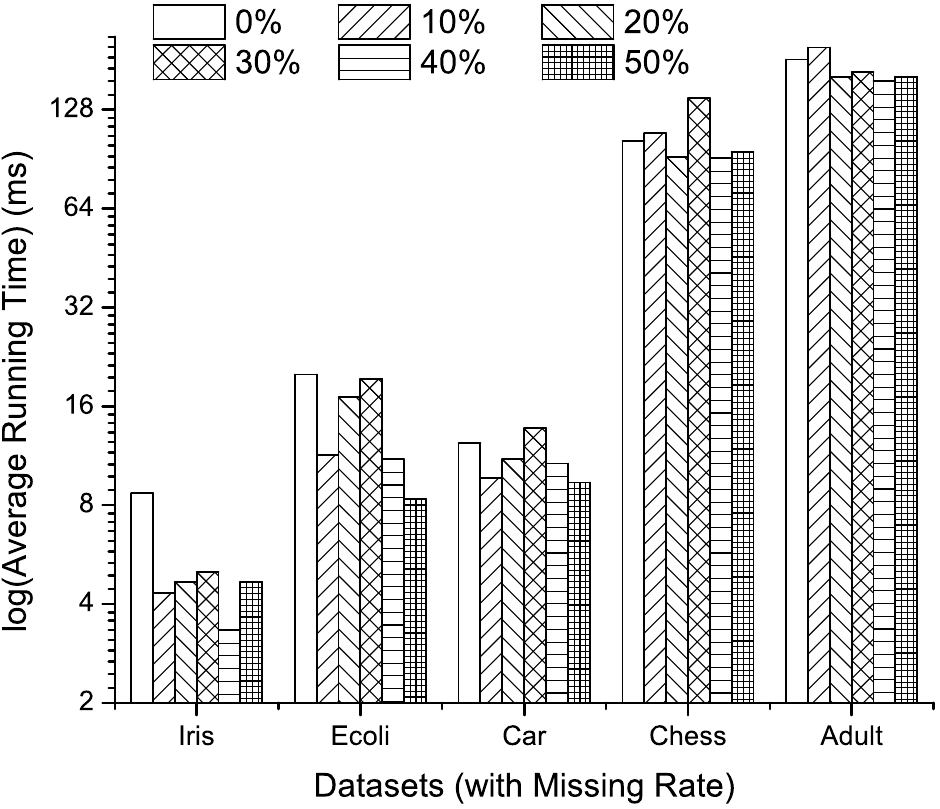}
\label{fig:bn-miss-t}
}
\vspace{-2mm}
\caption{Results on Classification for Bayesian Network Algorithm: Varying Missing Rate.}
\vspace{-2mm}
\label{fig:bn-miss}
\end{figure*}

\begin{figure*}[!htb]
\centering
\subfigure{
\includegraphics[width=1.4in,height=1.0in]{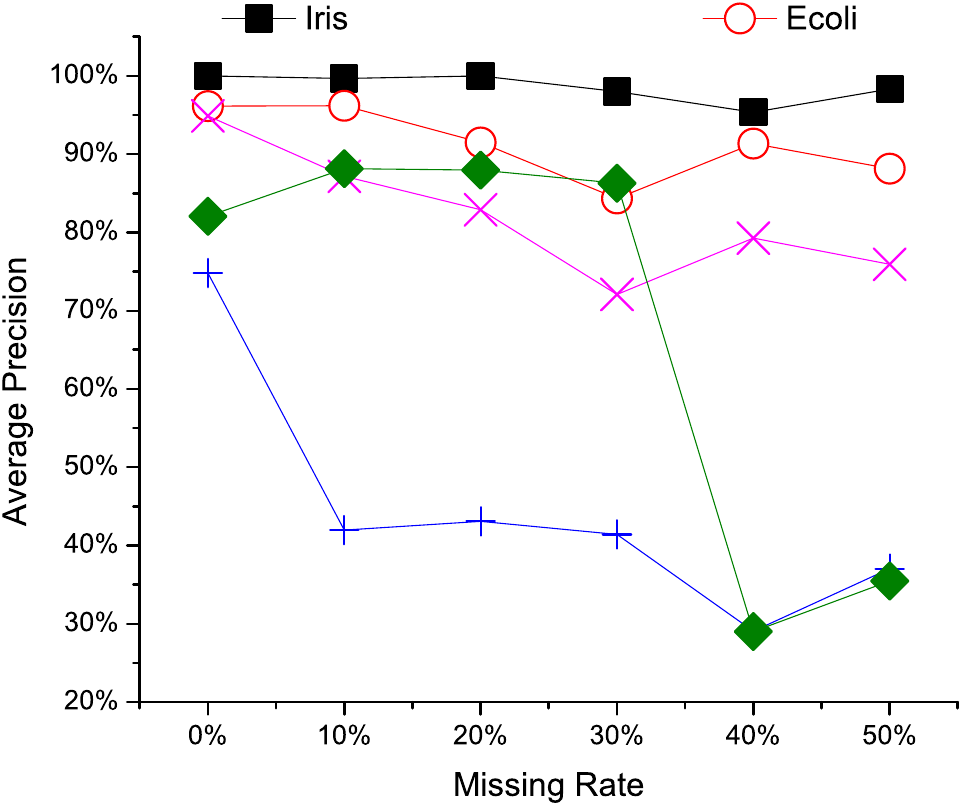}
\label{fig:log-miss-p}
}
\subfigure{
\includegraphics[width=1.4in,height=1.0in]{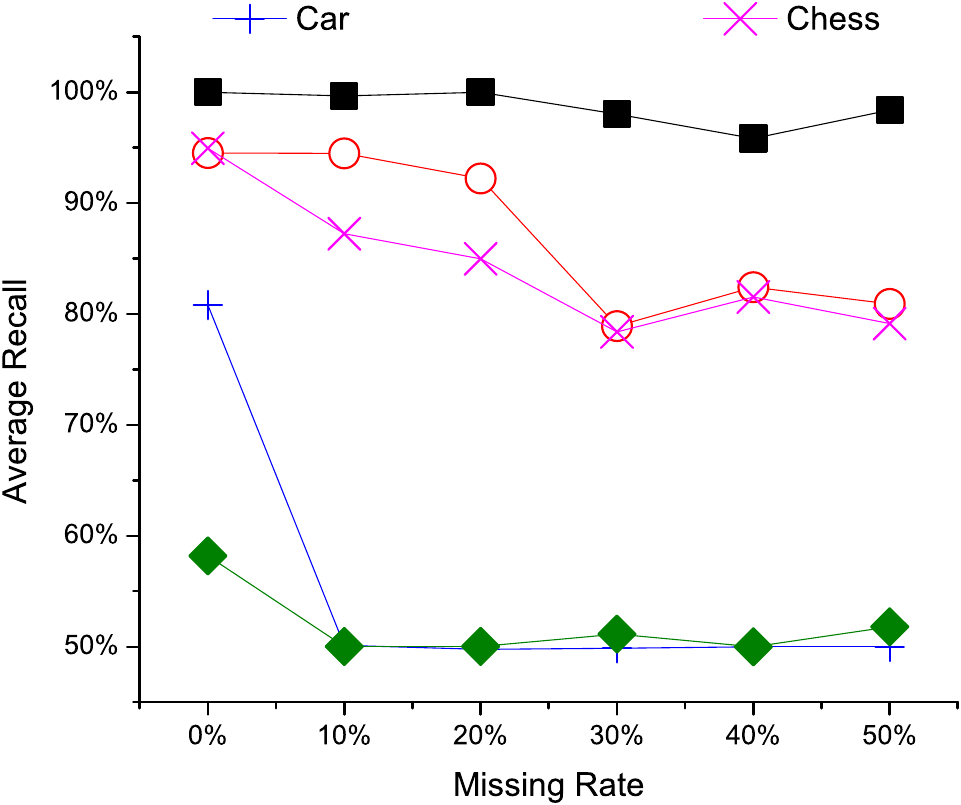}
\label{fig:log-miss-r}
}
\subfigure{
\includegraphics[width=1.4in,height=1.0in]{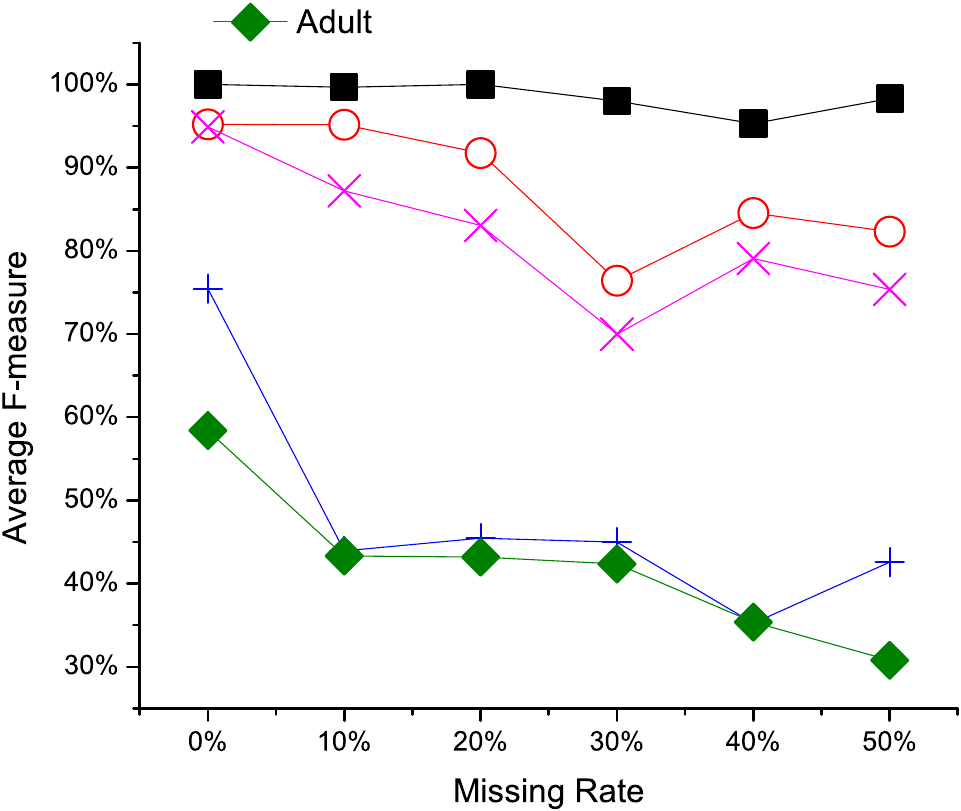}
\label{fig:log-miss-f}
}
\subfigure{
\includegraphics[width=1.4in,height=1.0in]{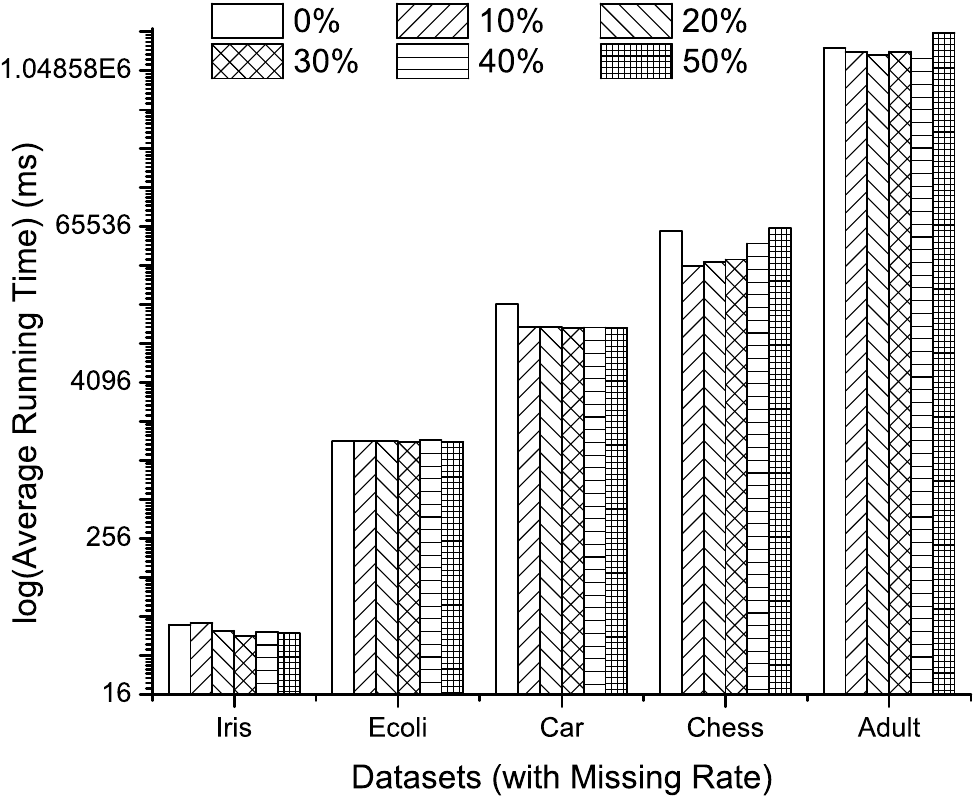}
\label{fig:log-miss-t}
}
\vspace{-2mm}
\caption{Results on Classification for Logistic Regression Algorithm: Varying Missing Rate.}
\vspace{-2mm}
\label{fig:log-miss}
\end{figure*}

\begin{figure*}[!htb]
\centering
\subfigure{
\includegraphics[width=1.4in,height=1.0in]{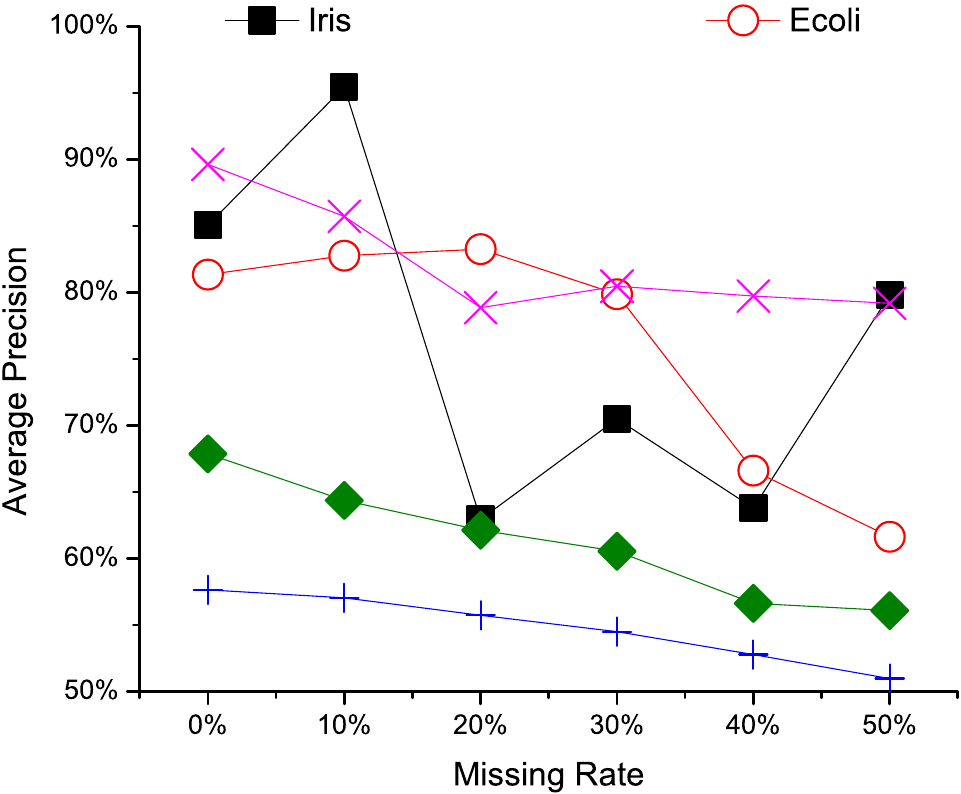}
\label{fig:ad-miss-p}
}
\subfigure{
\includegraphics[width=1.4in,height=1.0in]{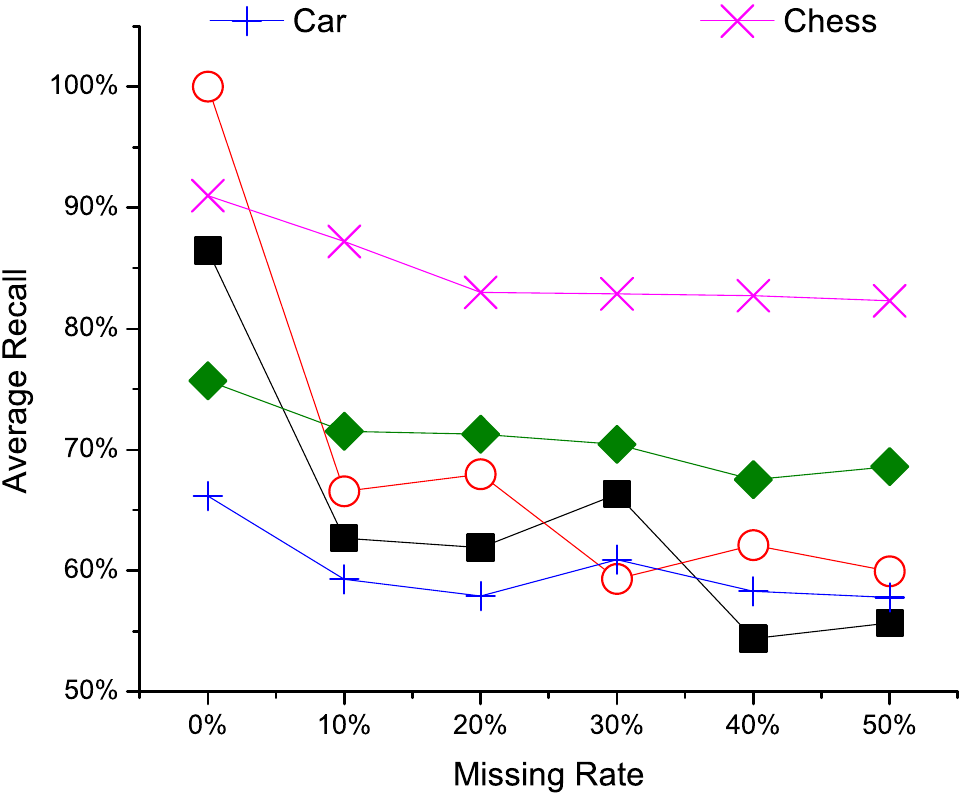}
\label{fig:ad-miss-r}
}
\subfigure{
\includegraphics[width=1.4in,height=1.0in]{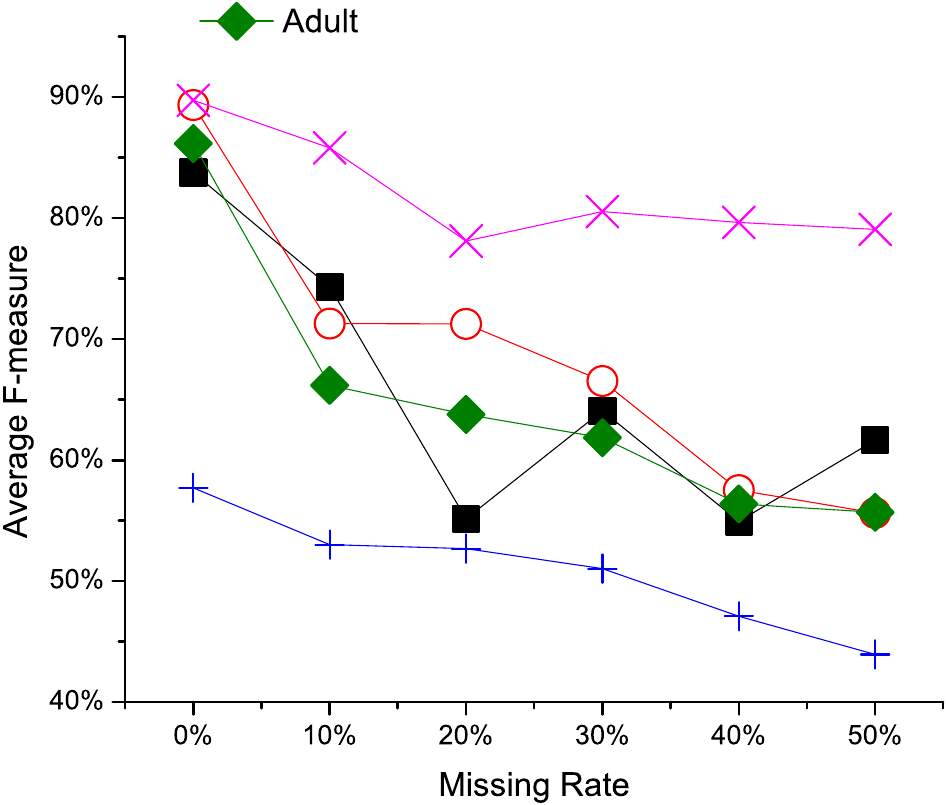}
\label{fig:ad-miss-f}
}
\subfigure{
\includegraphics[width=1.4in,height=1.0in]{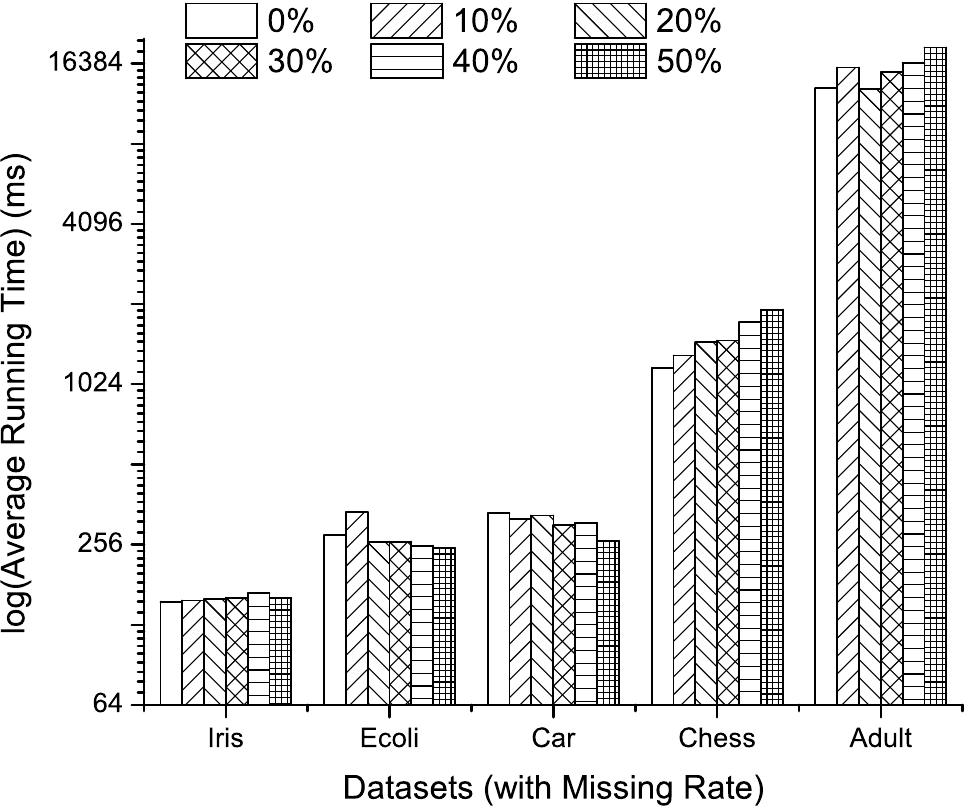}
\label{fig:ad-miss-t}
}
\vspace{-2mm}
\caption{Results on Classification for Random Forests Algorithm: Varying Missing Rate.}
\vspace{-2mm}
\label{fig:ad-miss}
\end{figure*}

\begin{figure*}[!htb]
\centering
\subfigure{
\includegraphics[width=1.4in,height=1.0in]{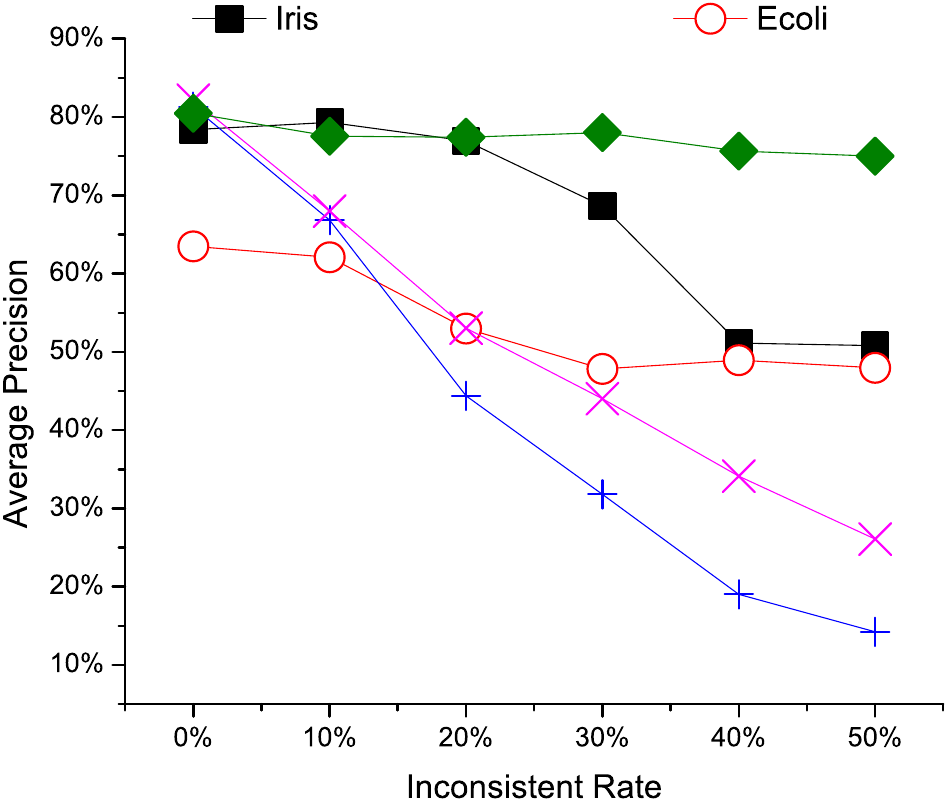}
\label{fig:dt-incons-p}
}
\subfigure{
\includegraphics[width=1.4in,height=1.0in]{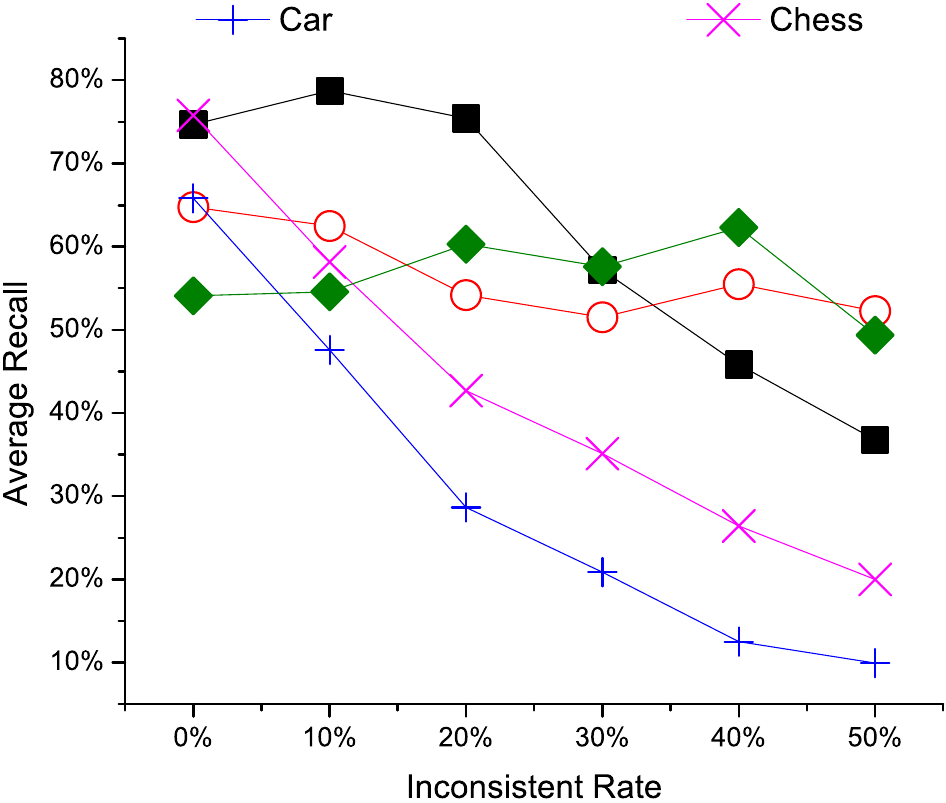}
\label{fig:dt-incons-r}
}
\subfigure{
\includegraphics[width=1.4in,height=1.0in]{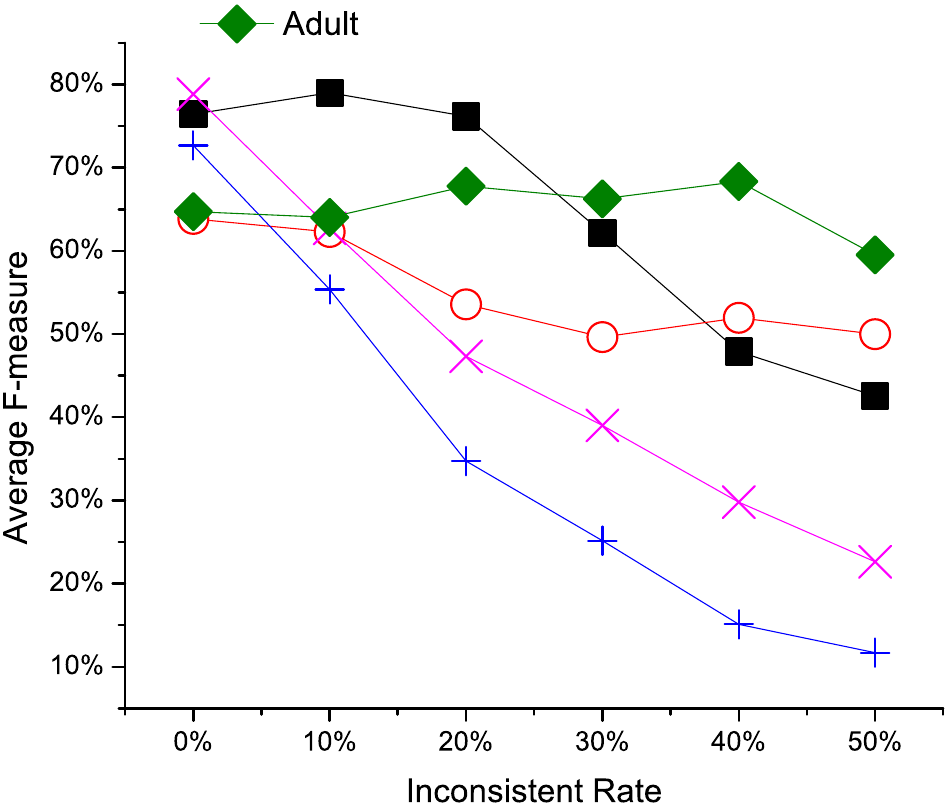}
\label{fig:dt-incons-f}
}
\subfigure{
\includegraphics[width=1.4in,height=1.0in]{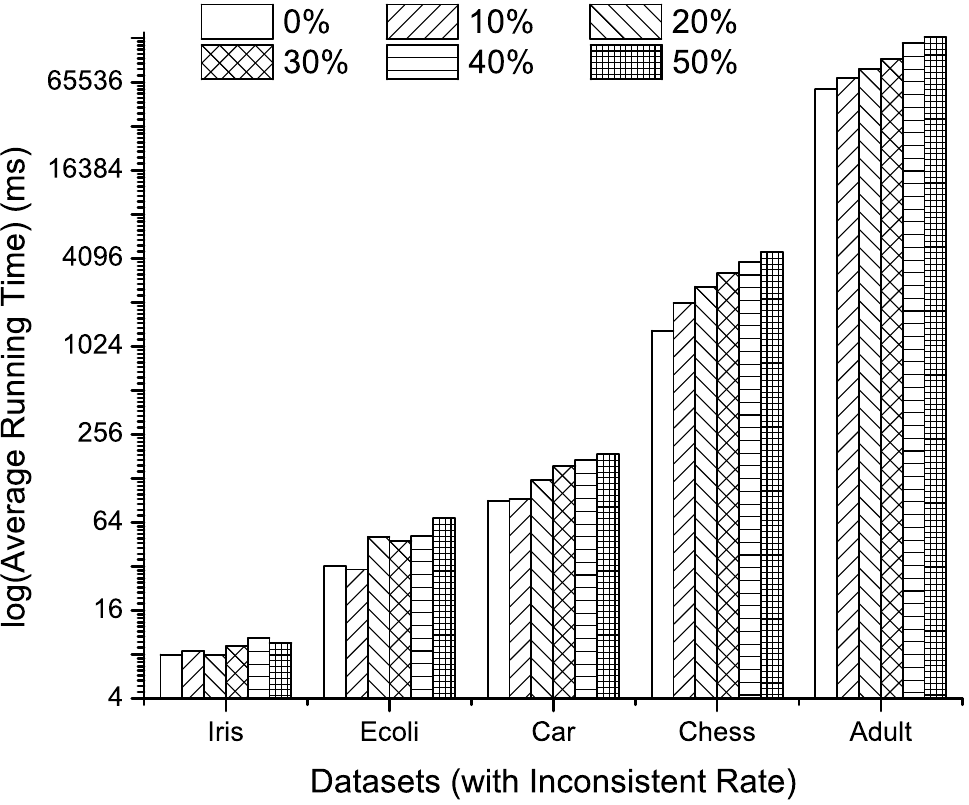}
\label{fig:dt-incons-t}
}
\vspace{-2mm}
\caption{Results on Classification for Decision Tree Algorithm: Varying Inconsistent Rate.}
\vspace{-2mm}
\label{fig:dt-incons}
\end{figure*}

\begin{figure*}[!htb]
\centering
\subfigure{
\includegraphics[width=1.4in,height=1.0in]{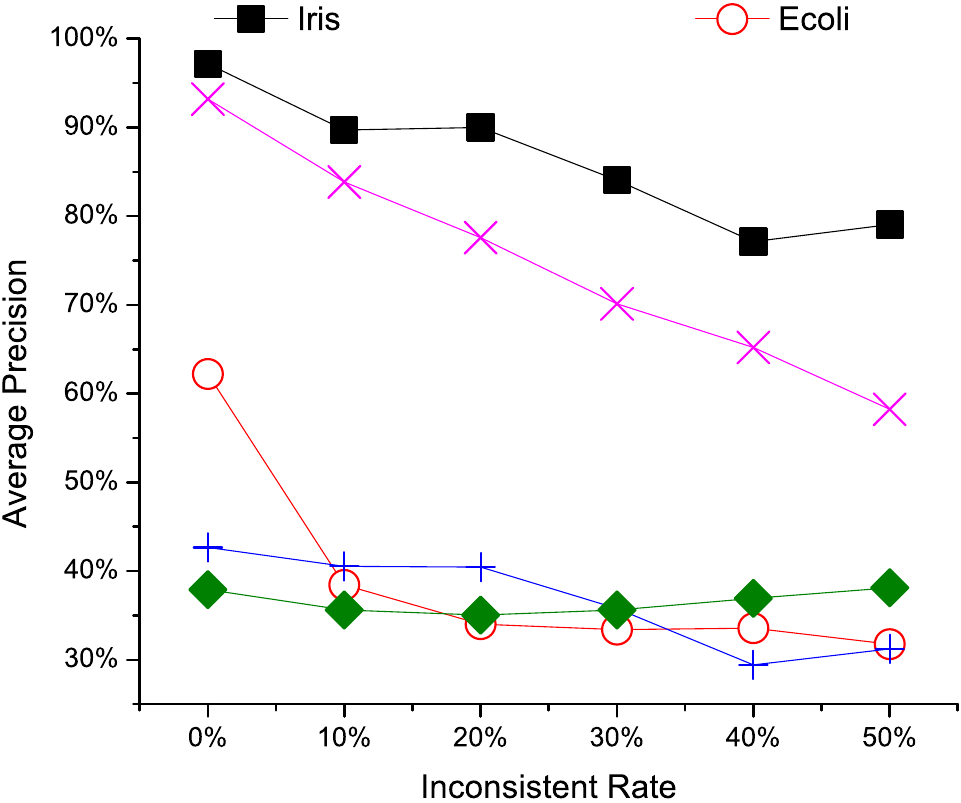}
\label{fig:knn-incons-p}
}
\subfigure{
\includegraphics[width=1.4in,height=1.0in]{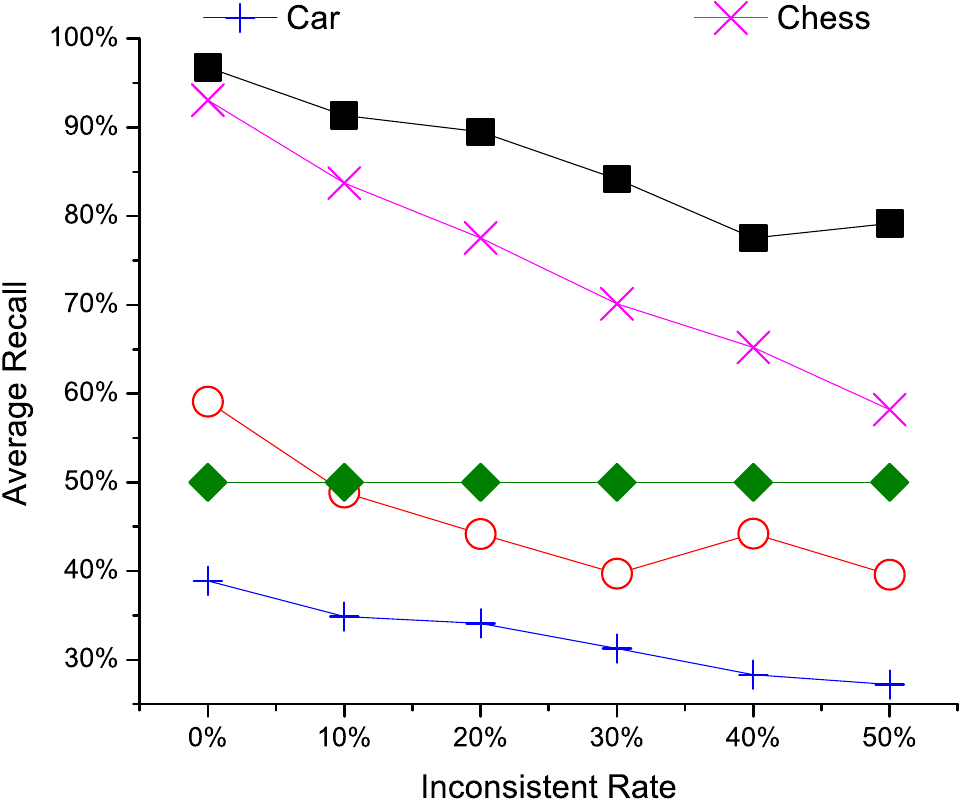}
\label{fig:knn-incons-r}
}
\subfigure{
\includegraphics[width=1.4in,height=1.0in]{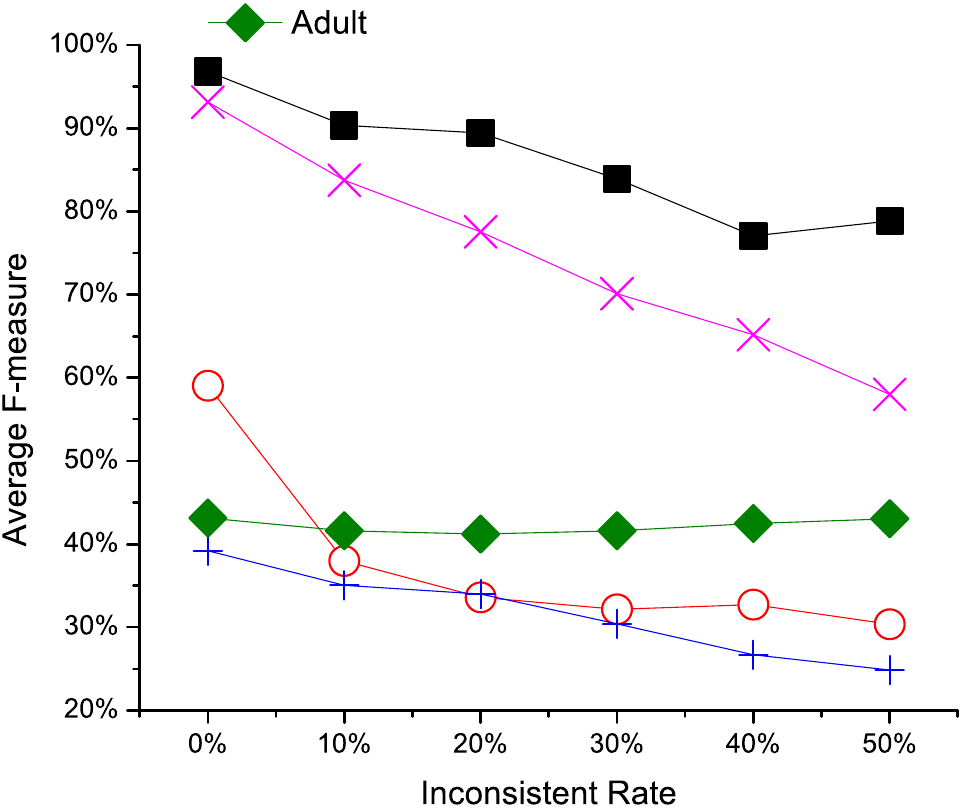}
\label{fig:knn-incons-f}
}
\subfigure{
\includegraphics[width=1.4in,height=1.0in]{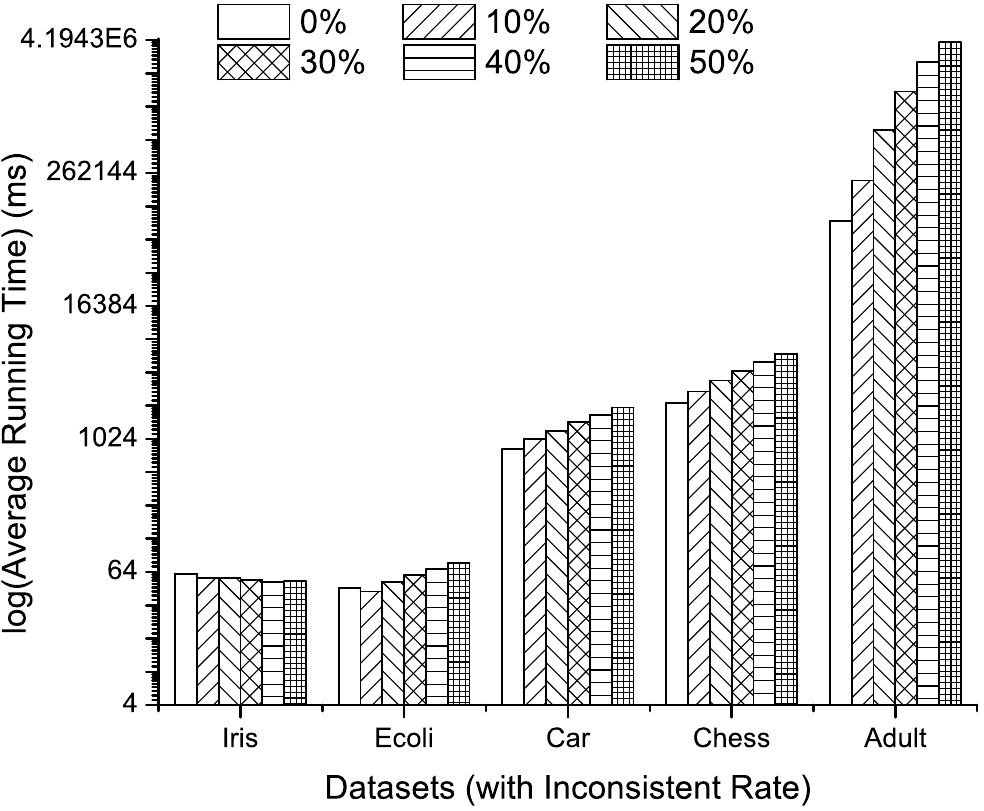}
\label{fig:knn-incons-t}
}
\vspace{-2mm}
\caption{Results on Classification for KNN Algorithm: Varying Inconsistent Rate.}
\vspace{-2mm}
\label{fig:knn-incons}
\end{figure*}

\begin{figure*}[!htb]
\centering
\subfigure{
\includegraphics[width=1.4in,height=1.0in]{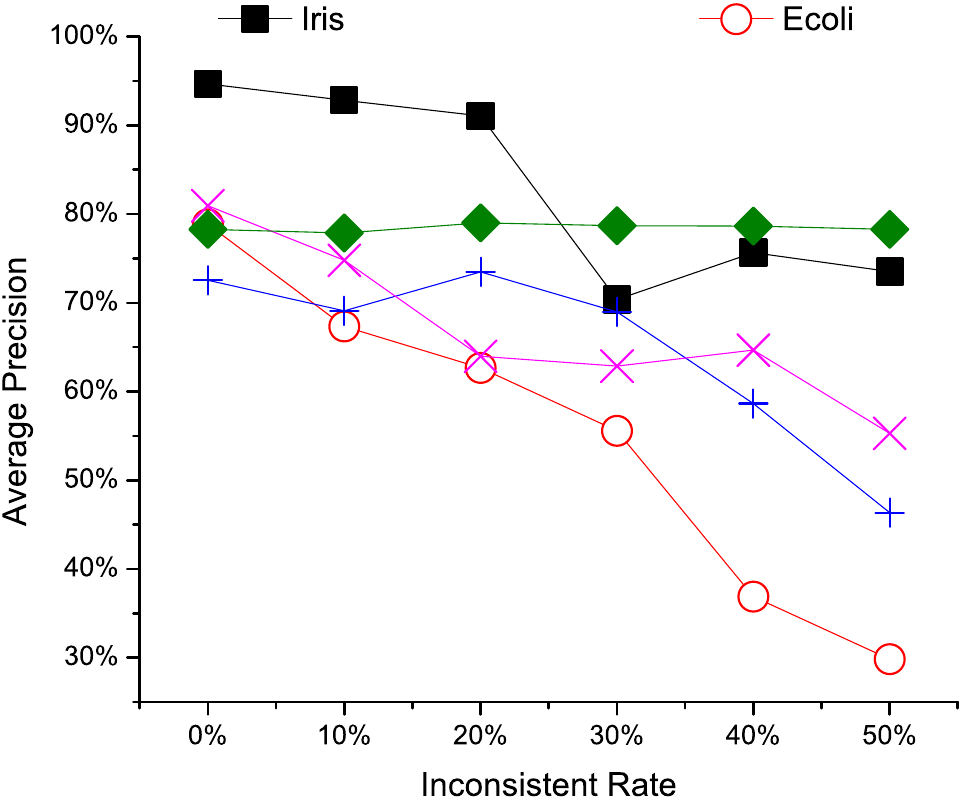}
\label{fig:nb-incons-p}
}
\subfigure{
\includegraphics[width=1.4in,height=1.0in]{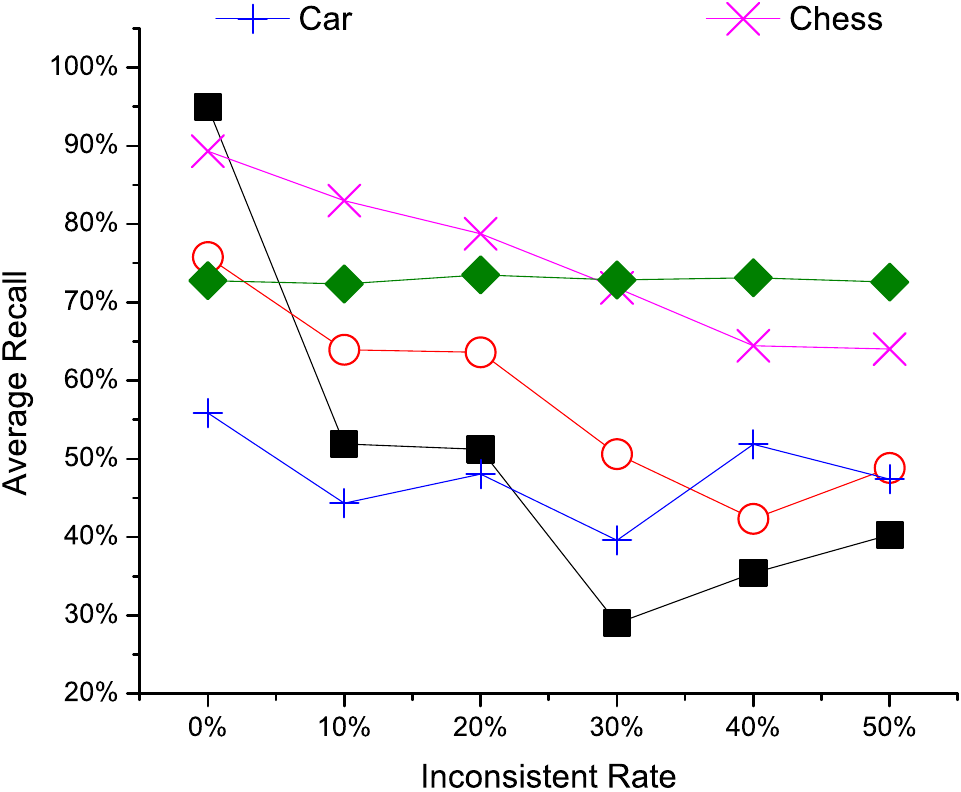}
\label{fig:nb-incons-r}
}
\subfigure{
\includegraphics[width=1.4in,height=1.0in]{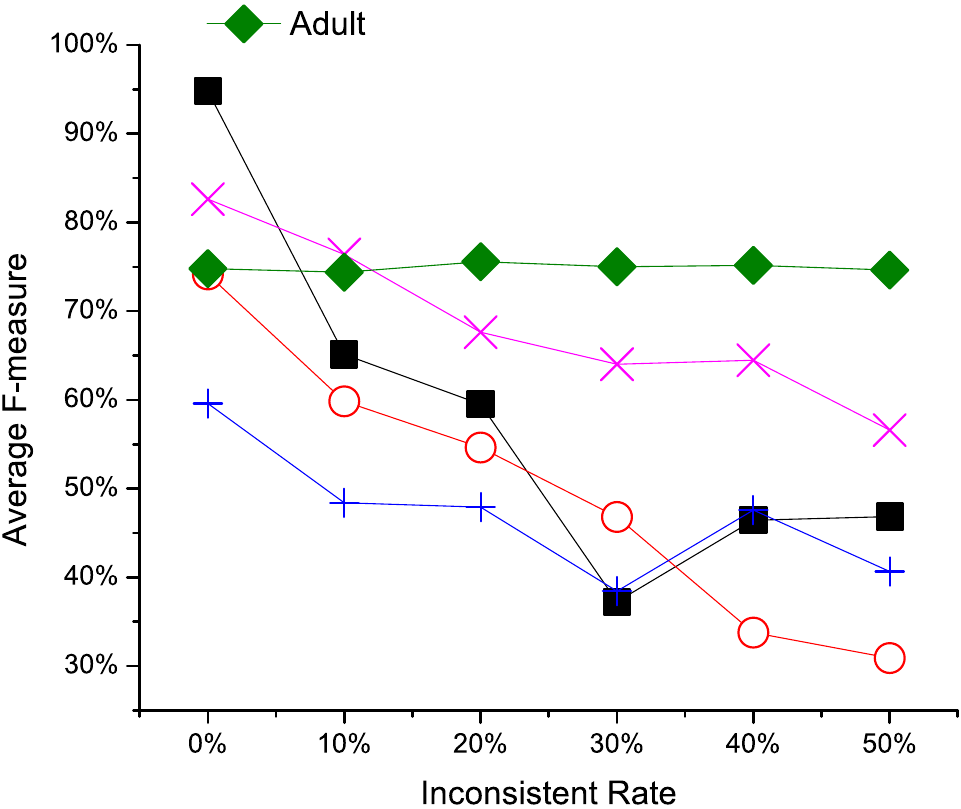}
\label{fig:nb-incons-f}
}
\subfigure{
\includegraphics[width=1.4in,height=1.0in]{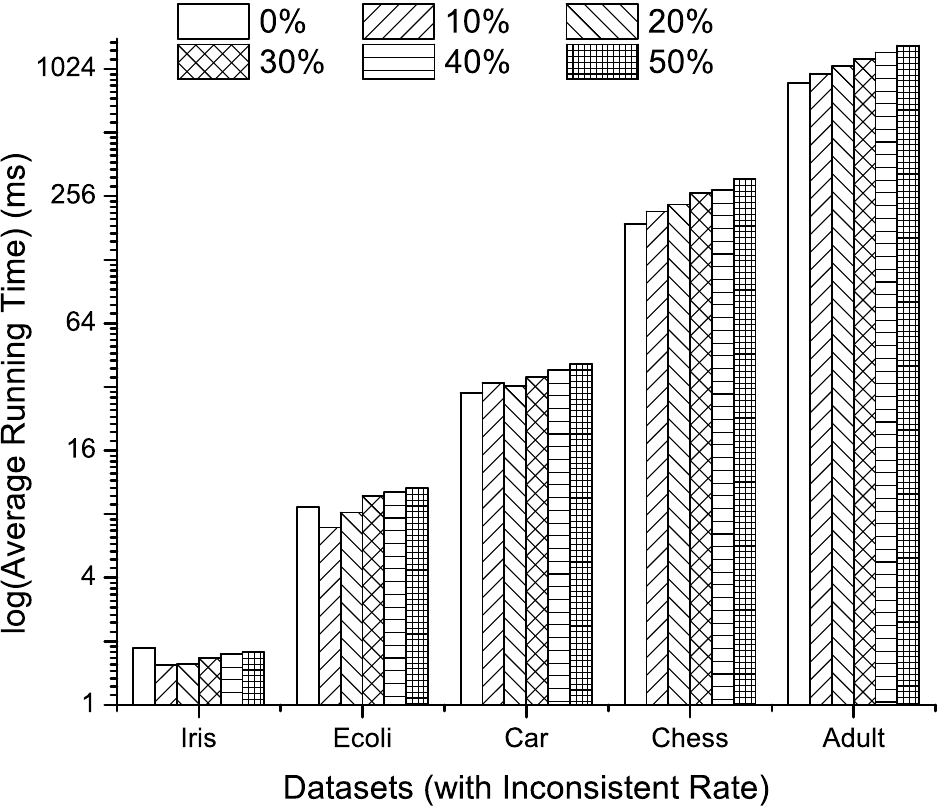}
\label{fig:nb-incons-t}
}
\vspace{-2mm}
\caption{Results on Classification for Naive Bayes Algorithm: Varying Inconsistent Rate.}
\vspace{-2mm}
\label{fig:nb-incons}
\end{figure*}

\begin{figure*}[!htb]
\centering
\subfigure{
\includegraphics[width=1.4in,height=1.0in]{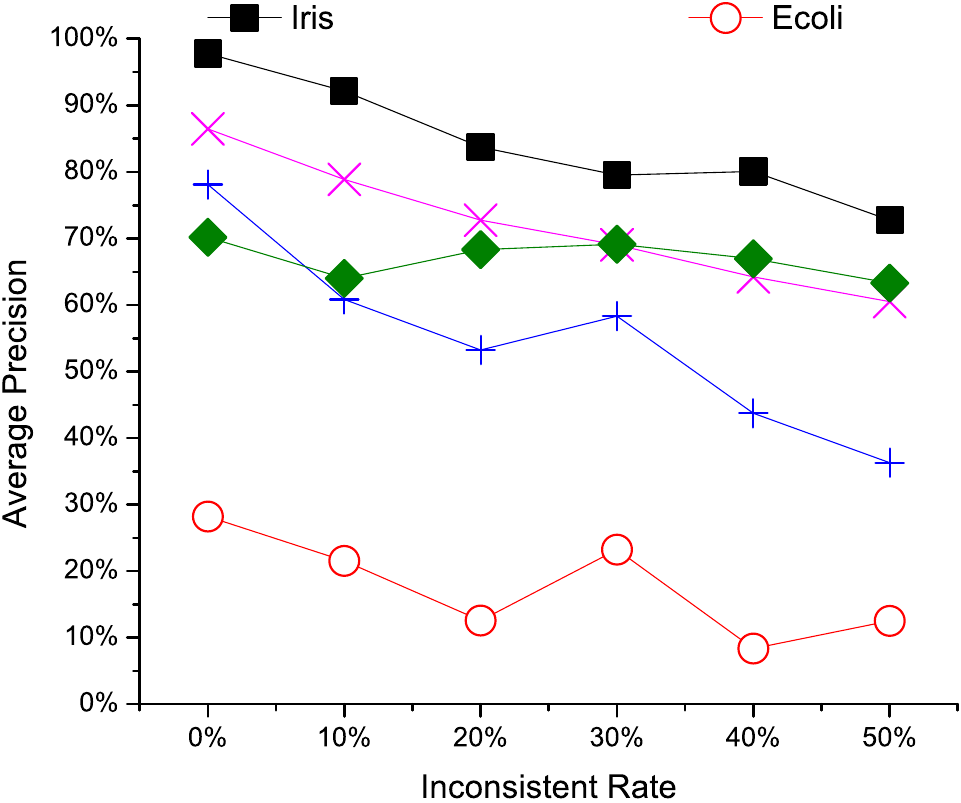}
\label{fig:bn-incons-p}
}
\subfigure{
\includegraphics[width=1.4in,height=1.0in]{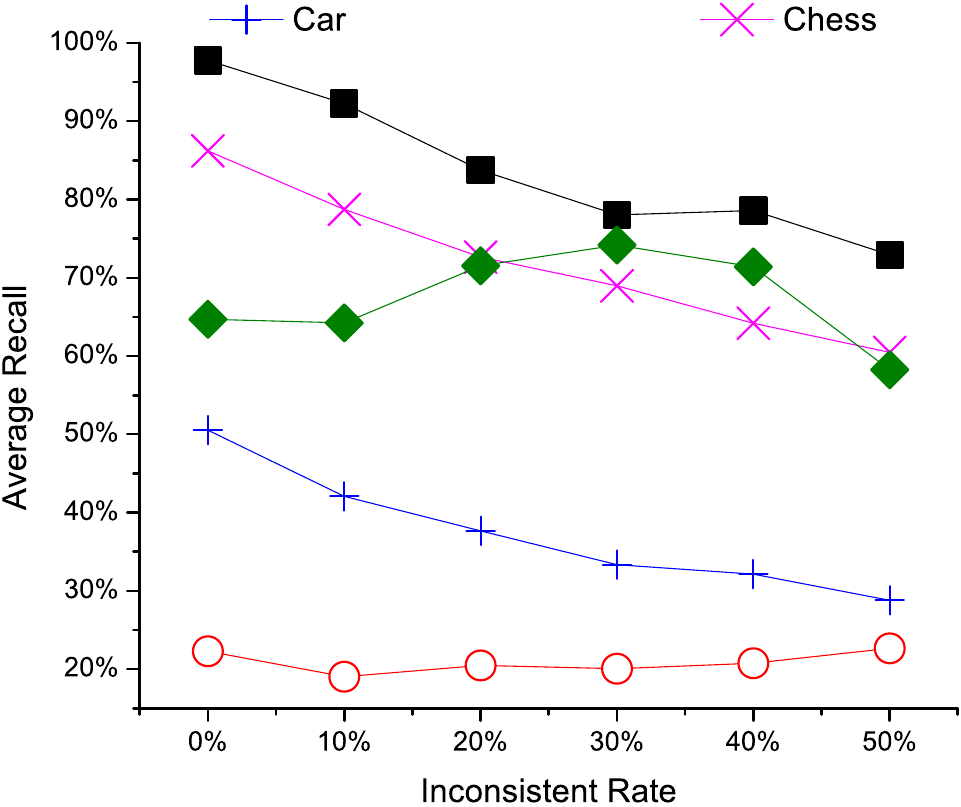}
\label{fig:bn-incons-r}
}
\subfigure{
\includegraphics[width=1.4in,height=1.0in]{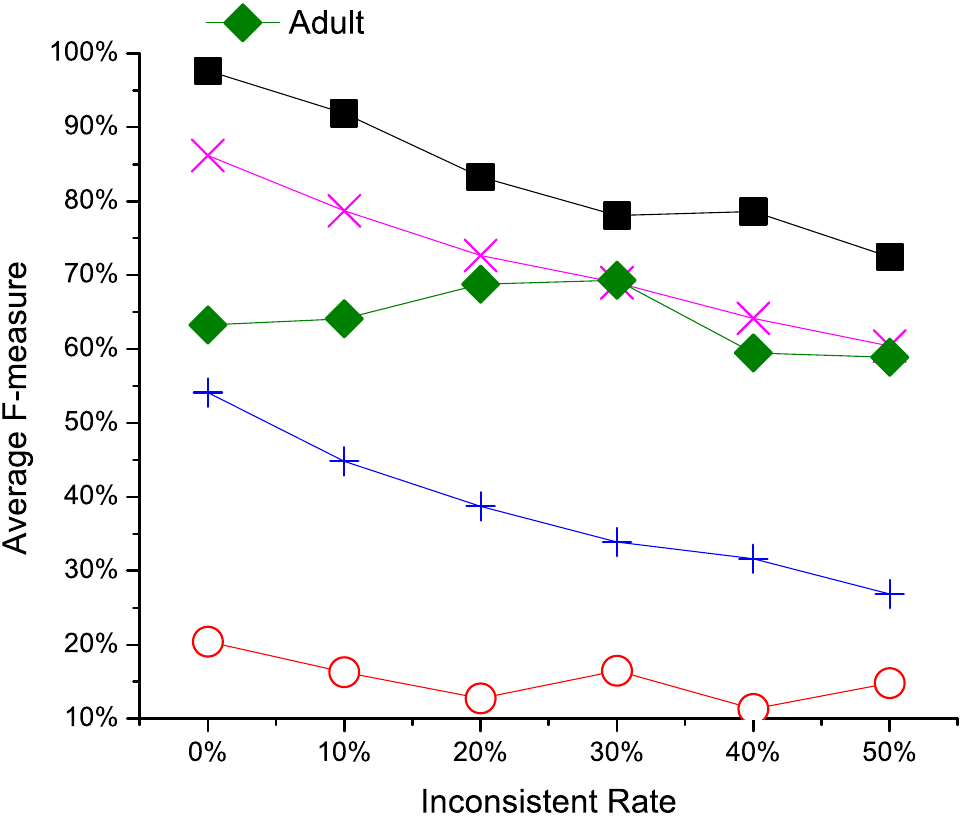}
\label{fig:bn-incons-f}
}
\subfigure{
\includegraphics[width=1.4in,height=1.0in]{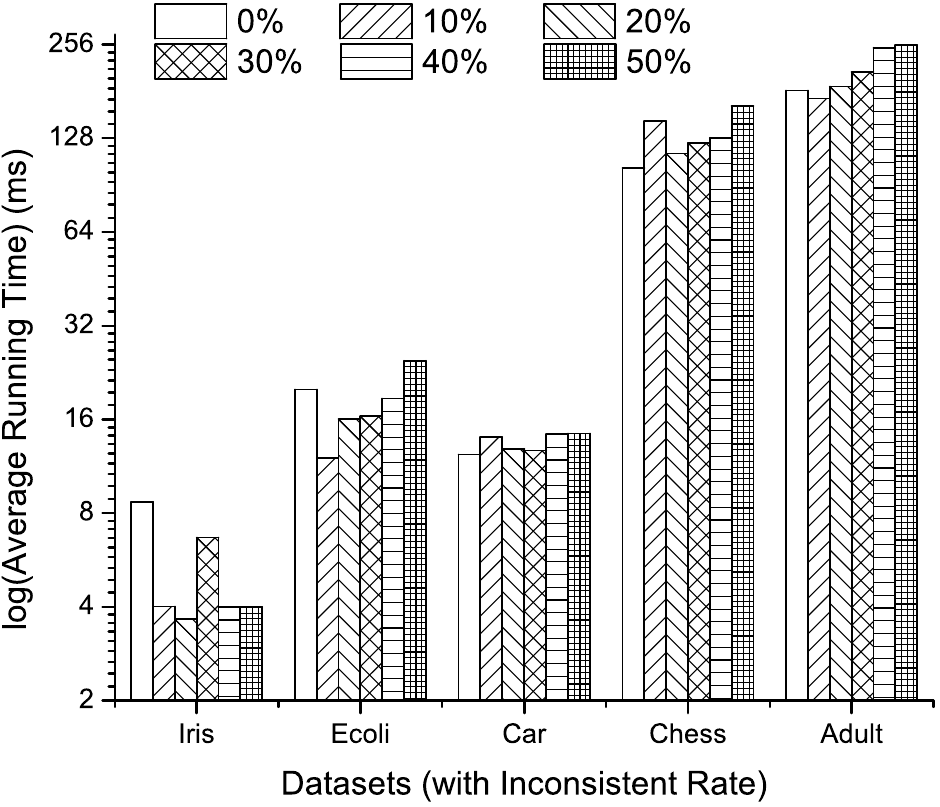}
\label{fig:bn-incons-t}
}
\vspace{-2mm}
\caption{Results on Classification for Bayesian Network Algorithm: Varying Inconsistent Rate.}
\vspace{-2mm}
\label{fig:bn-incons}
\end{figure*}

\begin{figure*}[!htb]
\centering
\subfigure{
\includegraphics[width=1.4in,height=1.0in]{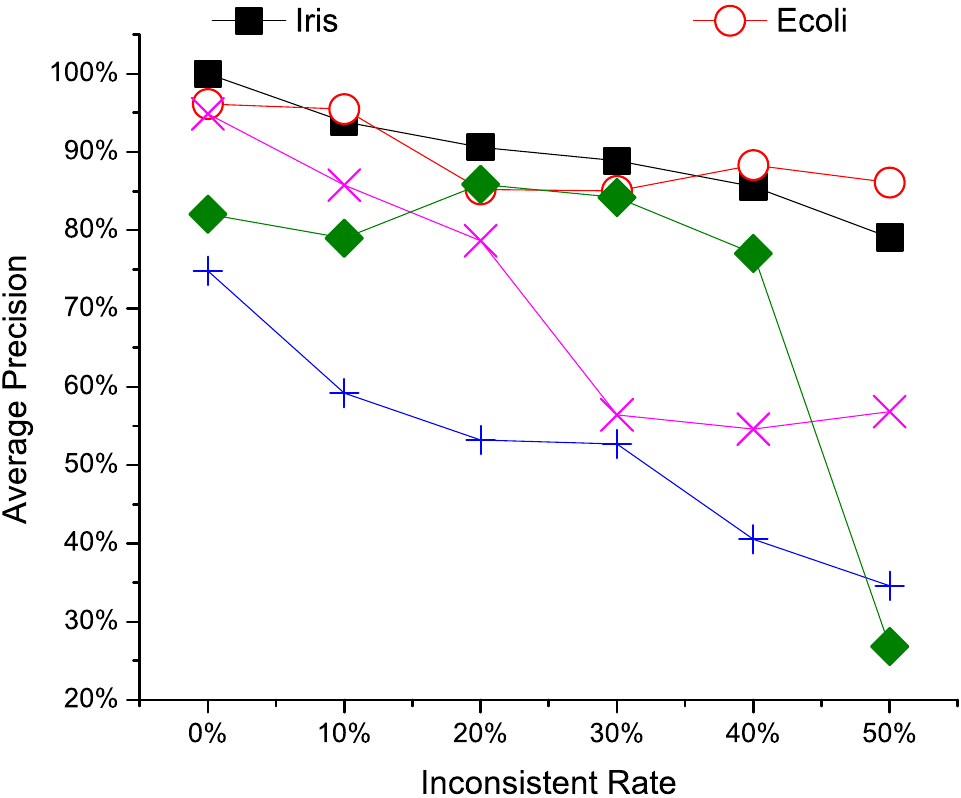}
\label{fig:log-incons-p}
}
\subfigure{
\includegraphics[width=1.4in,height=1.0in]{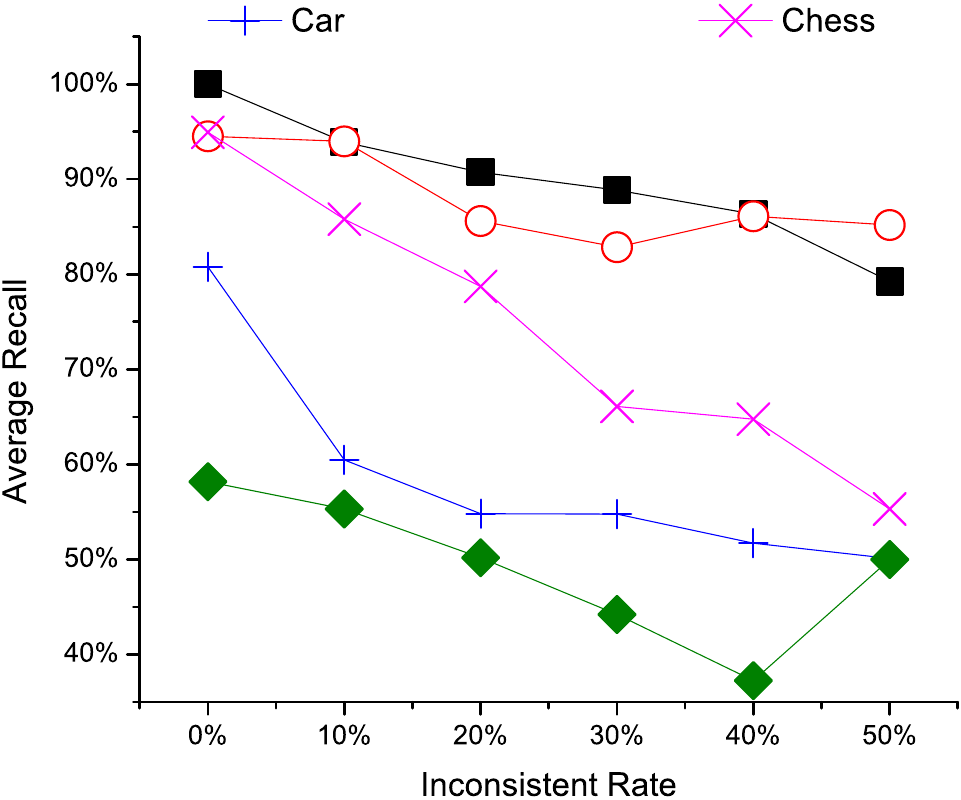}
\label{fig:log-incons-r}
}
\subfigure{
\includegraphics[width=1.4in,height=1.0in]{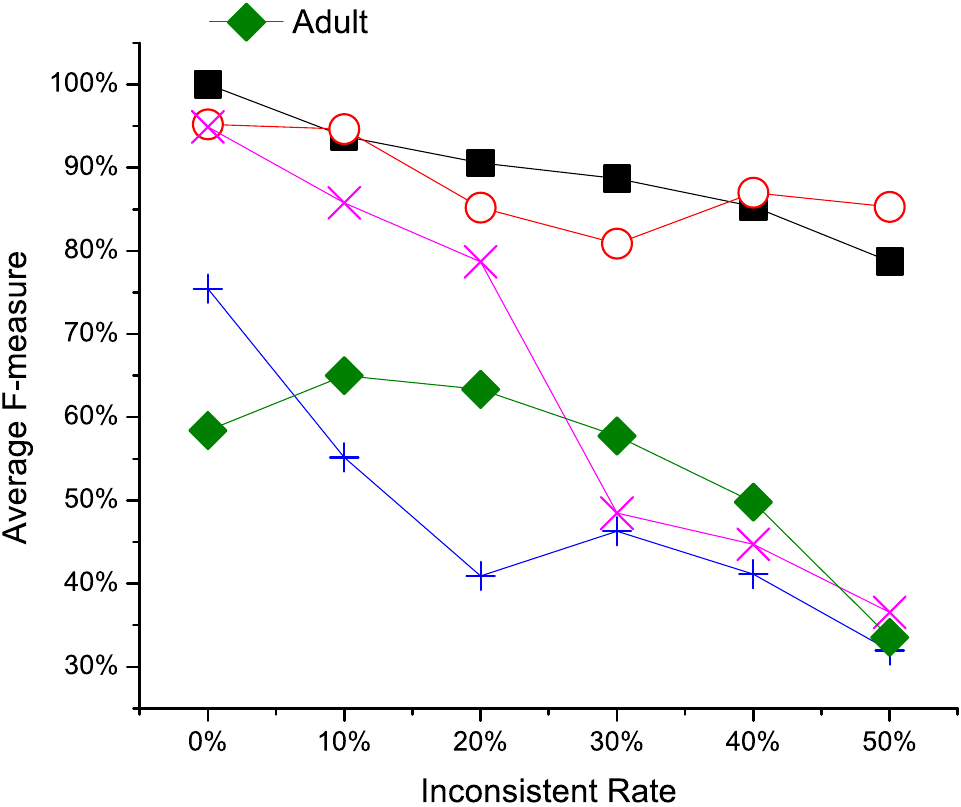}
\label{fig:log-incons-f}
}
\subfigure{
\includegraphics[width=1.4in,height=1.0in]{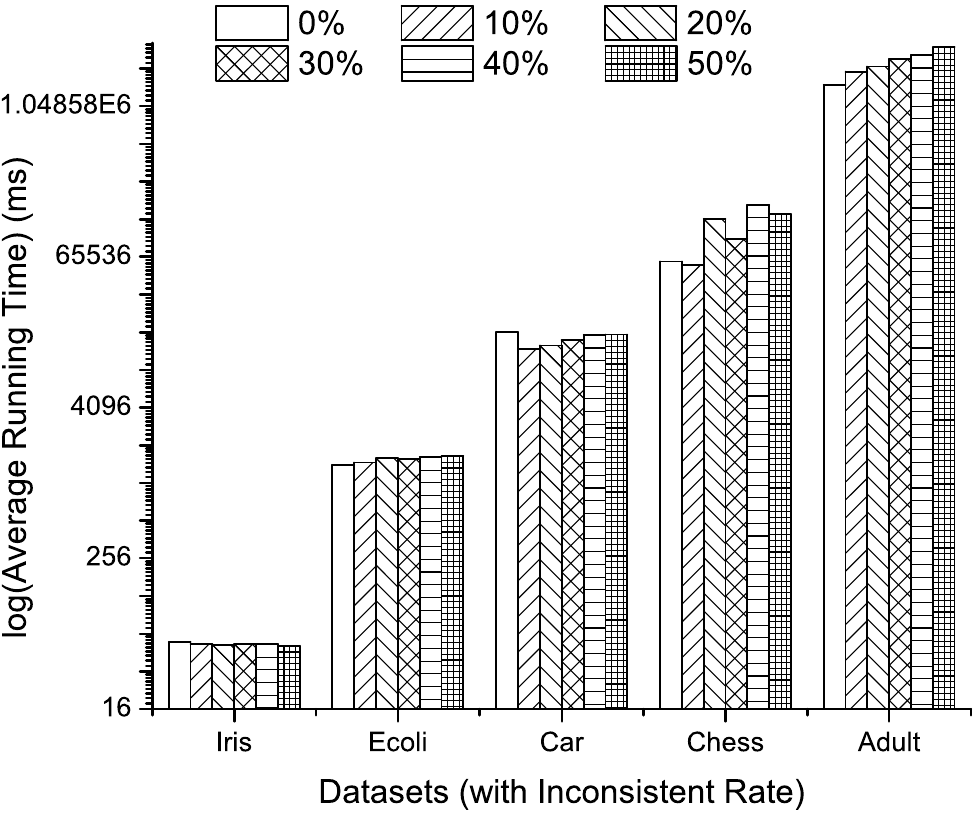}
\label{fig:log-incons-t}
}
\vspace{-2mm}
\caption{Results on Classification for Logistic Regression Algorithm: Varying Inconsistent Rate.}
\vspace{-2mm}
\label{fig:log-incons}
\end{figure*}

\begin{figure*}[!htb]
\centering
\subfigure{
\includegraphics[width=1.4in,height=1.0in]{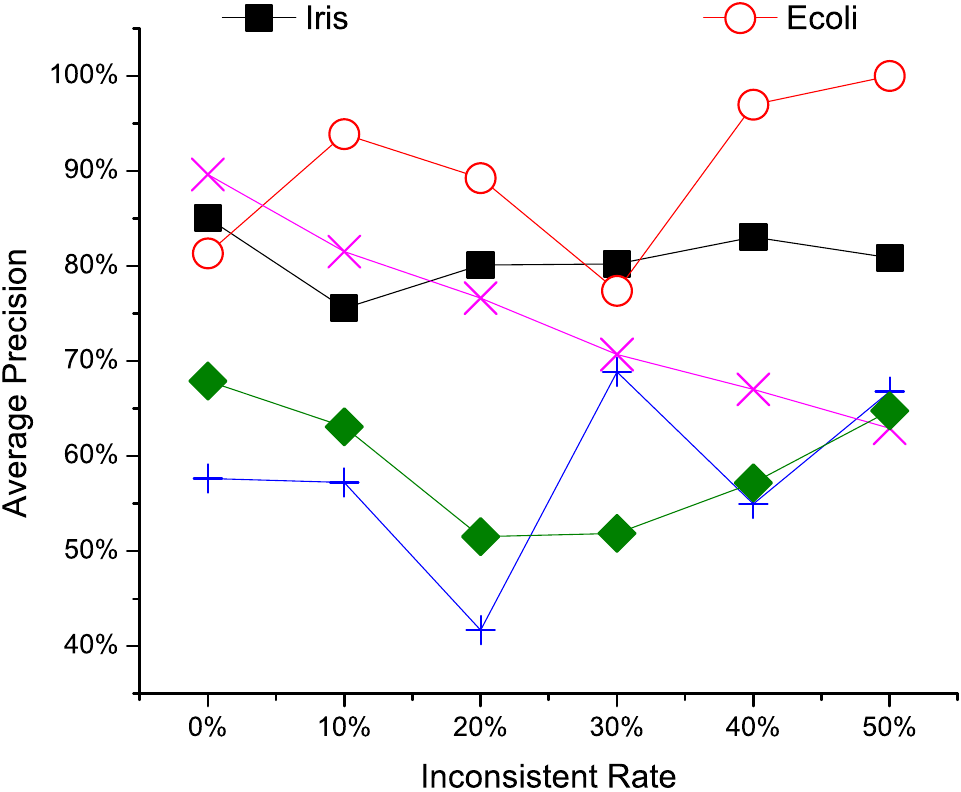}
\label{fig:ad-incons-p}
}
\subfigure{
\includegraphics[width=1.4in,height=1.0in]{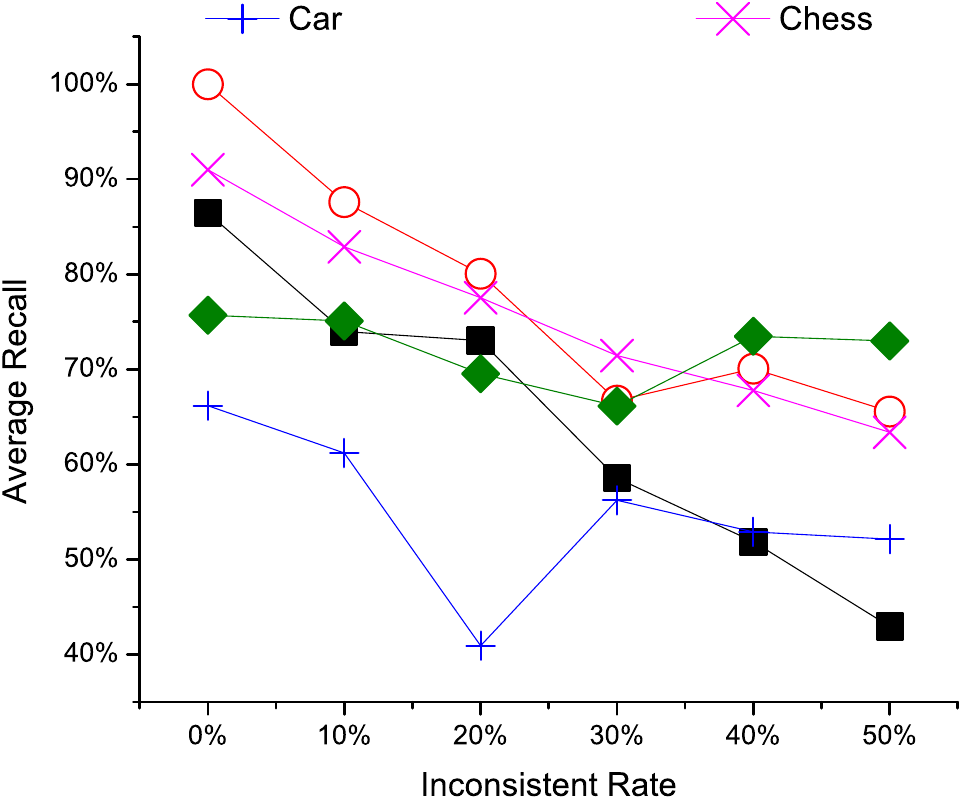}
\label{fig:ad-incons-r}
}
\subfigure{
\includegraphics[width=1.4in,height=1.0in]{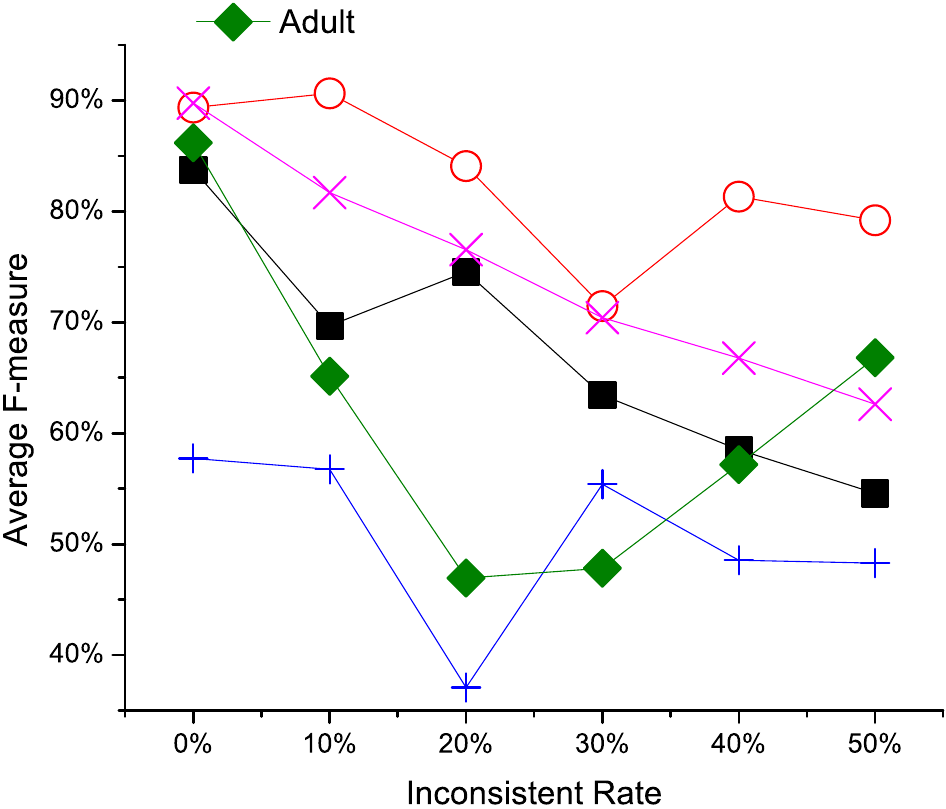}
\label{fig:ad-incons-f}
}
\subfigure{
\includegraphics[width=1.4in,height=1.0in]{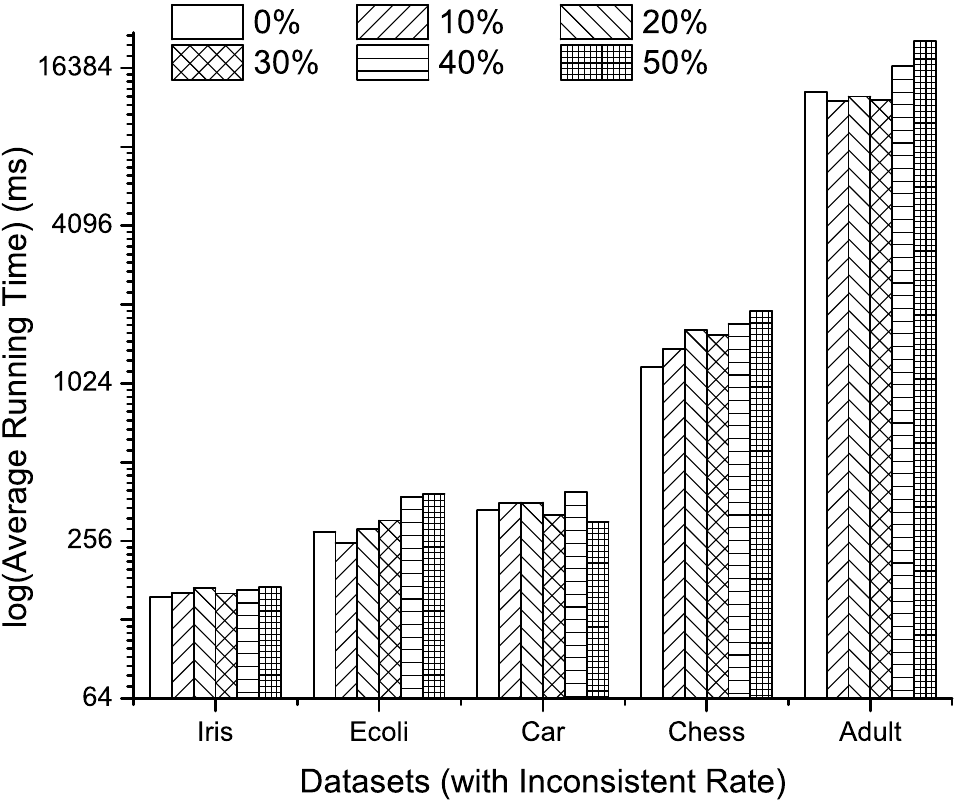}
\label{fig:ad-incons-t}
}
\vspace{-2mm}
\caption{Results on Classification for Random Forests Algorithm: Varying Inconsistent Rate.}
\vspace{-2mm}
\label{fig:ad-incons}
\end{figure*}

\begin{figure*}[!htb]
\centering
\subfigure{
\includegraphics[width=1.4in,height=1.0in]{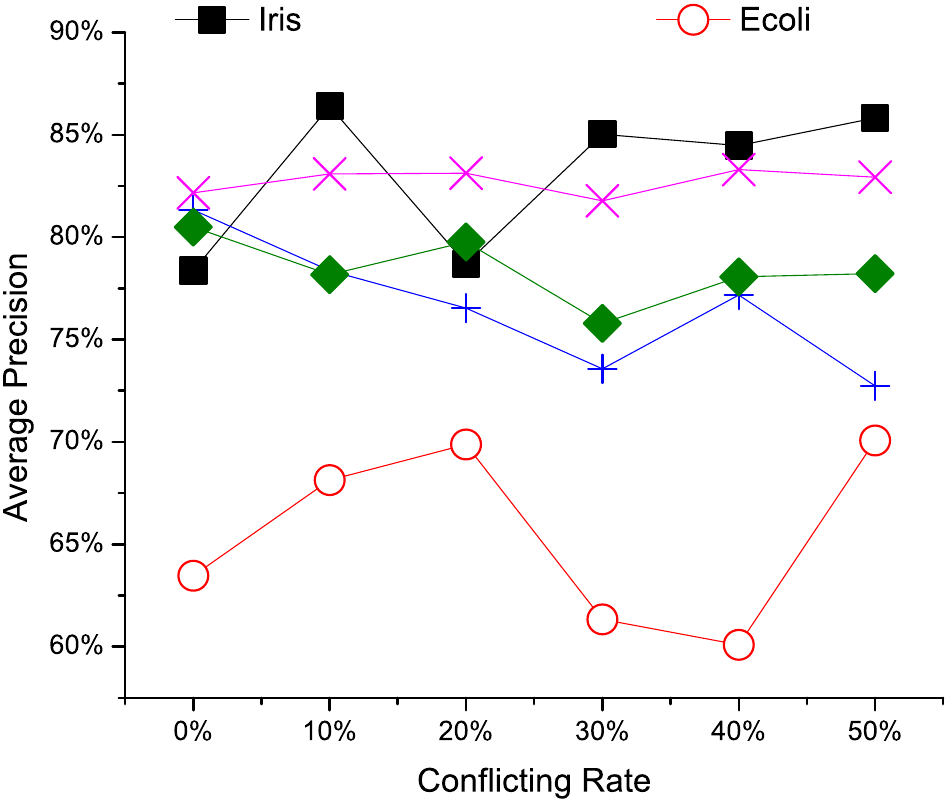}
\label{fig:dt-conf-p}
}
\subfigure{
\includegraphics[width=1.4in,height=1.0in]{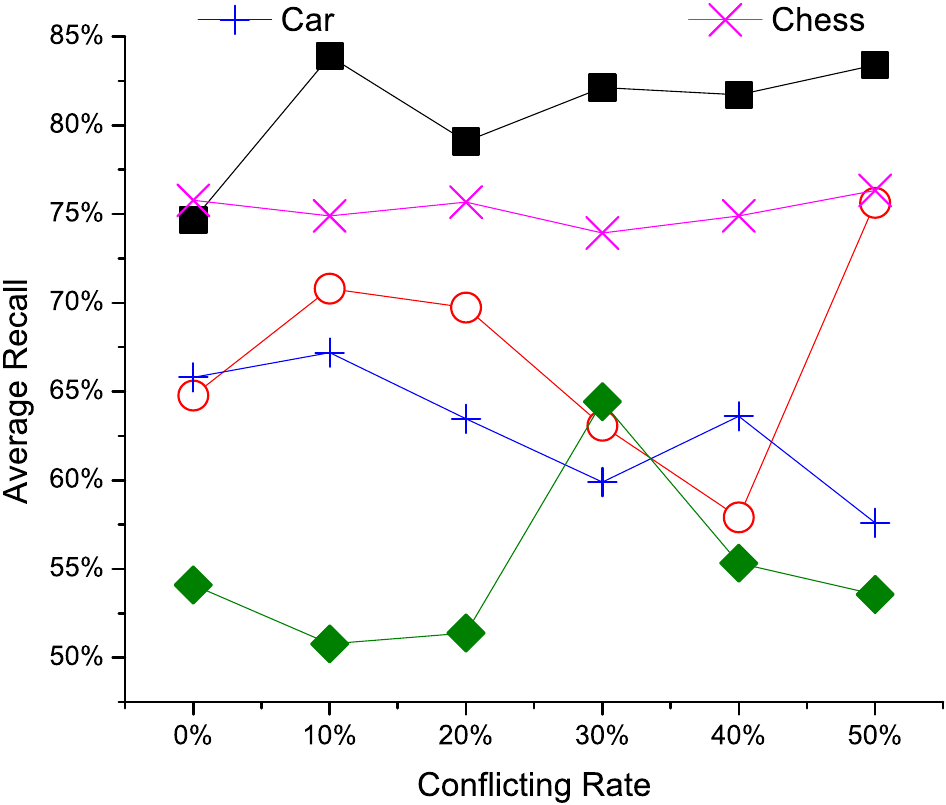}
\label{fig:dt-conf-r}
}
\subfigure{
\includegraphics[width=1.4in,height=1.0in]{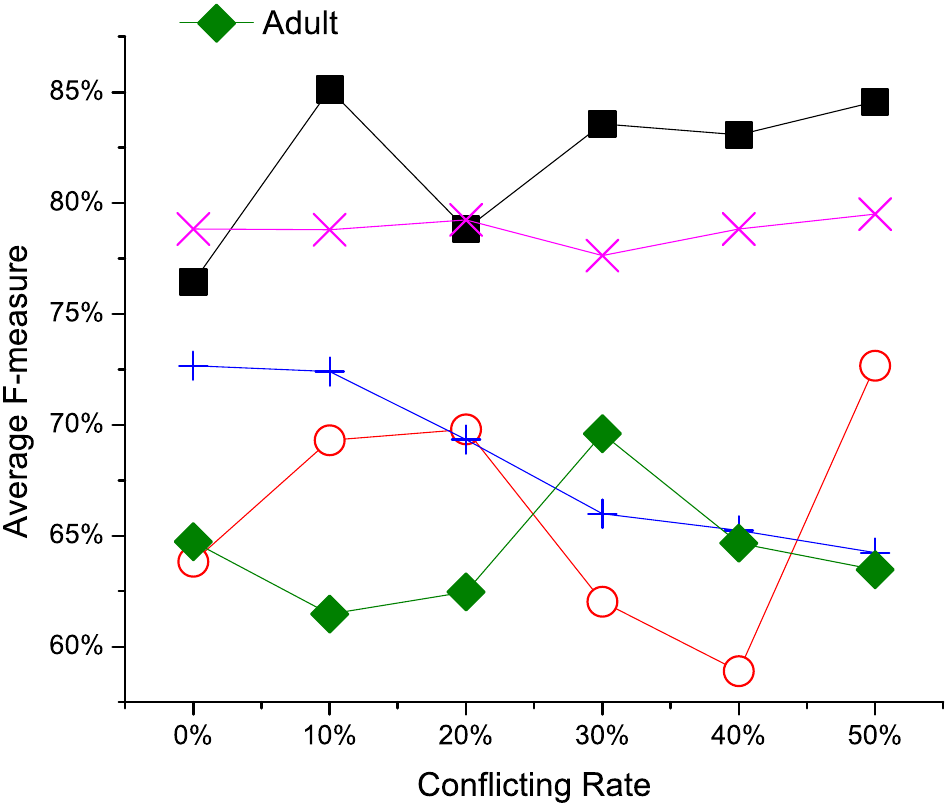}
\label{fig:dt-conf-f}
}
\subfigure{
\includegraphics[width=1.4in,height=1.0in]{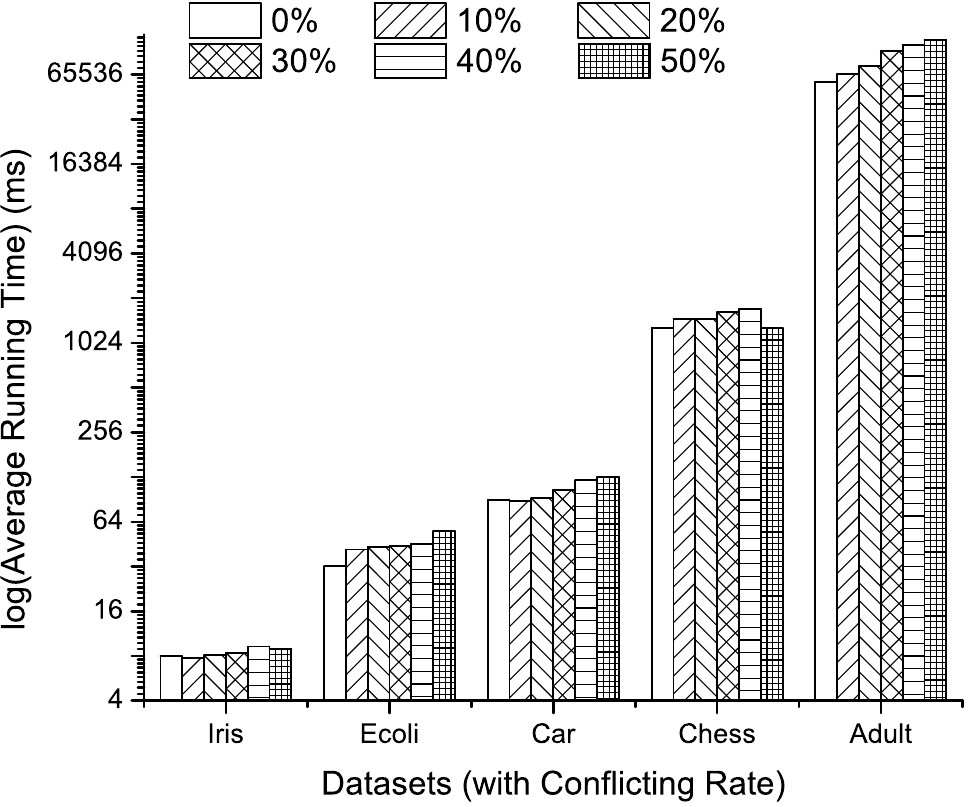}
\label{fig:dt-conf-t}
}
\vspace{-2mm}
\caption{Results on Classification for Decision Tree Algorithm: Varying Conflicting Rate.}
\vspace{-2mm}
\label{fig:dt-conf}
\end{figure*}

\begin{figure*}[!htb]
\centering
\subfigure{
\includegraphics[width=1.4in,height=1.0in]{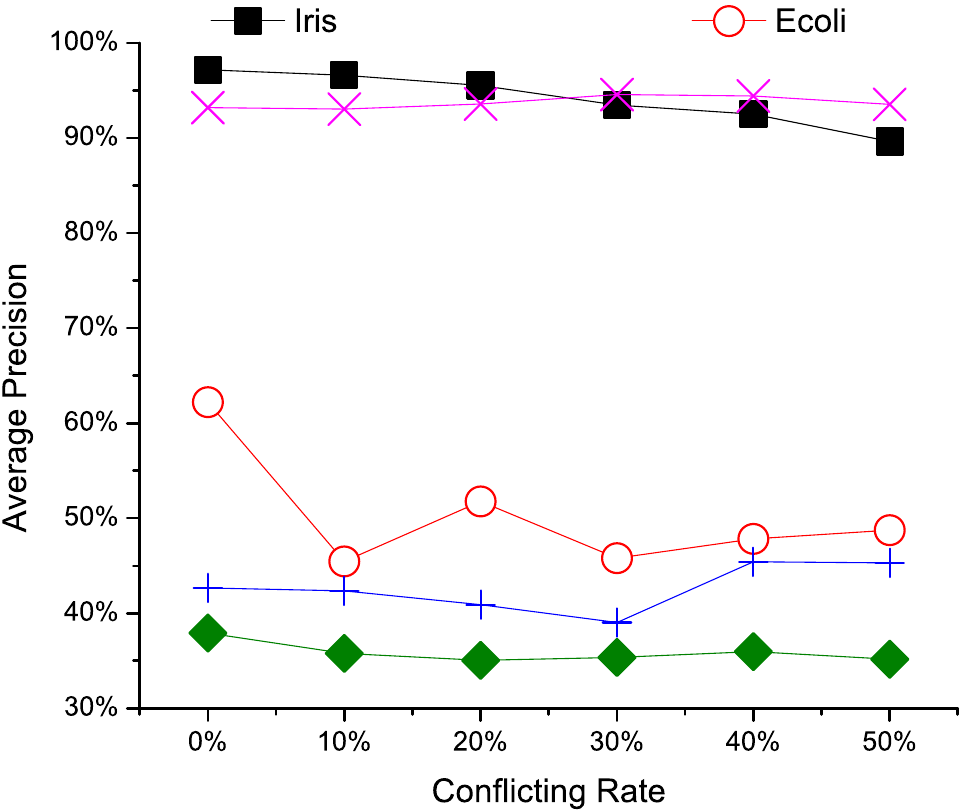}
\label{fig:knn-conf-p}
}
\subfigure{
\includegraphics[width=1.4in,height=1.0in]{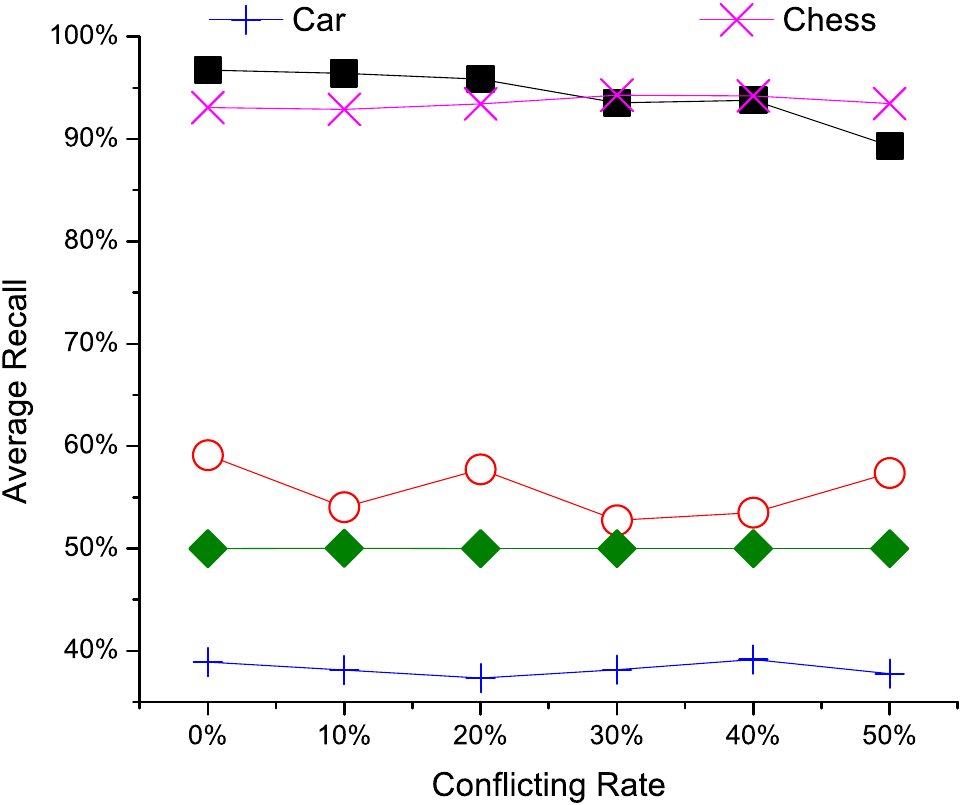}
\label{fig:knn-conf-r}
}
\subfigure{
\includegraphics[width=1.4in,height=1.0in]{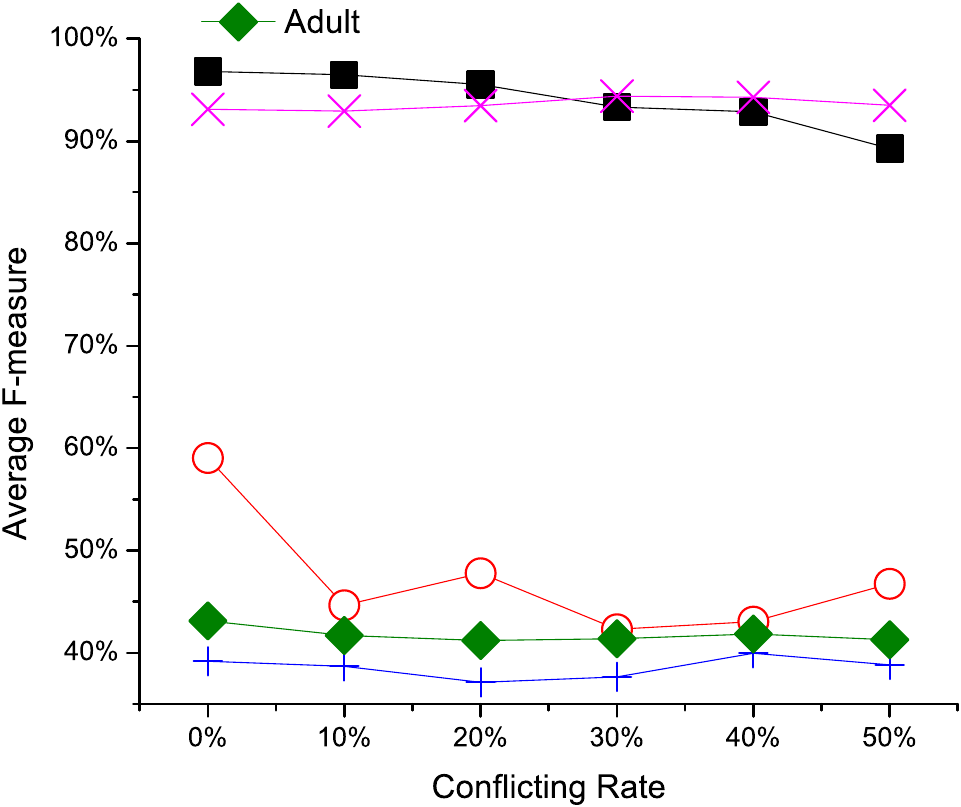}
\label{fig:knn-conf-f}
}
\subfigure{
\includegraphics[width=1.4in,height=1.0in]{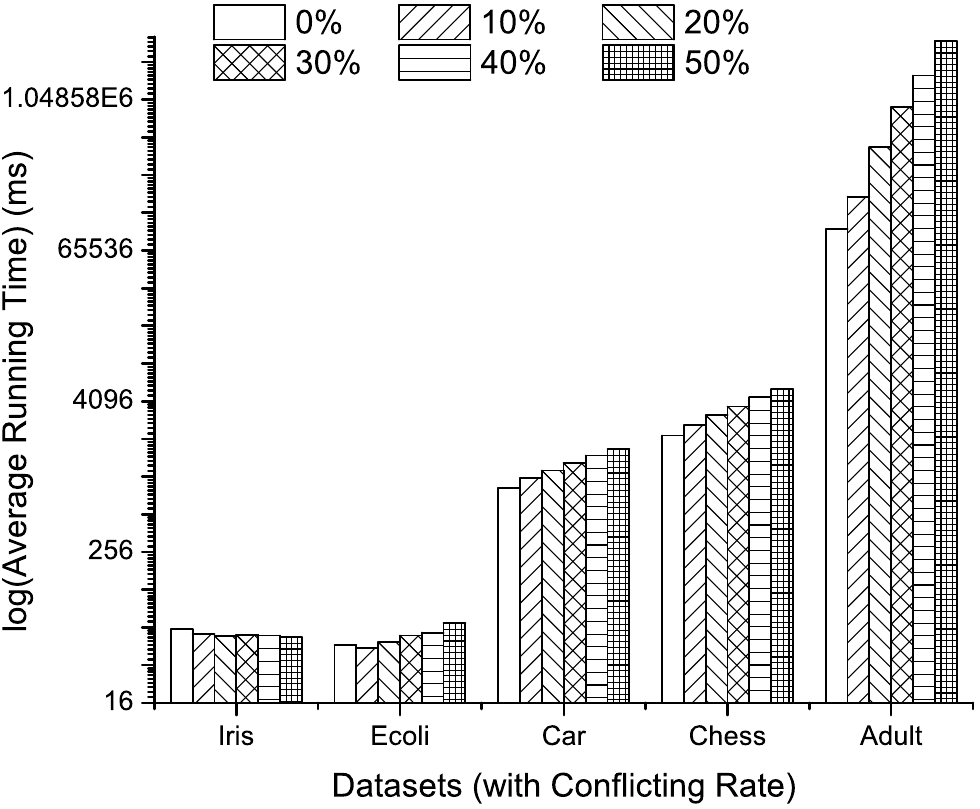}
\label{fig:knn-conf-t}
}
\vspace{-2mm}
\caption{Results on Classification for KNN Algorithm: Varying Conflicting Rate.}
\vspace{-2mm}
\label{fig:knn-conf}
\end{figure*}

\clearpage
\begin{figure*}[!htb]
\centering
\subfigure{
\includegraphics[width=1.4in,height=1.0in]{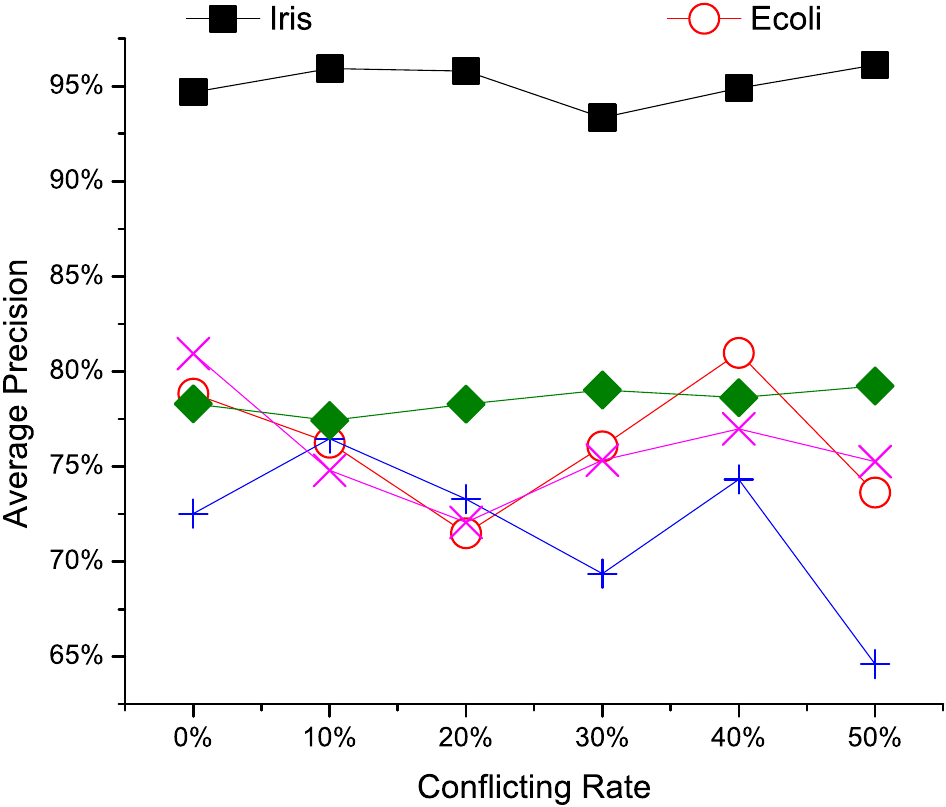}
\label{fig:nb-conf-p}
}
\subfigure{
\includegraphics[width=1.4in,height=1.0in]{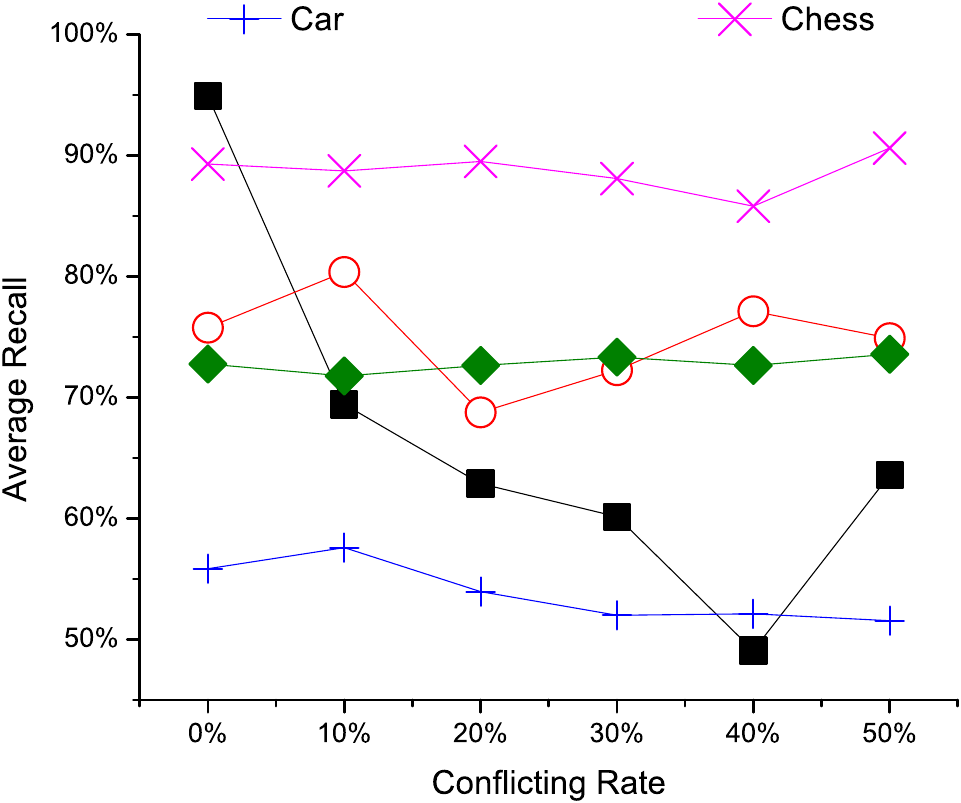}
\label{fig:nb-conf-r}
}
\subfigure{
\includegraphics[width=1.4in,height=1.0in]{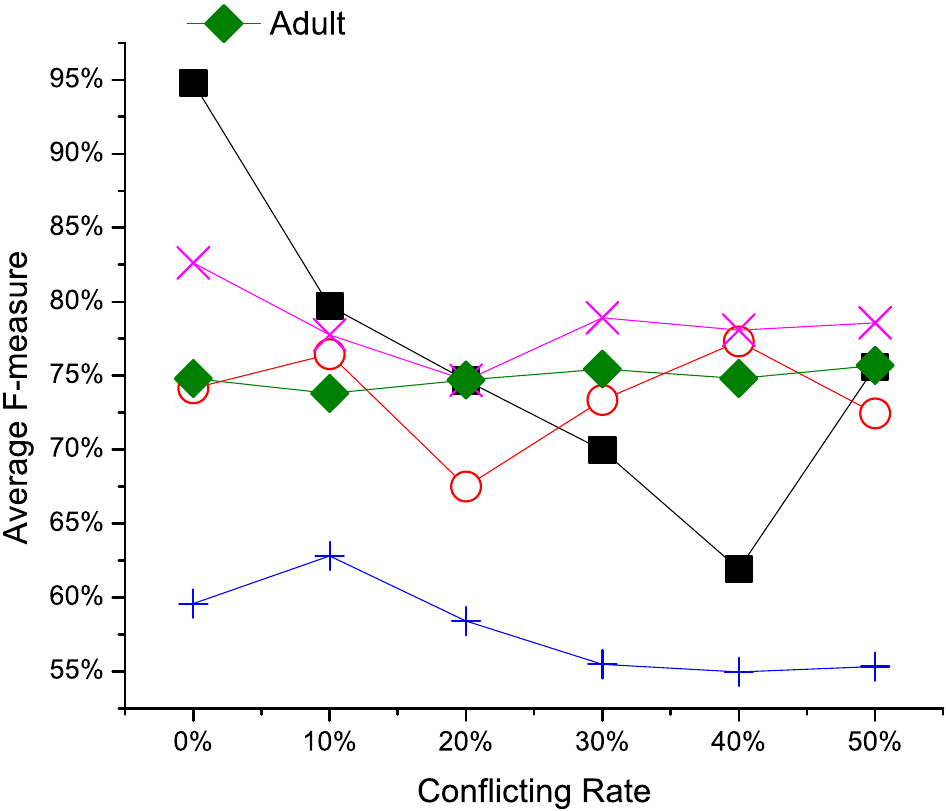}
\label{fig:nb-conf-f}
}
\subfigure{
\includegraphics[width=1.4in,height=1.0in]{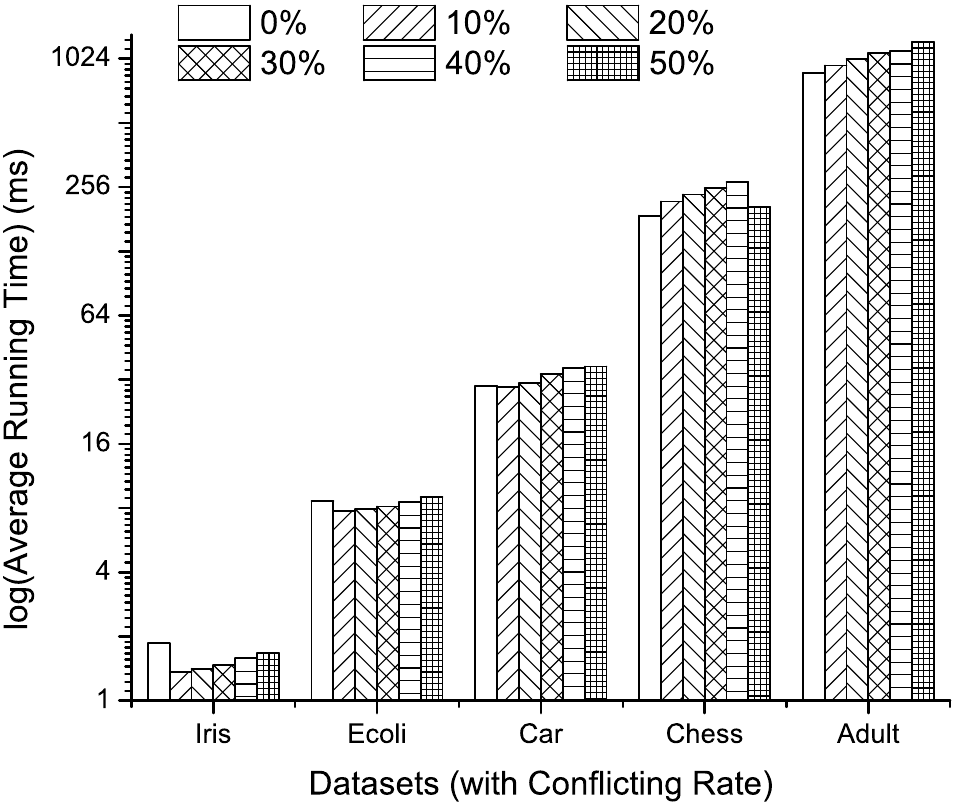}
\label{fig:nb-conf-t}
}
\vspace{-2mm}
\caption{Results on Classification for Naive Bayes Algorithm: Varying Conflicting Rate.}
\vspace{-2mm}
\label{fig:nb-conf}
\end{figure*}

\begin{figure*}[!htb]
\centering
\subfigure{
\includegraphics[width=1.4in,height=1.0in]{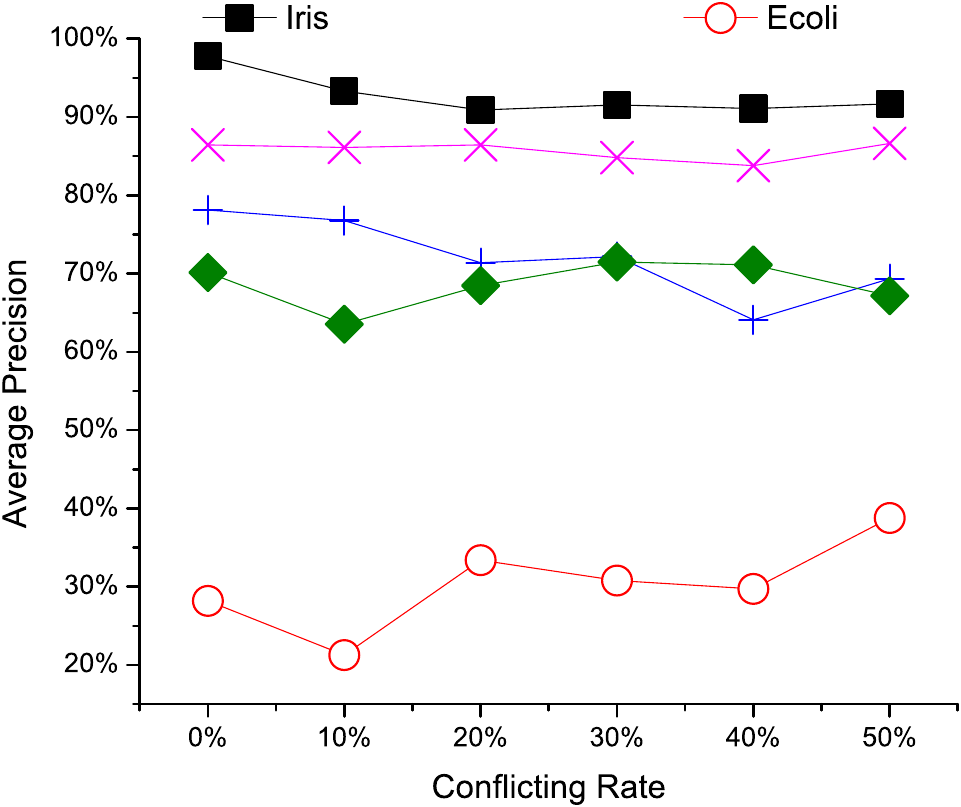}
\label{fig:bn-conf-p}
}
\subfigure{
\includegraphics[width=1.4in,height=1.0in]{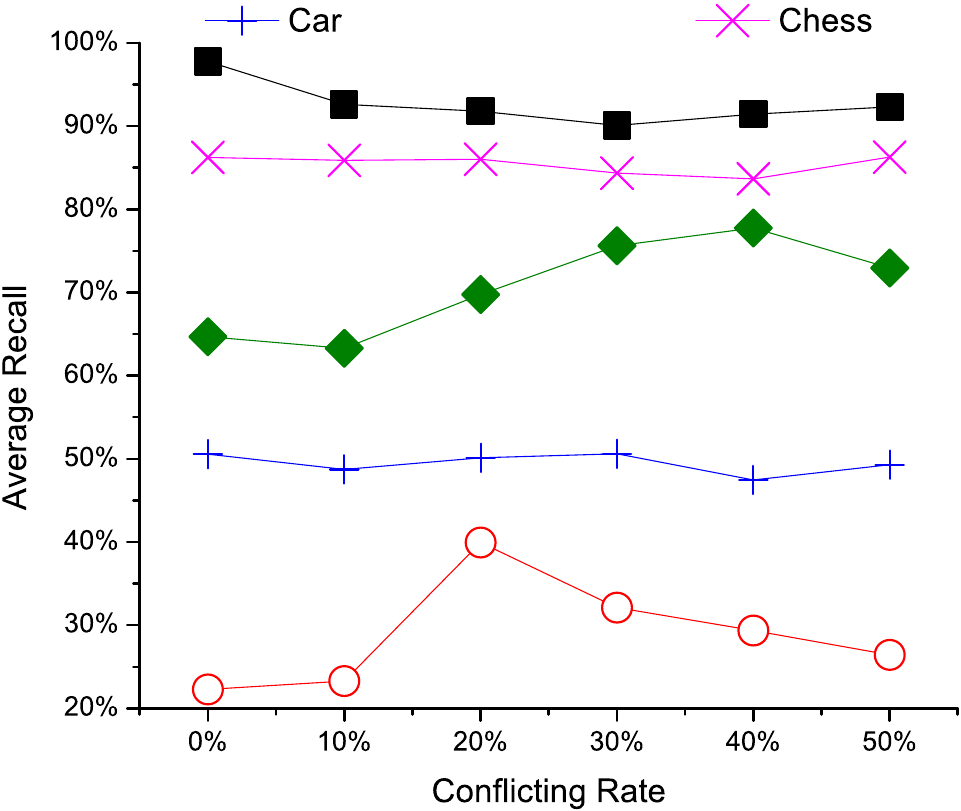}
\label{fig:bn-conf-r}
}
\subfigure{
\includegraphics[width=1.4in,height=1.0in]{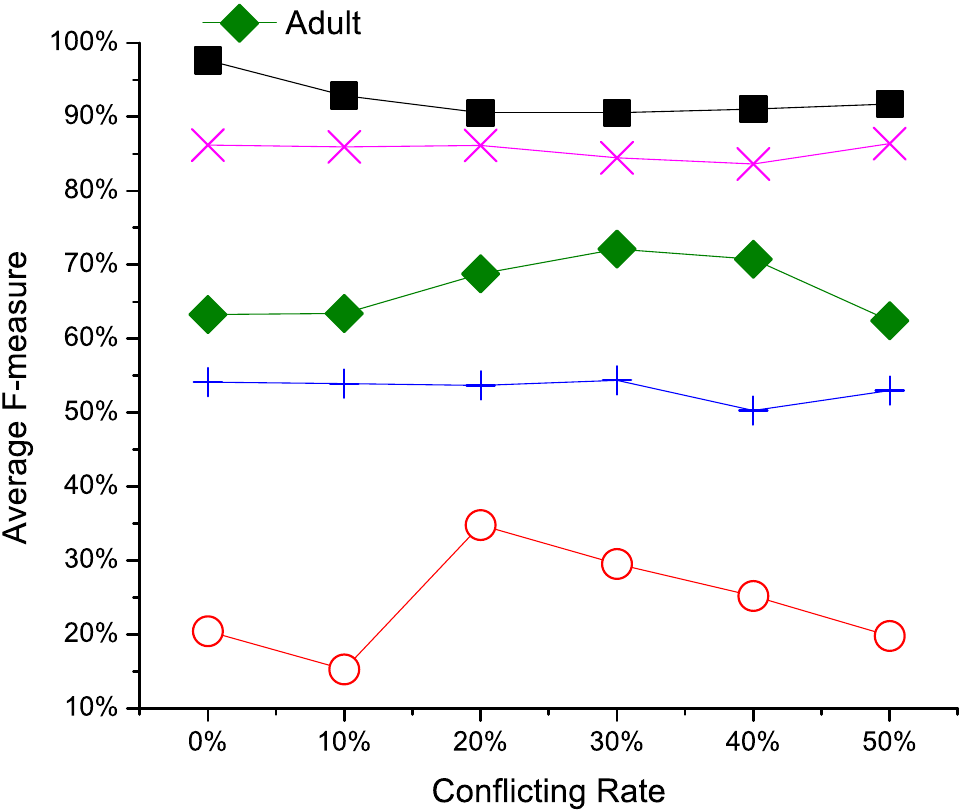}
\label{fig:bn-conf-f}
}
\subfigure{
\includegraphics[width=1.4in,height=1.0in]{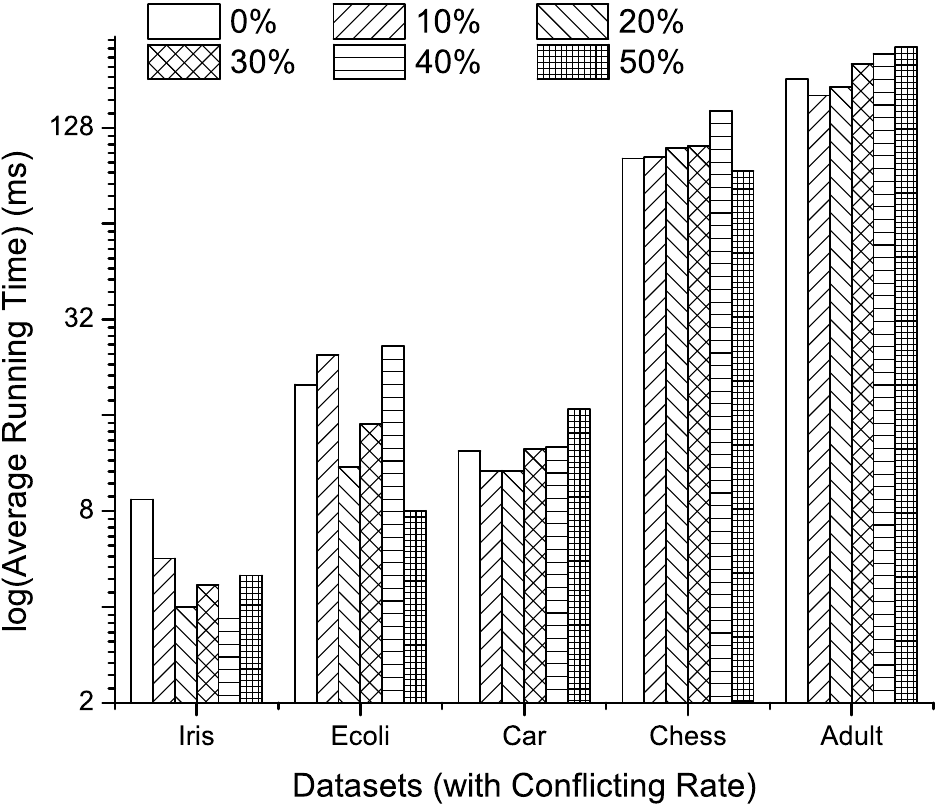}
\label{fig:bn-conf-t}
}
\vspace{-2mm}
\caption{Results on Classification for Bayesian Network Algorithm: Varying Conflicting Rate.}
\vspace{-2mm}
\label{fig:bn-conf}
\end{figure*}

\begin{figure*}[!htb]
\centering
\subfigure{
\includegraphics[width=1.4in,height=1.0in]{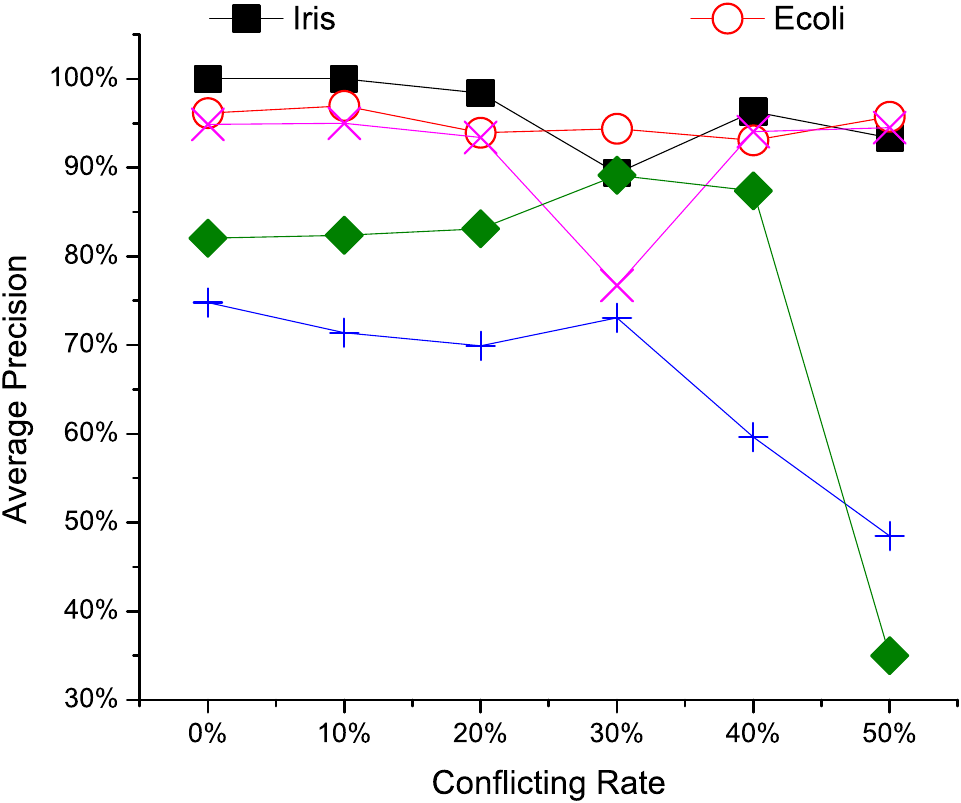}
\label{fig:log-conf-p}
}
\subfigure{
\includegraphics[width=1.4in,height=1.0in]{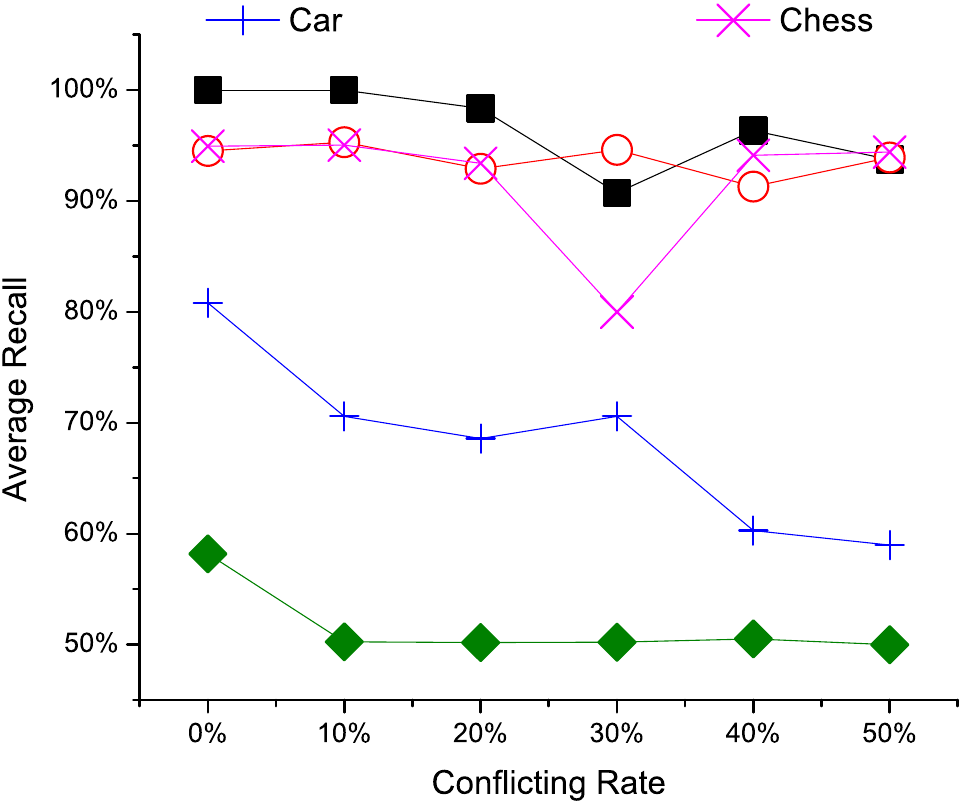}
\label{fig:log-conf-r}
}
\subfigure{
\includegraphics[width=1.4in,height=1.0in]{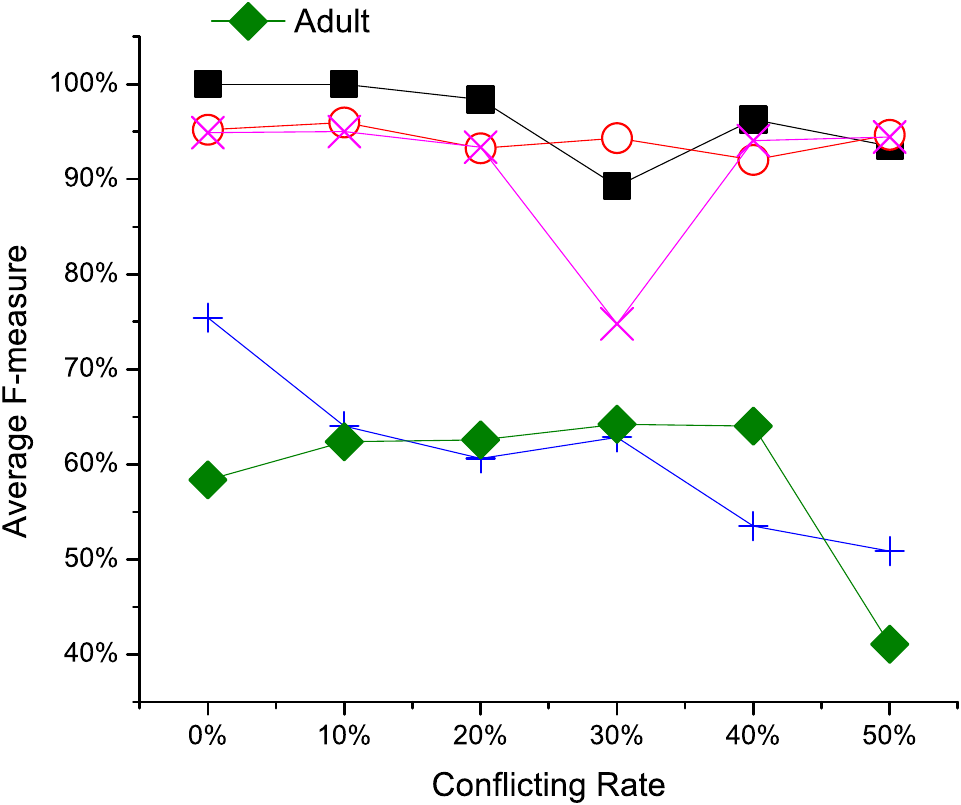}
\label{fig:log-conf-f}
}
\subfigure{
\includegraphics[width=1.4in,height=1.0in]{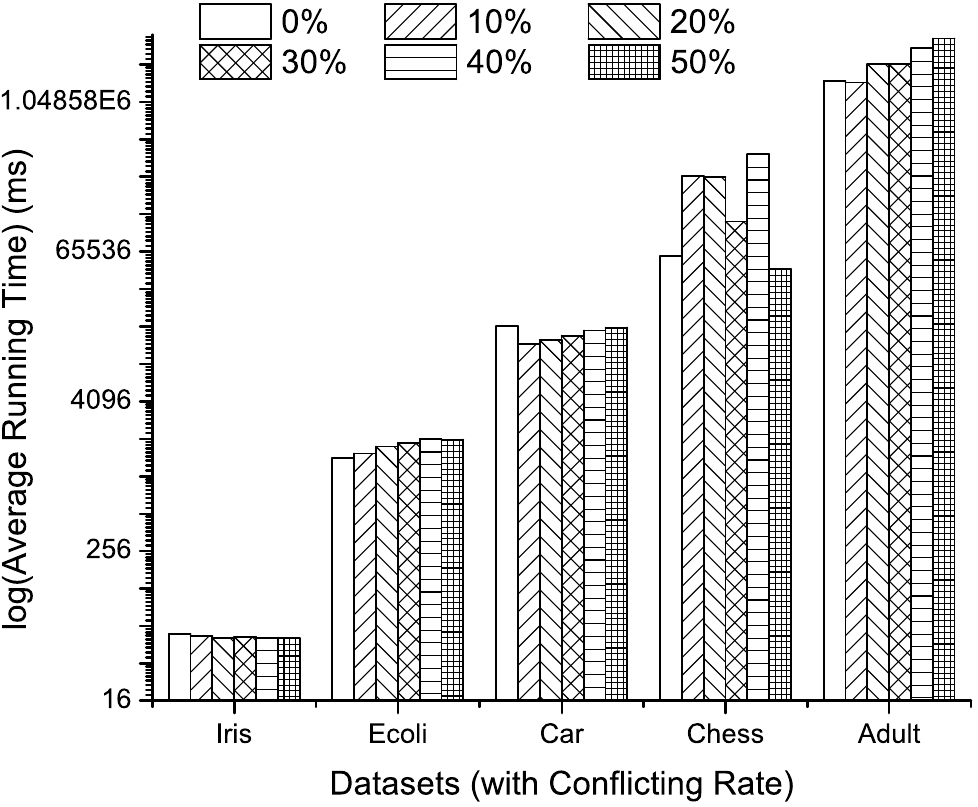}
\label{fig:log-conf-t}
}
\vspace{-2mm}
\caption{Results on Classification for Logistic Regression Algorithm: Varying Conflicting Rate.}
\vspace{-2mm}
\label{fig:log-conf}
\end{figure*}

\begin{figure*}[!htb]
\centering
\subfigure{
\includegraphics[width=1.4in,height=1.0in]{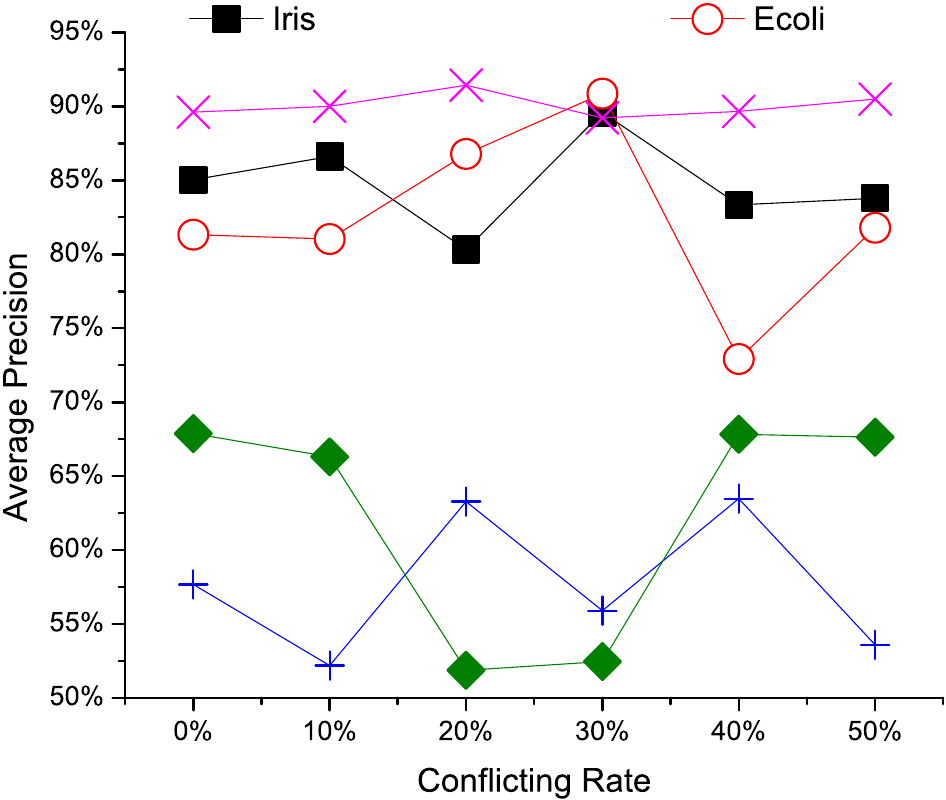}
\label{fig:ad-conf-p}
}
\subfigure{
\includegraphics[width=1.4in,height=1.0in]{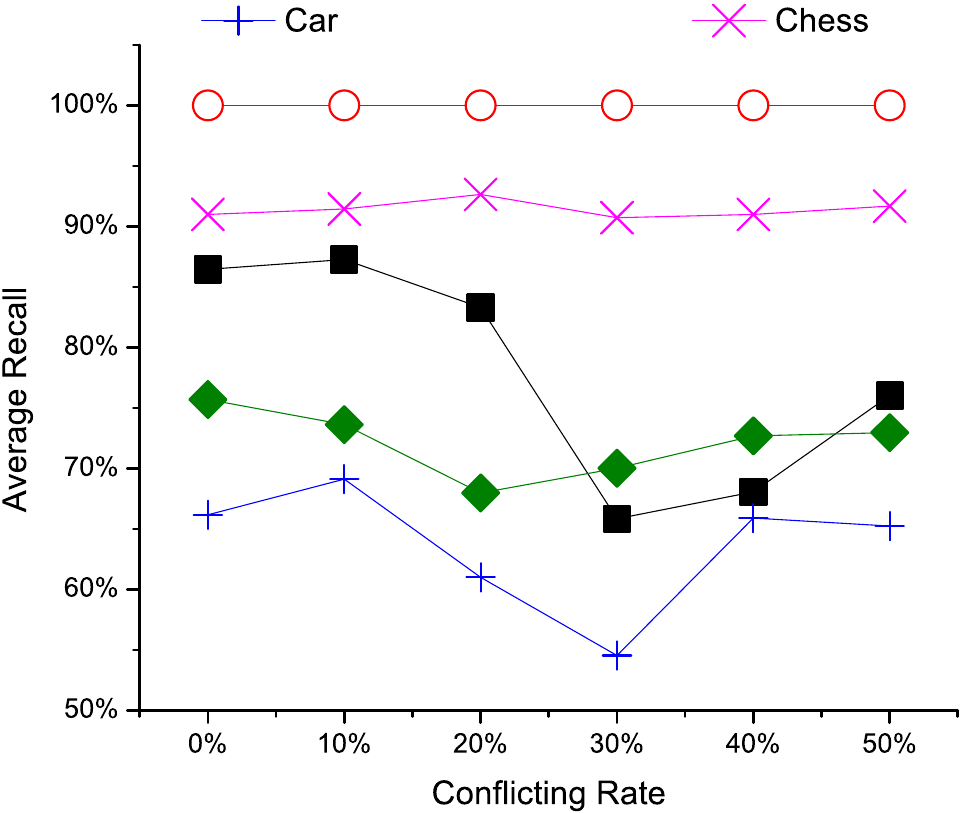}
\label{fig:ad-conf-r}
}
\subfigure{
\includegraphics[width=1.4in,height=1.0in]{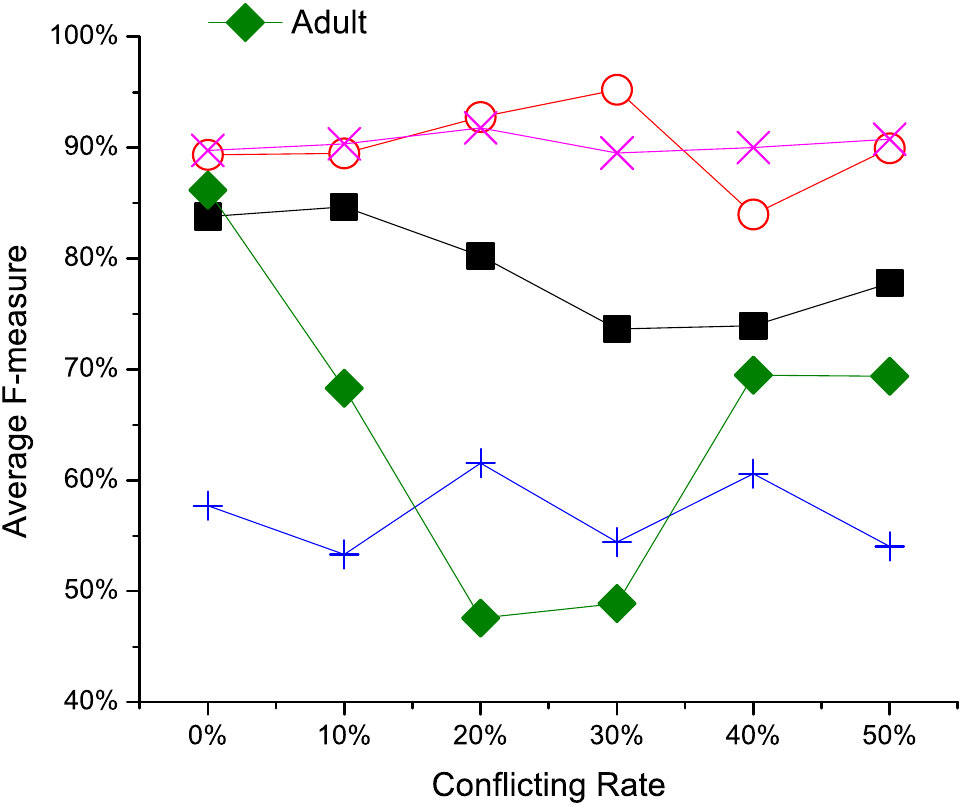}
\label{fig:ad-conf-f}
}
\subfigure{
\includegraphics[width=1.4in,height=1.0in]{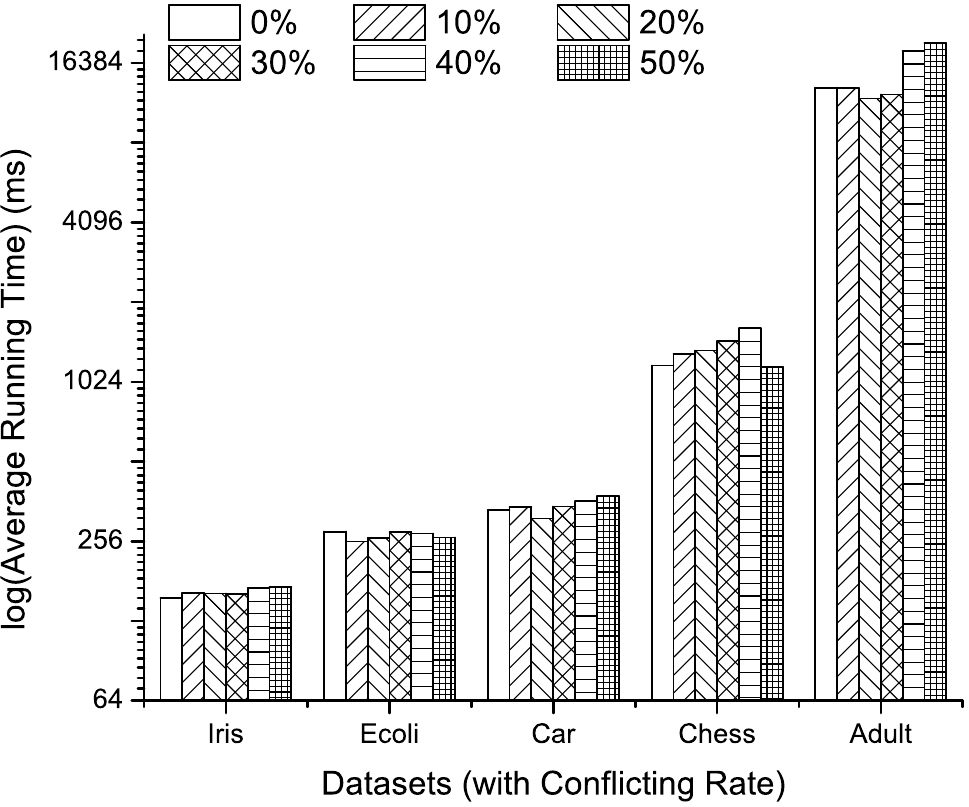}
\label{fig:ad-conf-t}
}
\vspace{-2mm}
\caption{Results on Classification for Random Forests Algorithm: Varying Conflicting Rate.}
\vspace{-2mm}
\label{fig:ad-conf}
\end{figure*}

\begin{figure*}[!htb]
\centering
\subfigure{
\includegraphics[width=1.4in,height=1.0in]{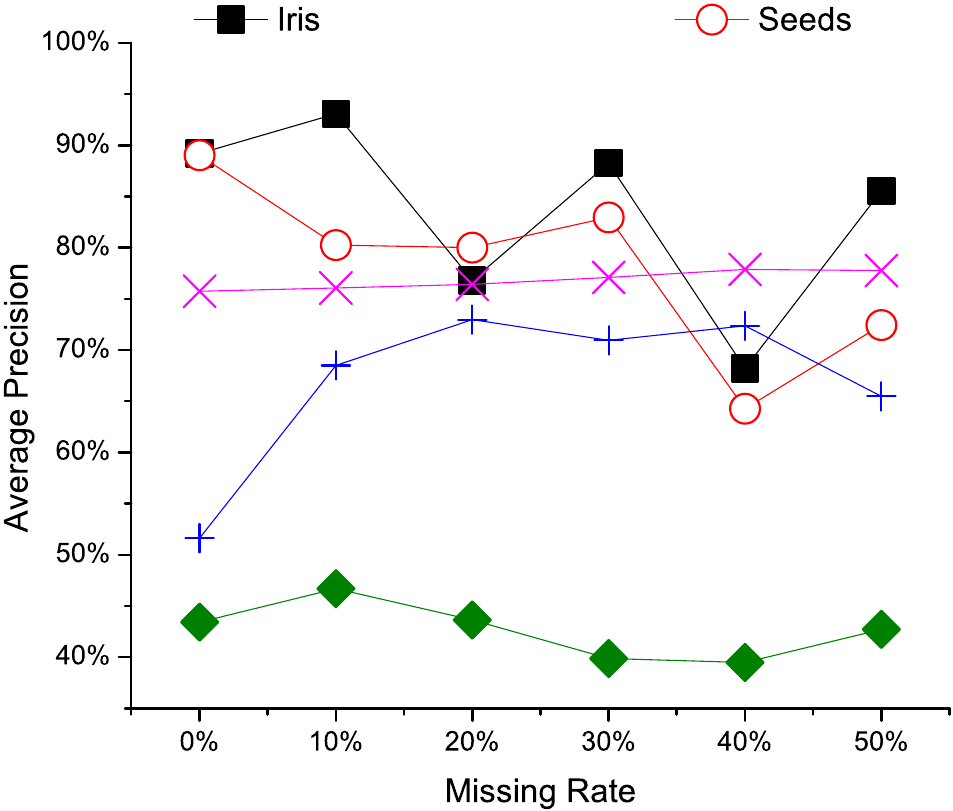}
\label{fig:km-miss-p}
}
\subfigure{
\includegraphics[width=1.4in,height=1.0in]{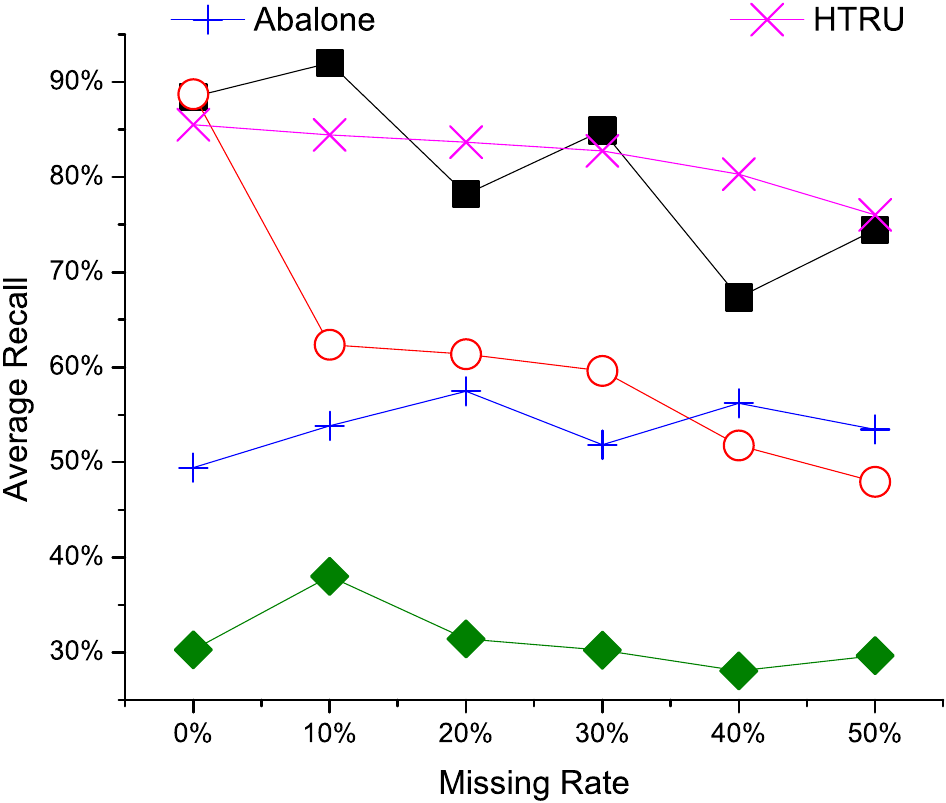}
\label{fig:km-miss-r}
}
\subfigure{
\includegraphics[width=1.4in,height=1.0in]{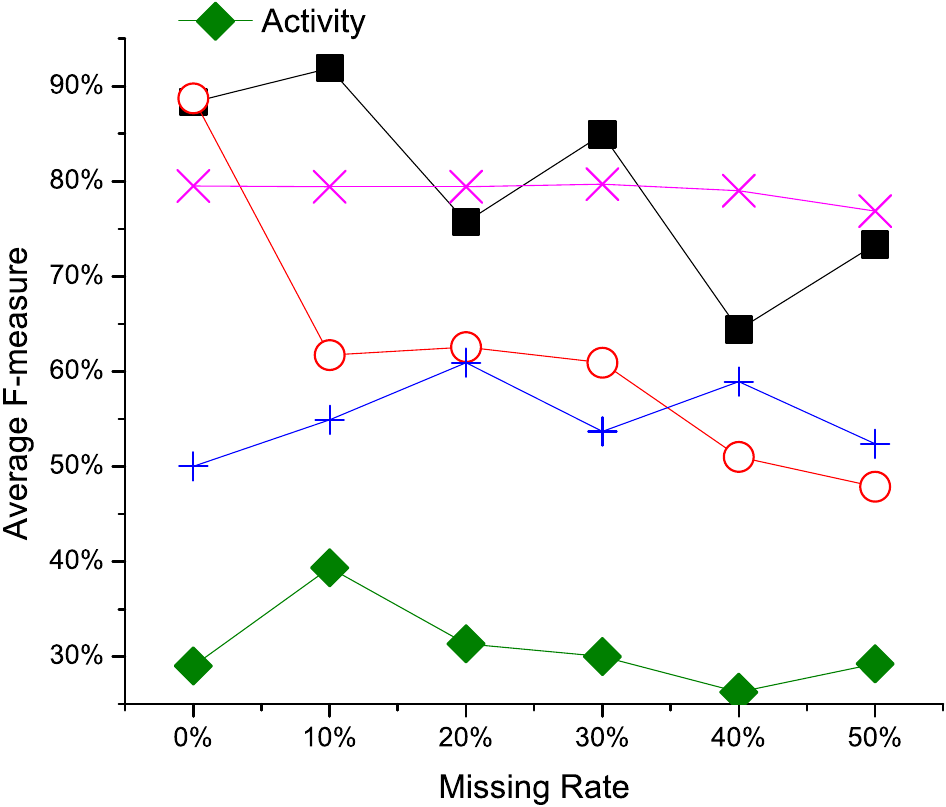}
\label{fig:km-miss-f}
}
\subfigure{
\includegraphics[width=1.4in,height=1.0in]{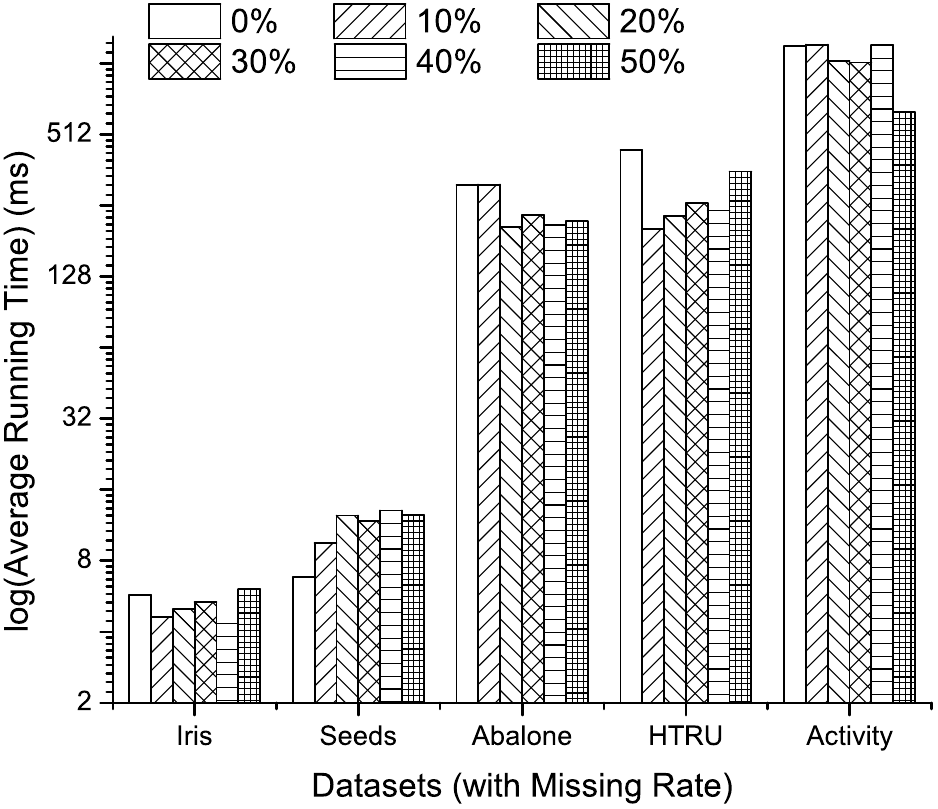}
\label{fig:km-miss-t}
}
\vspace{-2mm}
\caption{Results on Clustering for K-Means Algorithm: Varying Missing Rate.}
\vspace{-2mm}
\label{fig:km-miss}
\end{figure*}

\begin{figure*}[!htb]
\centering
\subfigure{
\includegraphics[width=1.4in,height=1.0in]{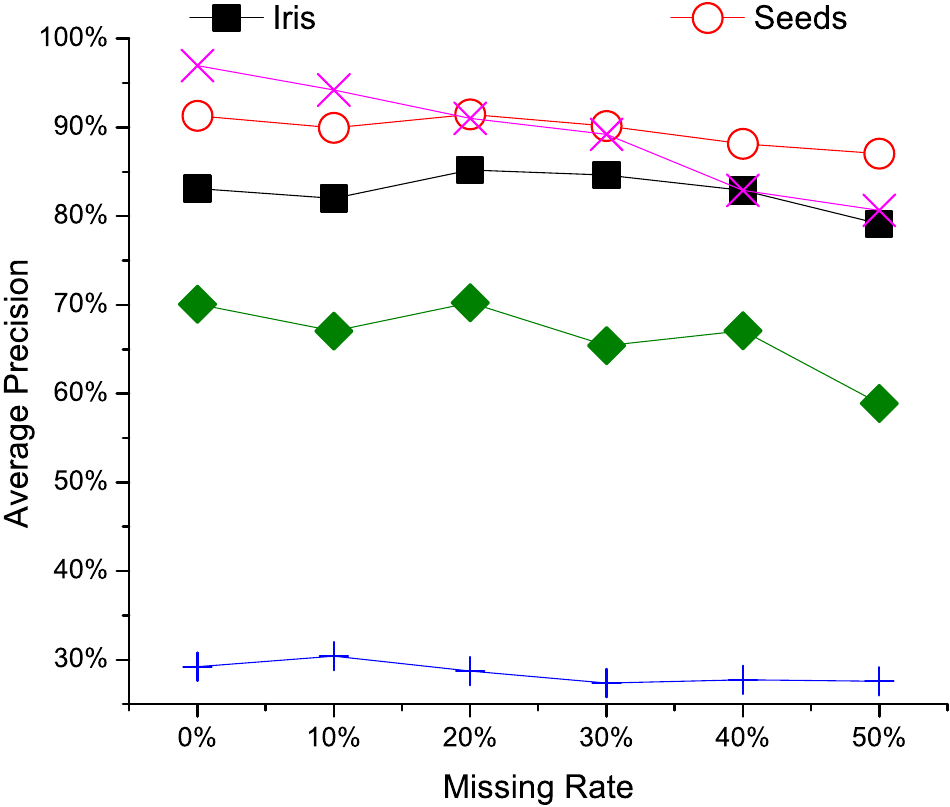}
\label{fig:lvq-miss-p}
}
\subfigure{
\includegraphics[width=1.4in,height=1.0in]{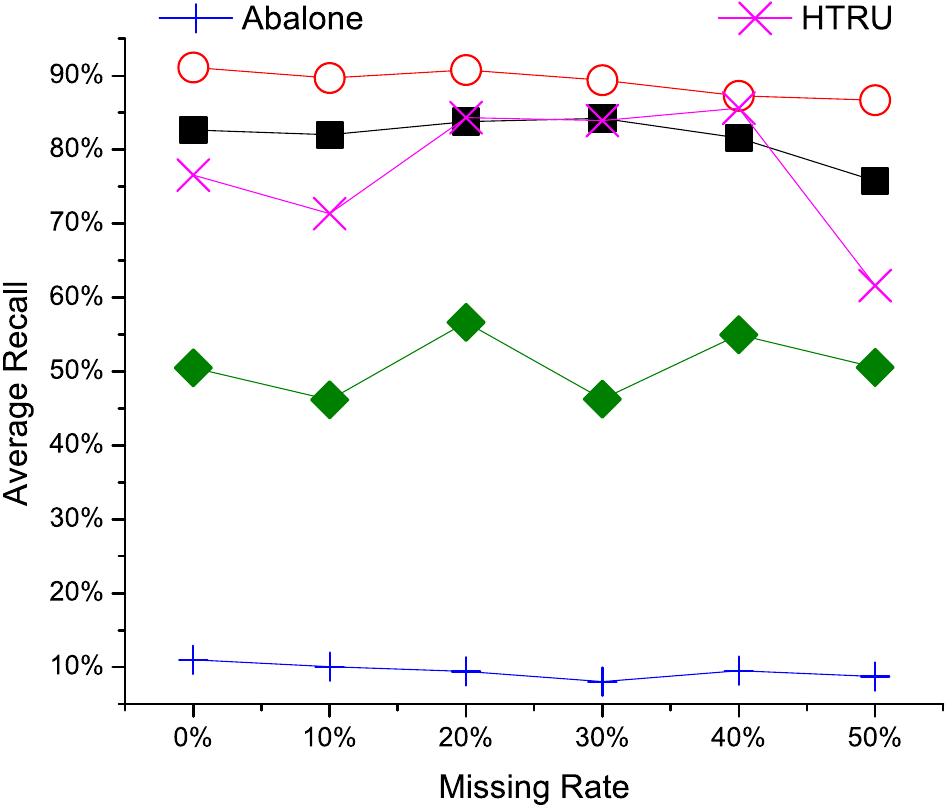}
\label{fig:lvq-miss-r}
}
\subfigure{
\includegraphics[width=1.4in,height=1.0in]{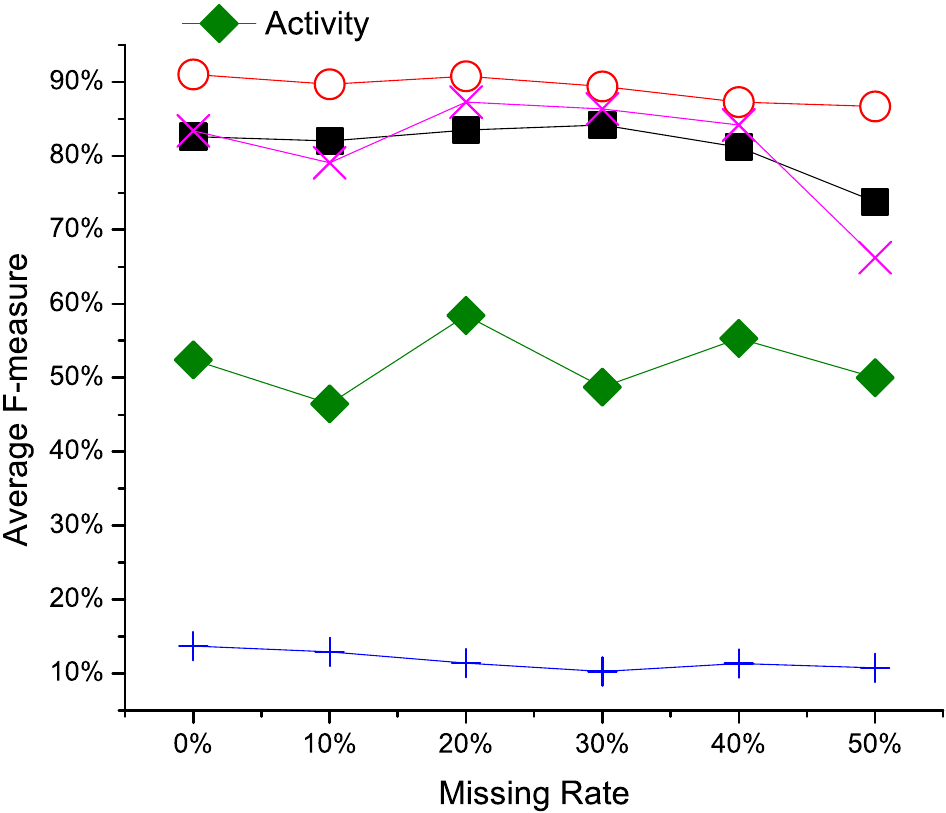}
\label{fig:lvq-miss-f}
}
\subfigure{
\includegraphics[width=1.4in,height=1.0in]{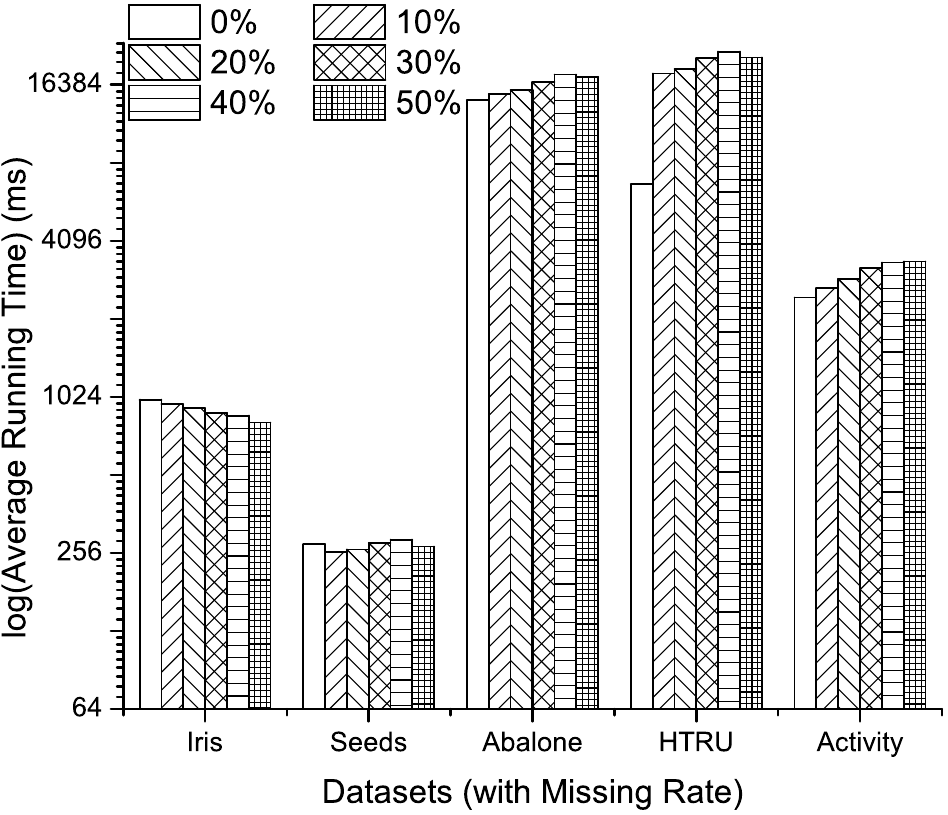}
\label{fig:lvq-miss-t}
}
\vspace{-2mm}
\caption{Results on Clustering for LVQ Algorithm: Varying Missing Rate.}
\vspace{-2mm}
\label{fig:lvq-miss}
\end{figure*}

\begin{figure*}[!htb]
\centering
\subfigure{
\includegraphics[width=1.4in,height=1.0in]{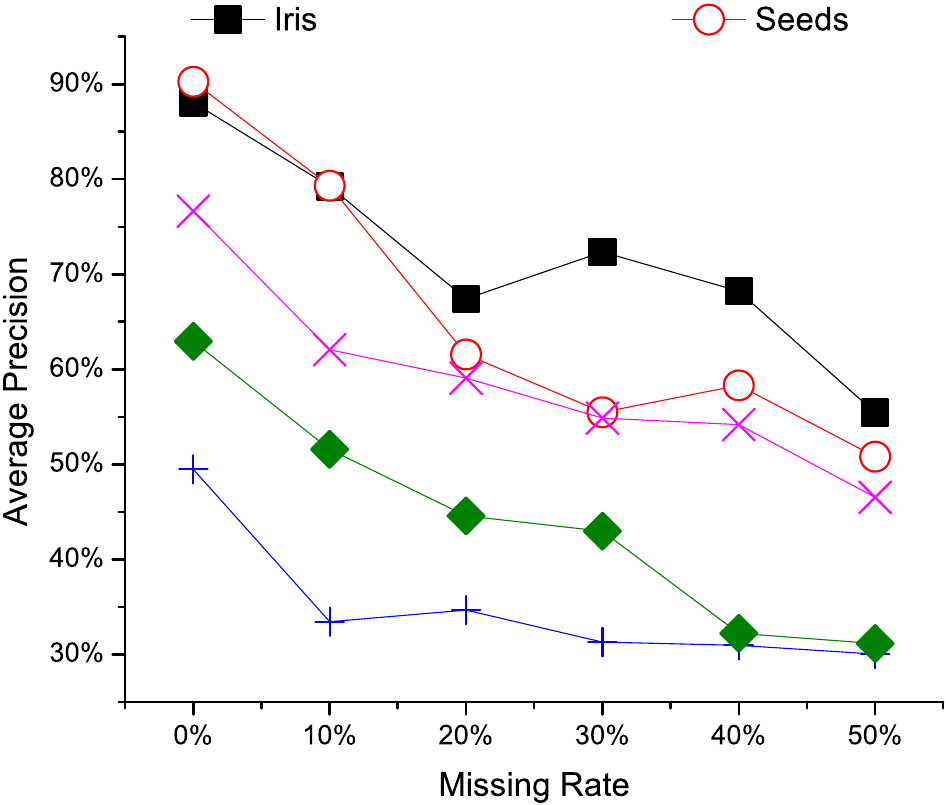}
\label{fig:clr-miss-p}
}
\subfigure{
\includegraphics[width=1.4in,height=1.0in]{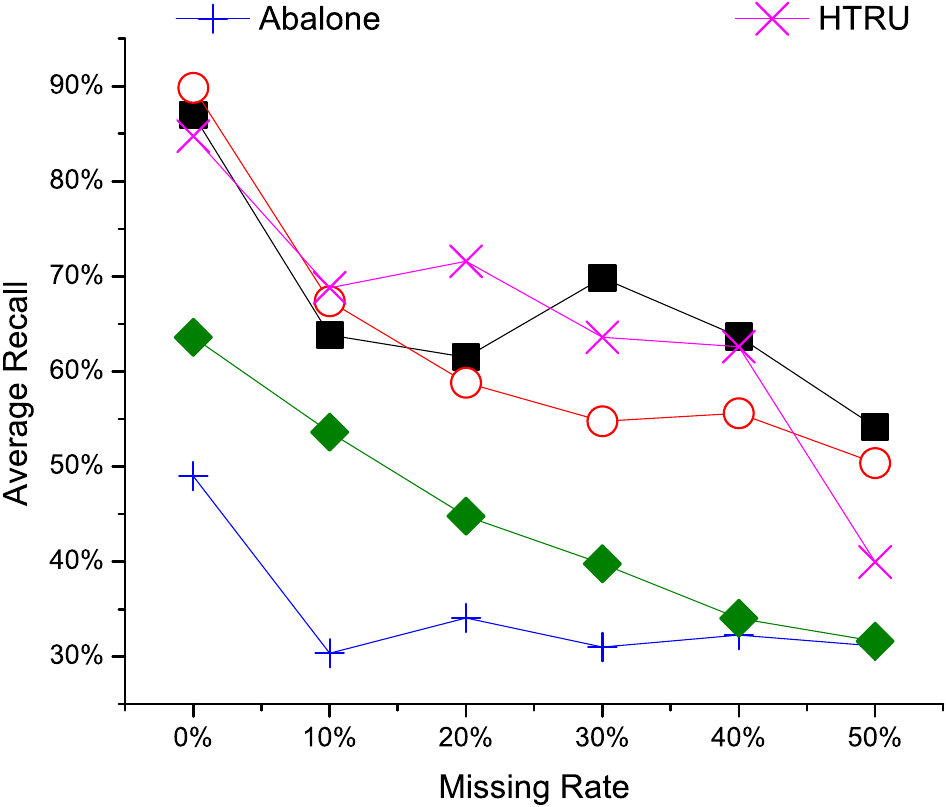}
\label{fig:clr-miss-r}
}
\subfigure{
\includegraphics[width=1.4in,height=1.0in]{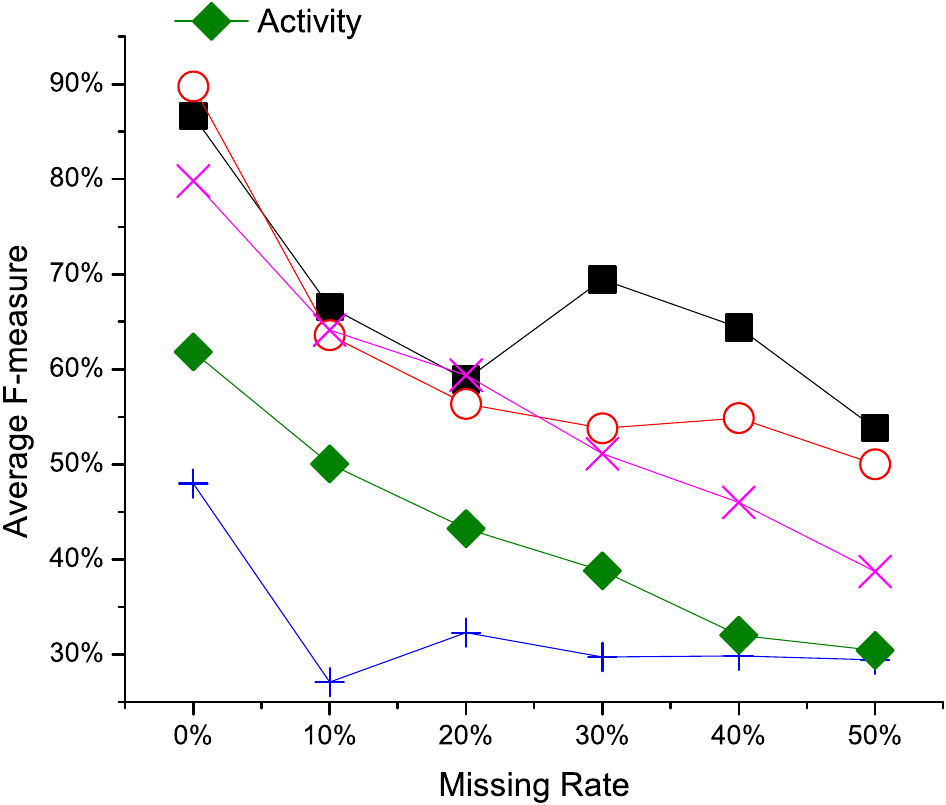}
\label{fig:clr-miss-f}
}
\subfigure{
\includegraphics[width=1.4in,height=1.0in]{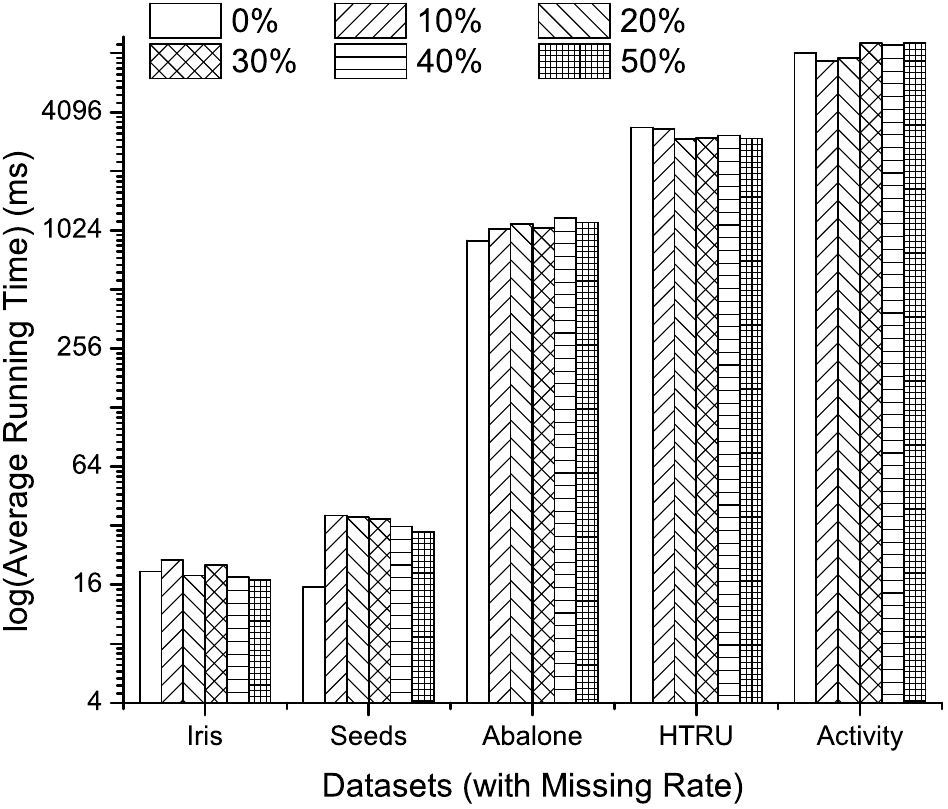}
\label{fig:clr-miss-t}
}
\vspace{-2mm}
\caption{Results on Clustering for CLARANS Algorithm: Varying Missing Rate.}
\vspace{-2mm}
\label{fig:clr-miss}
\end{figure*}

\begin{figure*}[!htb]
\centering
\subfigure{
\includegraphics[width=1.4in,height=1.0in]{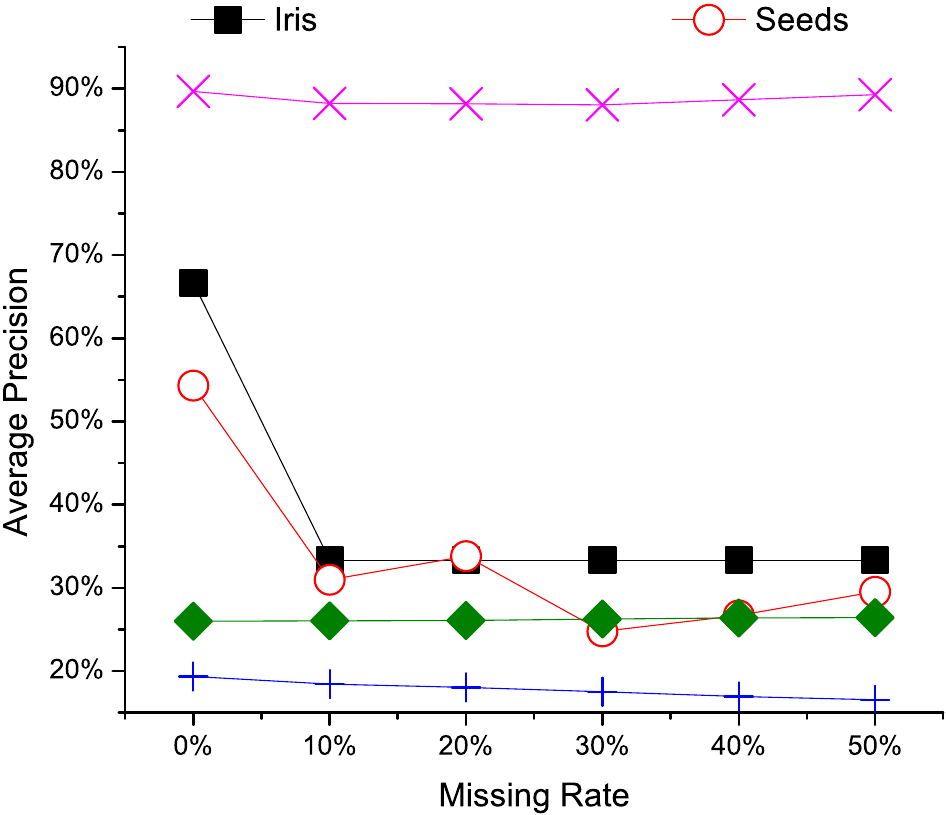}
\label{fig:db-miss-p}
}
\subfigure{
\includegraphics[width=1.4in,height=1.0in]{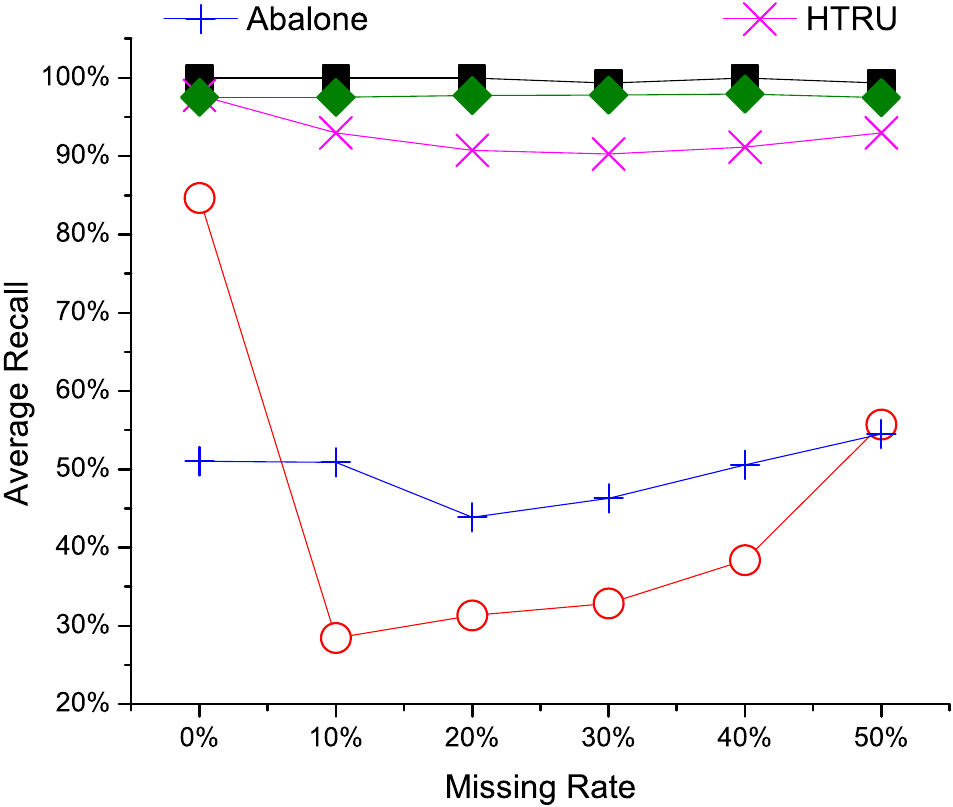}
\label{fig:db-miss-r}
}
\subfigure{
\includegraphics[width=1.4in,height=1.0in]{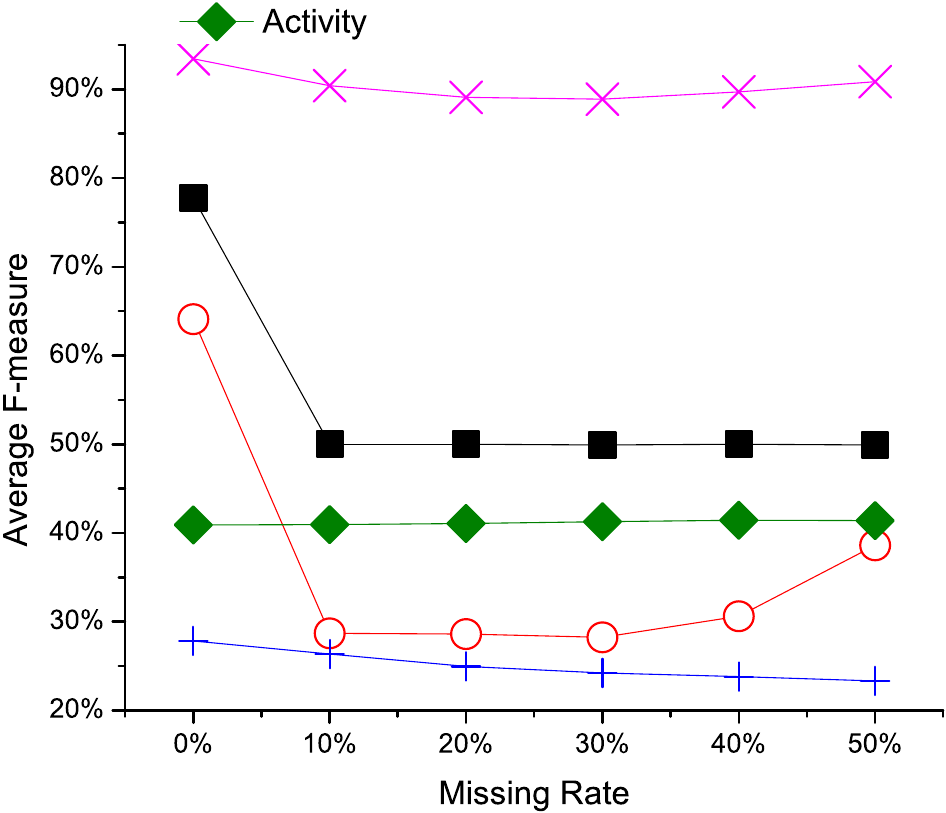}
\label{fig:db-miss-f}
}
\subfigure{
\includegraphics[width=1.4in,height=1.0in]{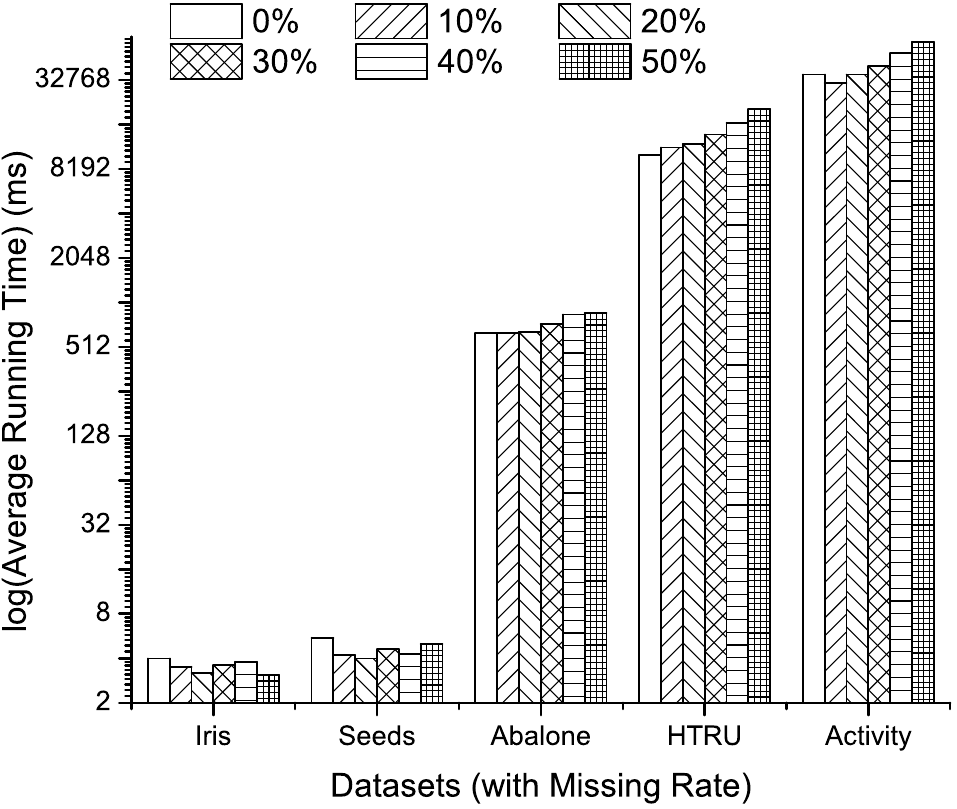}
\label{fig:db-miss-t}
}
\vspace{-2mm}
\caption{Results on Clustering for DBSCAN Algorithm: Varying Missing Rate.}
\vspace{-2mm}
\label{fig:db-miss}
\end{figure*}

\begin{figure*}[!htb]
\centering
\subfigure{
\includegraphics[width=1.4in,height=1.0in]{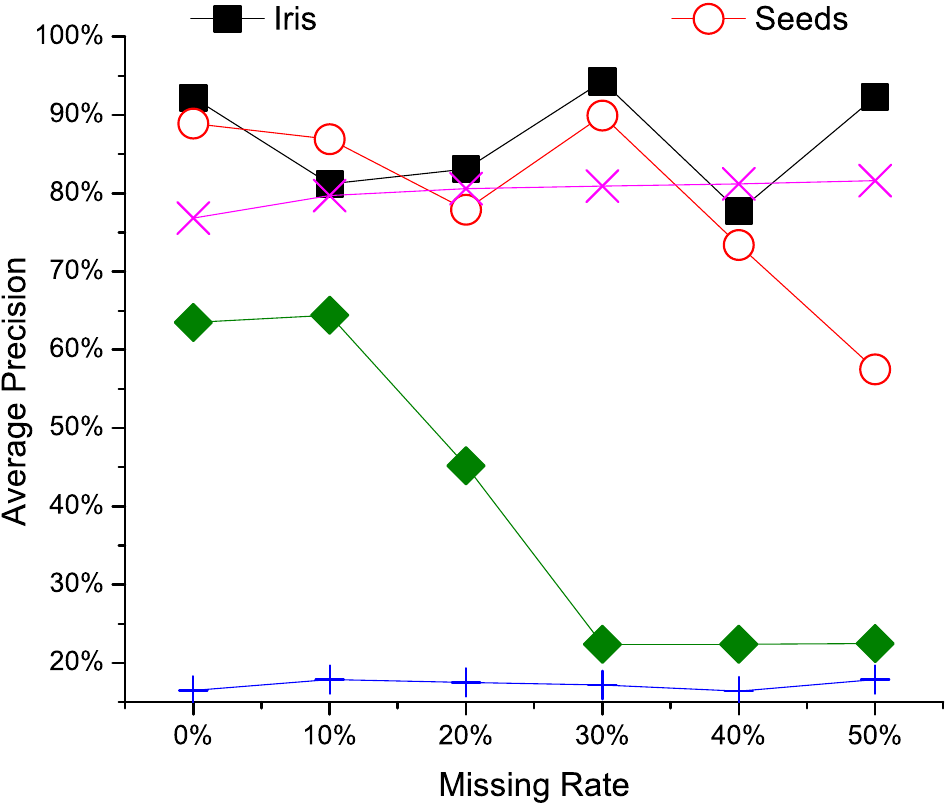}
\label{fig:bir-miss-p}
}
\subfigure{
\includegraphics[width=1.4in,height=1.0in]{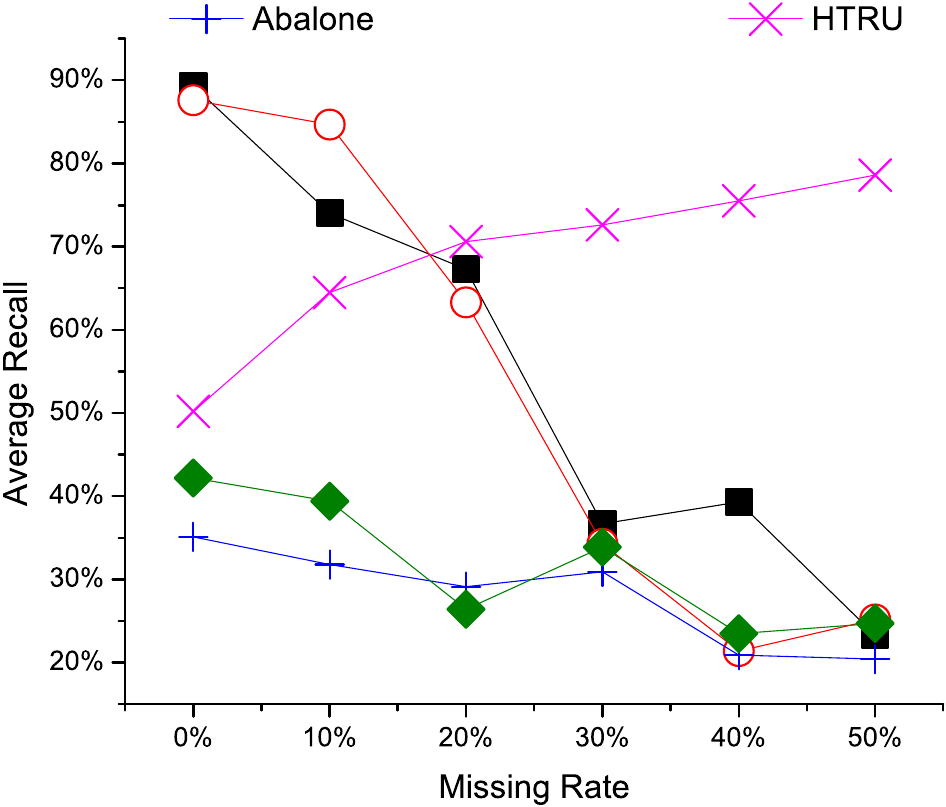}
\label{fig:bir-miss-r}
}
\subfigure{
\includegraphics[width=1.4in,height=1.0in]{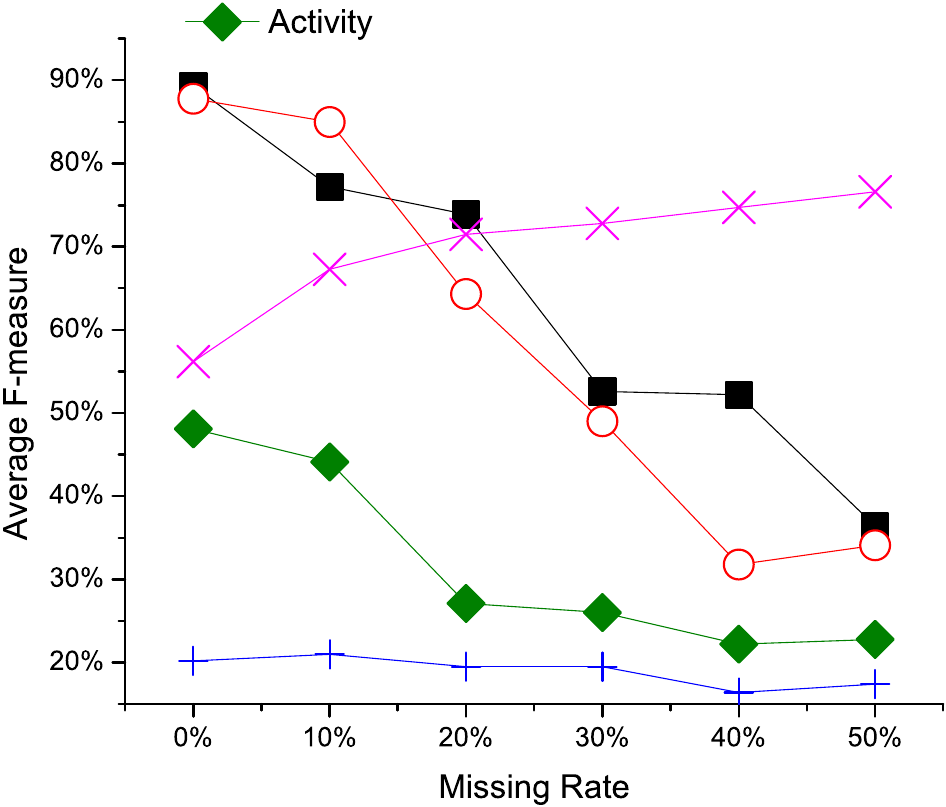}
\label{fig:bir-miss-f}
}
\subfigure{
\includegraphics[width=1.4in,height=1.0in]{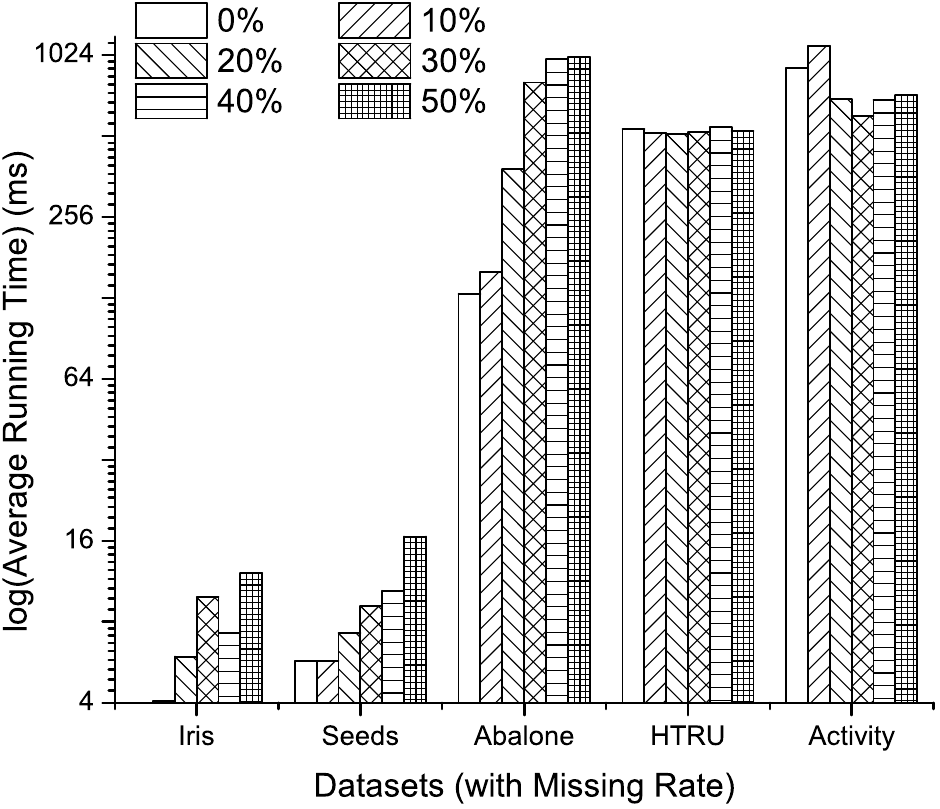}
\label{fig:bir-miss-t}
}
\vspace{-2mm}
\caption{Results on Clustering for BIRCH Algorithm: Varying Missing Rate.}
\vspace{-2mm}
\label{fig:bir-miss}
\end{figure*}

\begin{figure*}[!htb]
\centering
\subfigure{
\includegraphics[width=1.4in,height=1.0in]{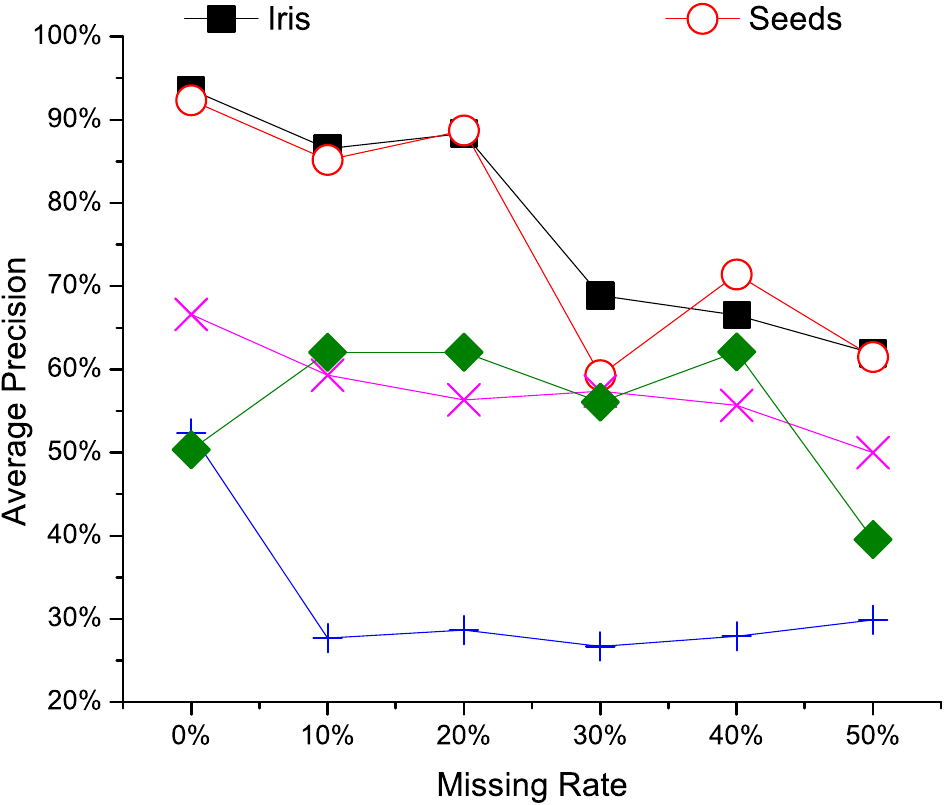}
\label{fig:cure-miss-p}
}
\subfigure{
\includegraphics[width=1.4in,height=1.0in]{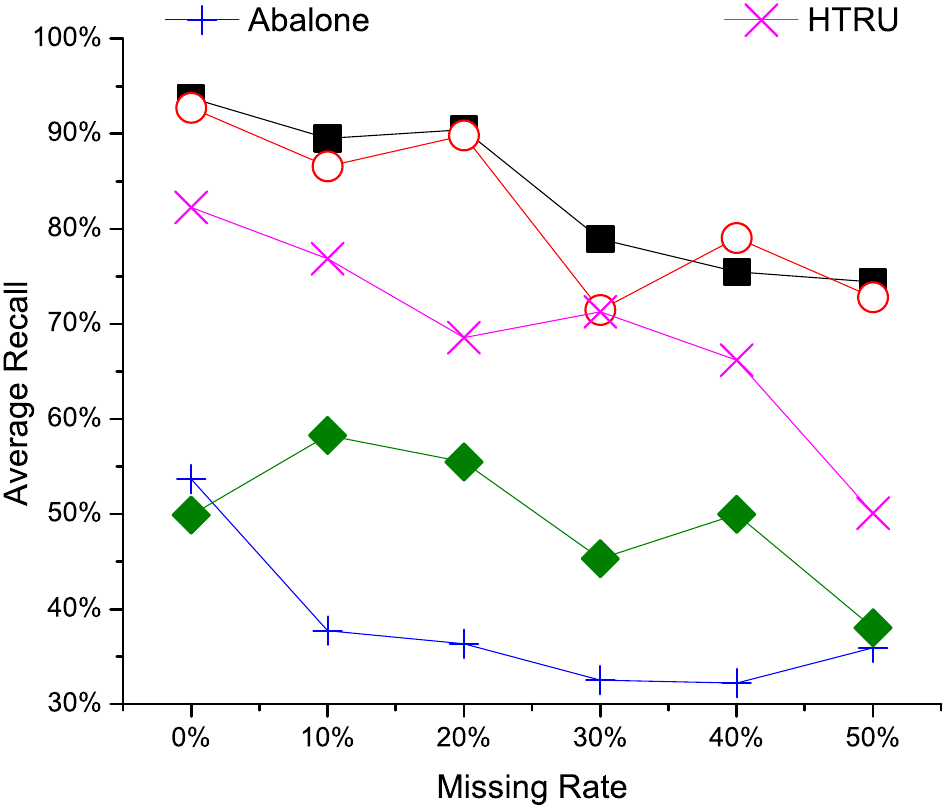}
\label{fig:cure-miss-r}
}
\subfigure{
\includegraphics[width=1.4in,height=1.0in]{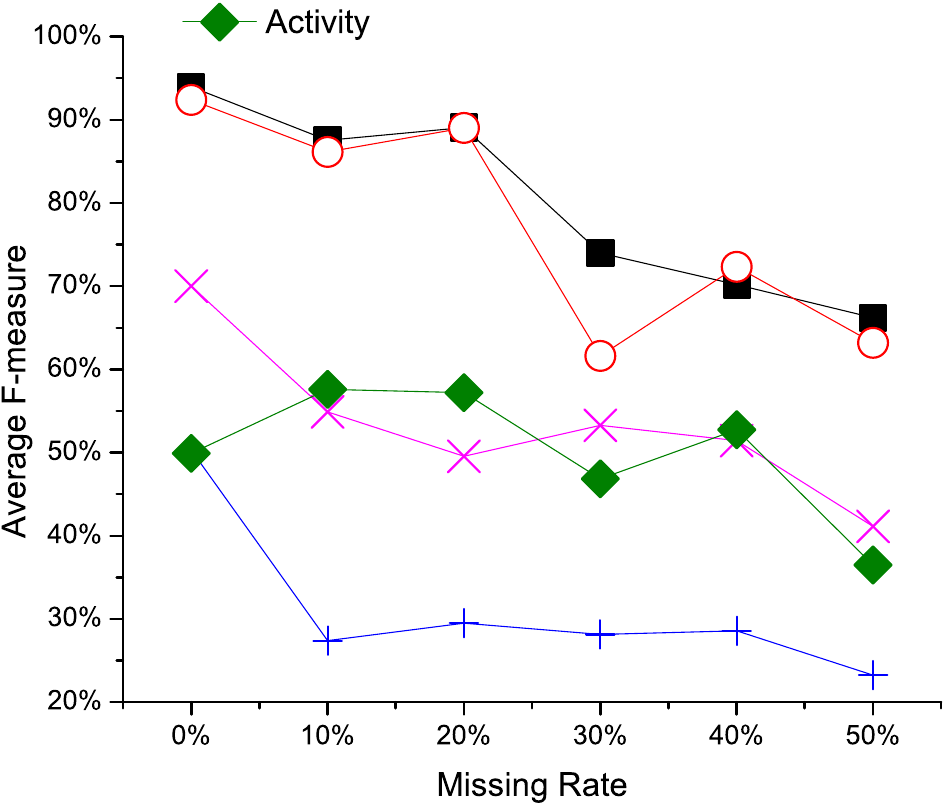}
\label{fig:cure-miss-f}
}
\subfigure{
\includegraphics[width=1.4in,height=1.0in]{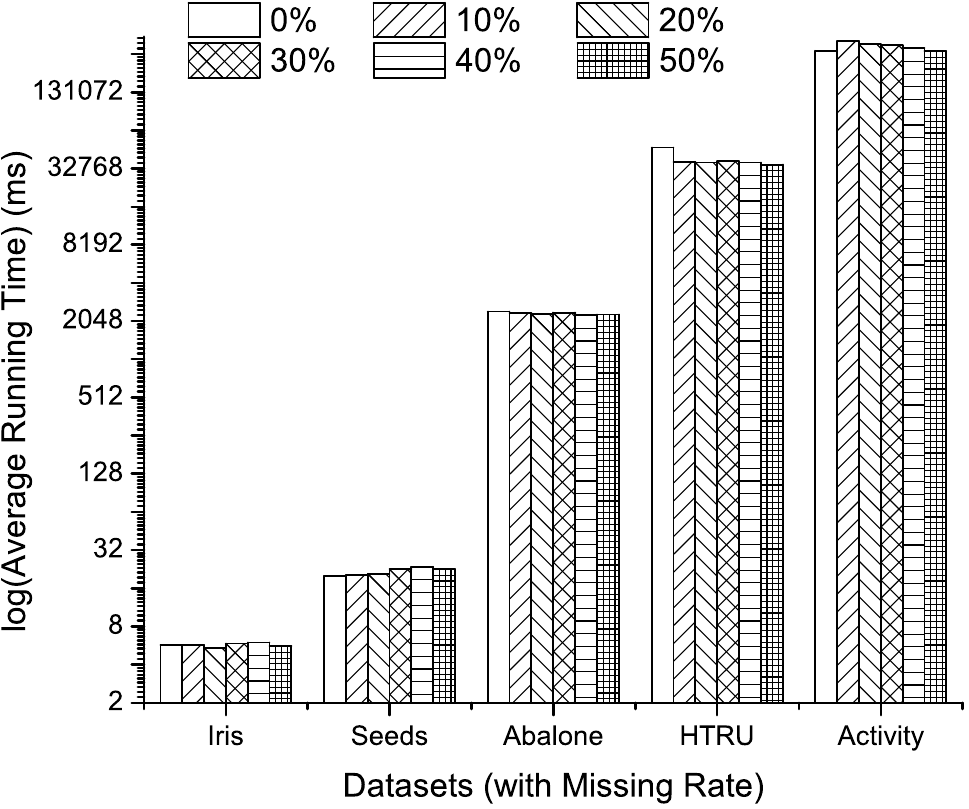}
\label{fig:cure-miss-t}
}
\vspace{-2mm}
\caption{Results on Clustering for CURE Algorithm: Varying Missing Rate.}
\vspace{-2mm}
\label{fig:cure-miss}
\end{figure*}

\begin{figure*}[!htb]
\centering
\subfigure{
\includegraphics[width=1.4in,height=1.0in]{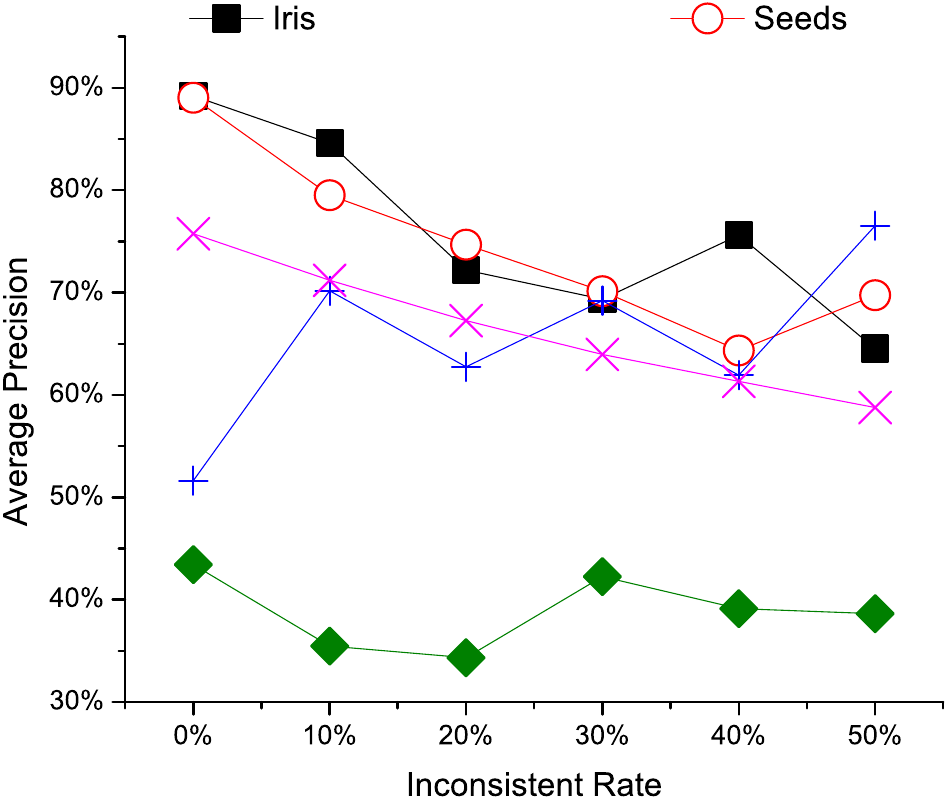}
\label{fig:km-incons-p}
}
\subfigure{
\includegraphics[width=1.4in,height=1.0in]{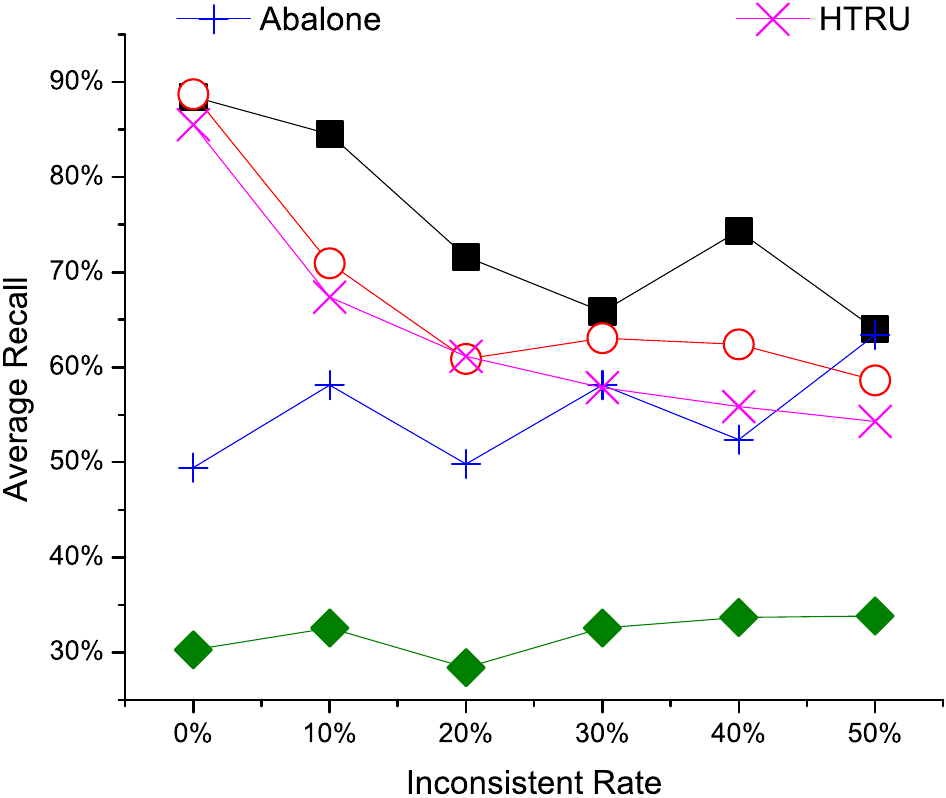}
\label{fig:km-incons-r}
}
\subfigure{
\includegraphics[width=1.4in,height=1.0in]{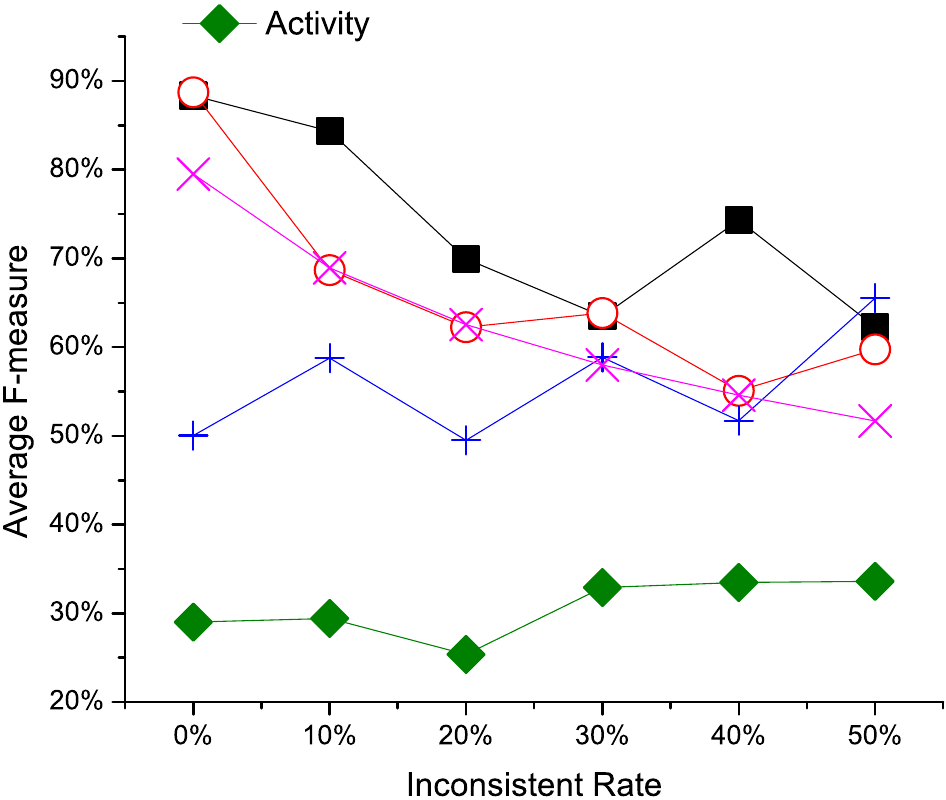}
\label{fig:km-incons-f}
}
\subfigure{
\includegraphics[width=1.4in,height=1.0in]{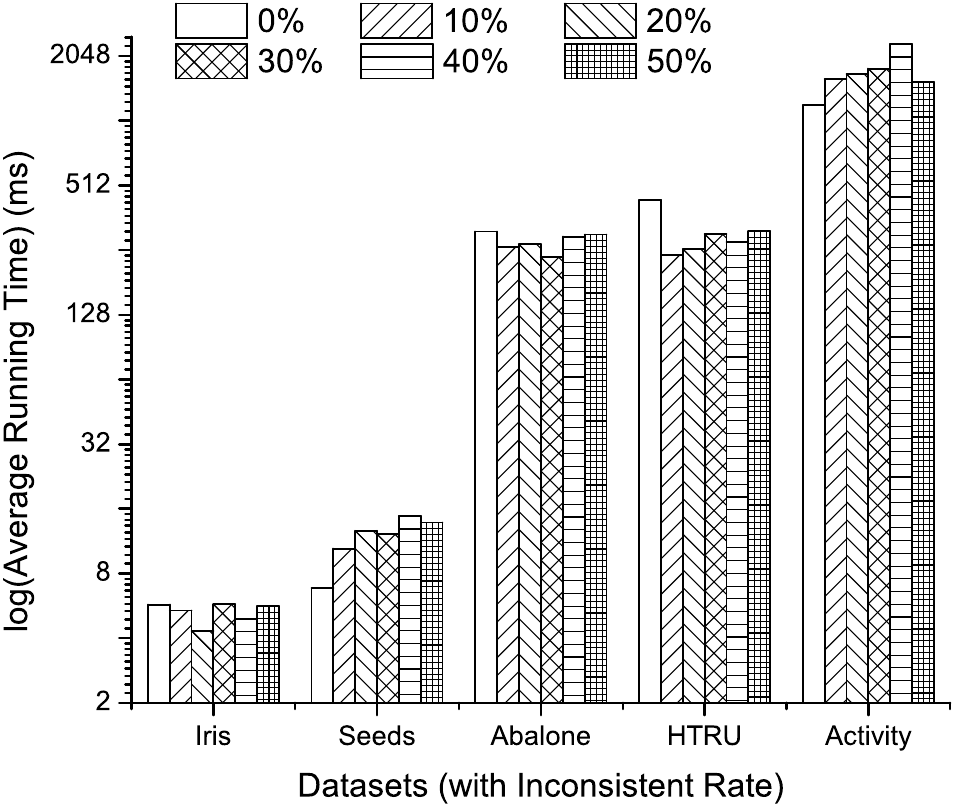}
\label{fig:km-incons-t}
}
\vspace{-2mm}
\caption{Results on Clustering for K-Means Algorithm: Varying Inconsistent Rate.}
\vspace{-2mm}
\label{fig:km-incons}
\end{figure*}

\begin{figure*}[!htb]
\centering
\subfigure{
\includegraphics[width=1.4in,height=1.0in]{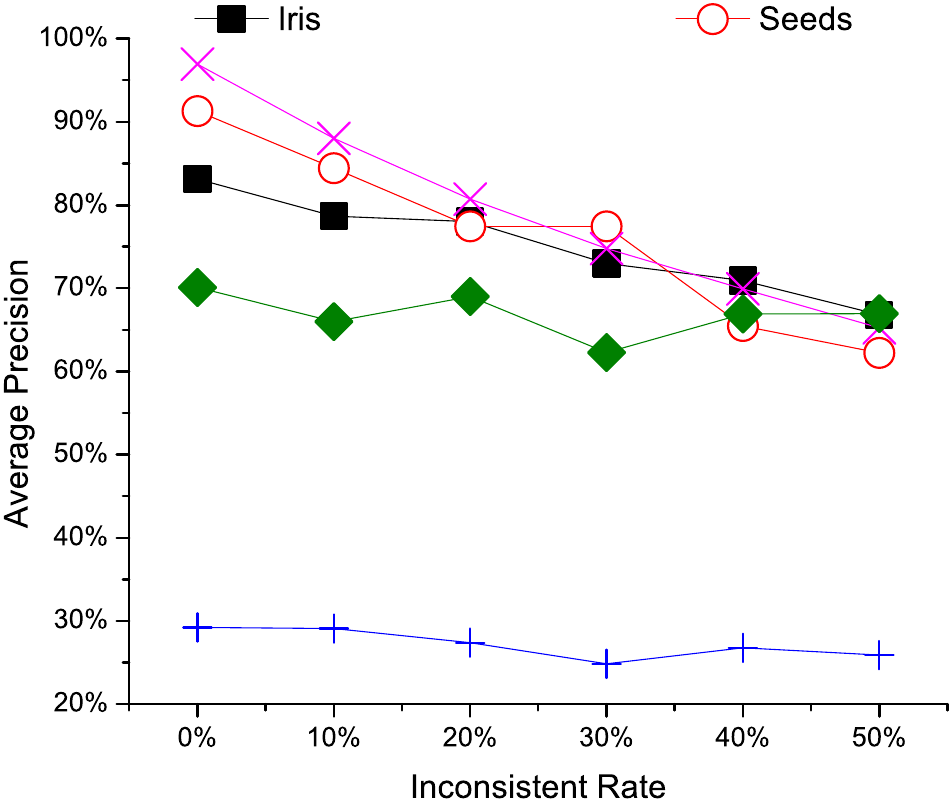}
\label{fig:lvq-incons-p}
}
\subfigure{
\includegraphics[width=1.4in,height=1.0in]{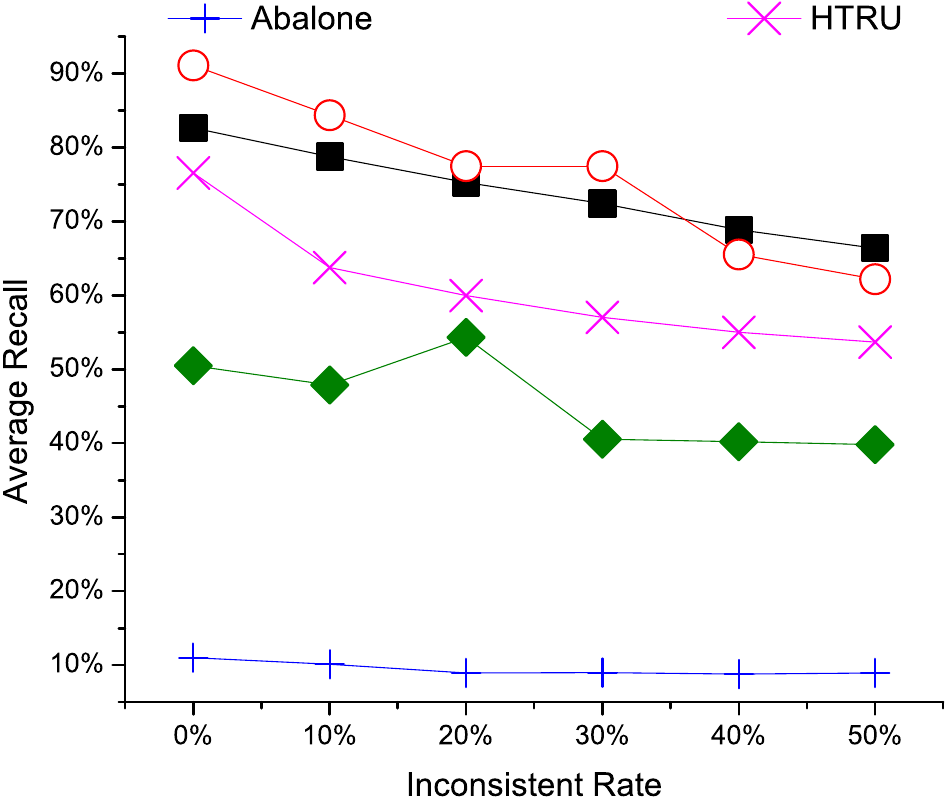}
\label{fig:lvq-incons-r}
}
\subfigure{
\includegraphics[width=1.4in,height=1.0in]{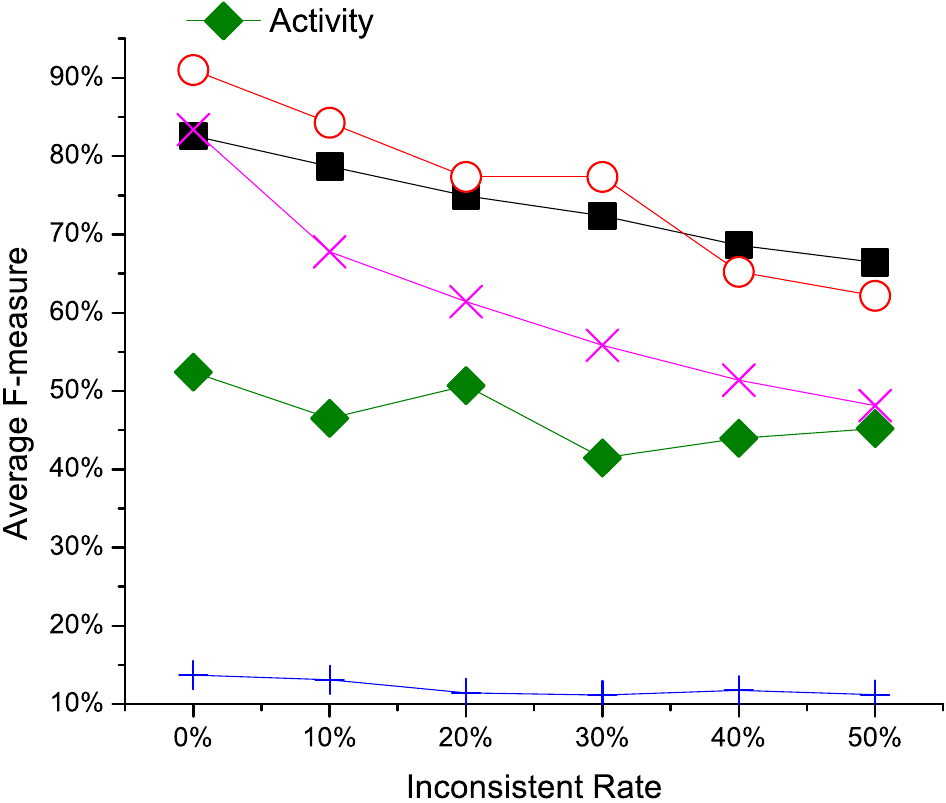}
\label{fig:lvq-incons-f}
}
\subfigure{
\includegraphics[width=1.4in,height=1.0in]{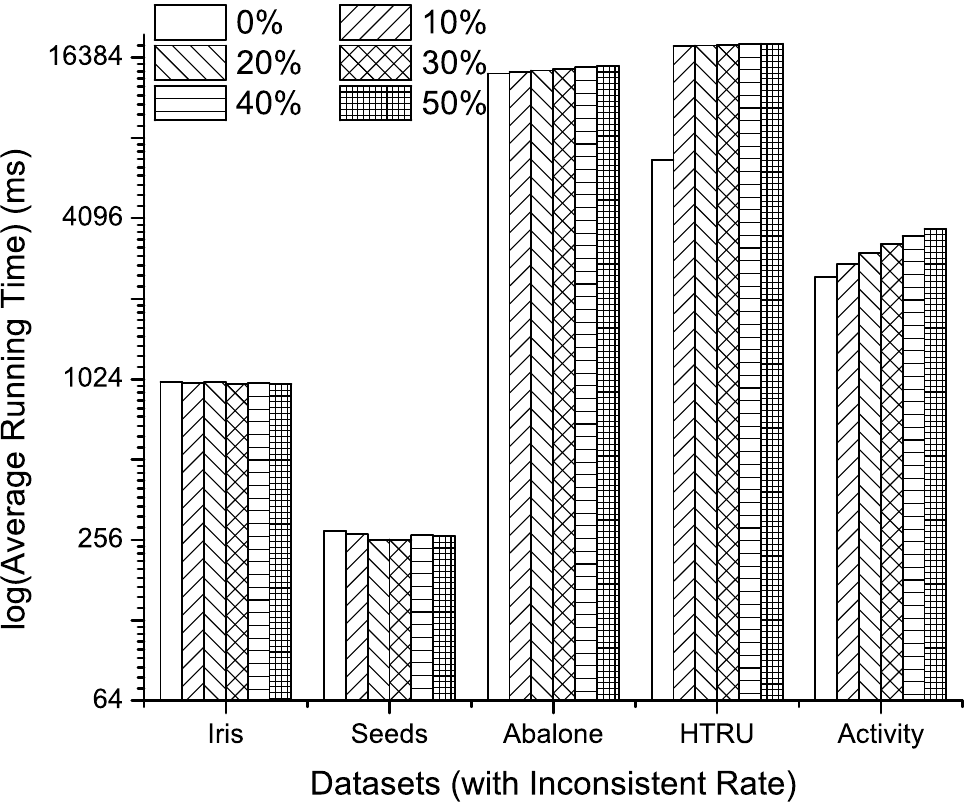}
\label{fig:lvq-incons-t}
}
\vspace{-2mm}
\caption{Results on Clustering for LVQ Algorithm: Varying Inconsistent Rate.}
\vspace{-2mm}
\label{fig:lvq-incons}
\end{figure*}

\begin{figure*}[!htb]
\centering
\subfigure{
\includegraphics[width=1.4in,height=1.0in]{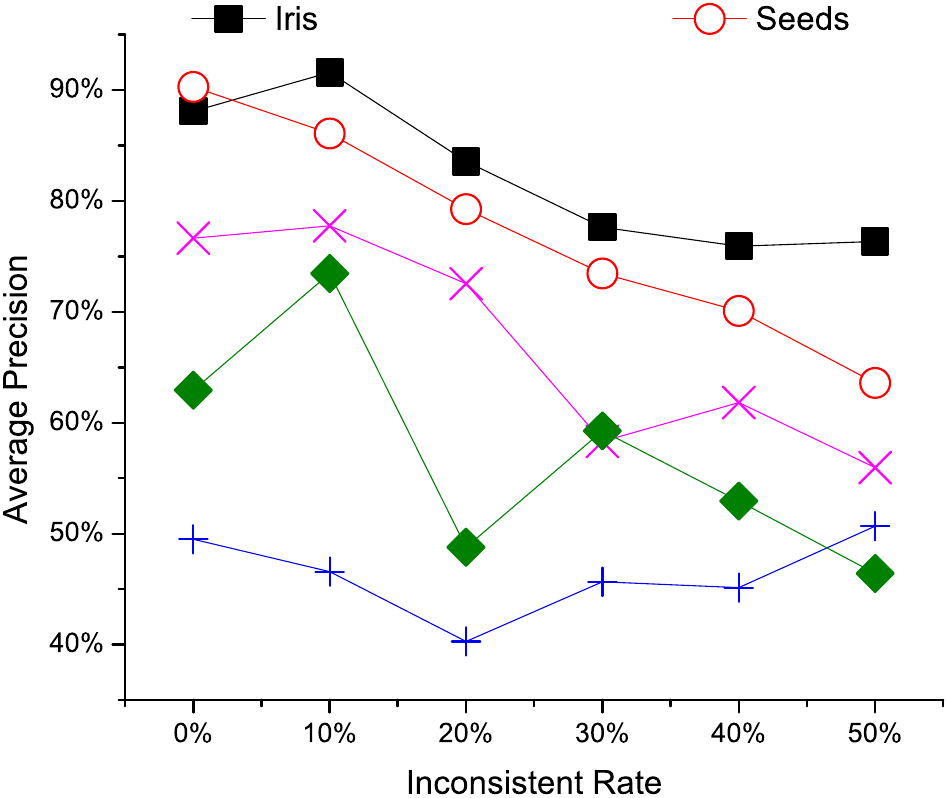}
\label{fig:clr-incons-p}
}
\subfigure{
\includegraphics[width=1.4in,height=1.0in]{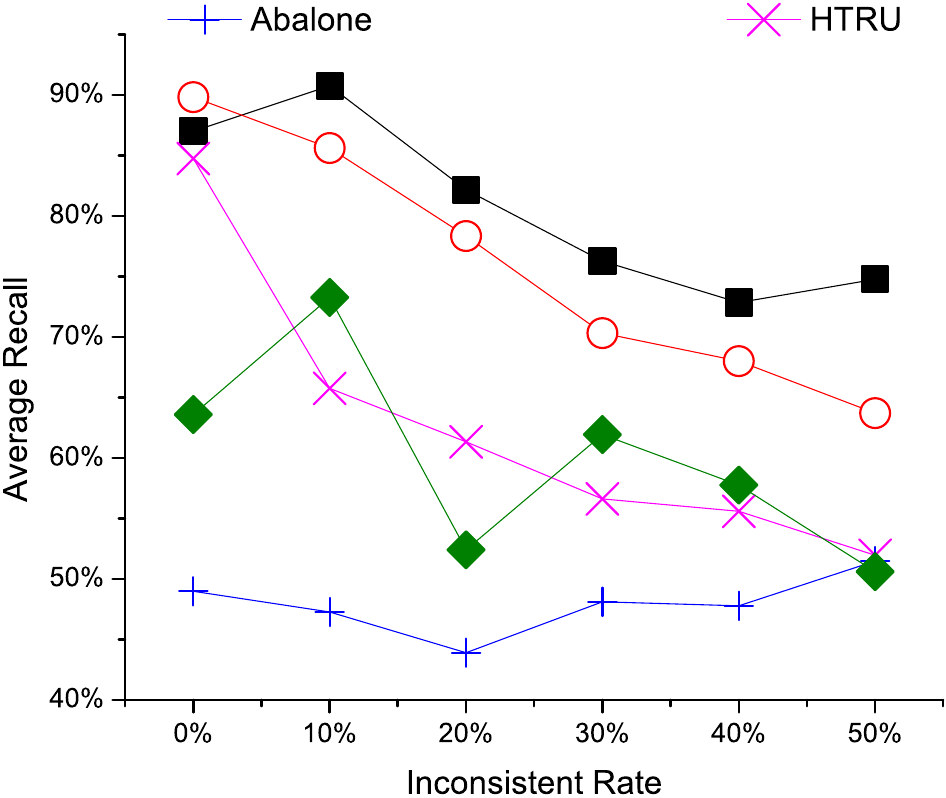}
\label{fig:clr-incons-r}
}
\subfigure{
\includegraphics[width=1.4in,height=1.0in]{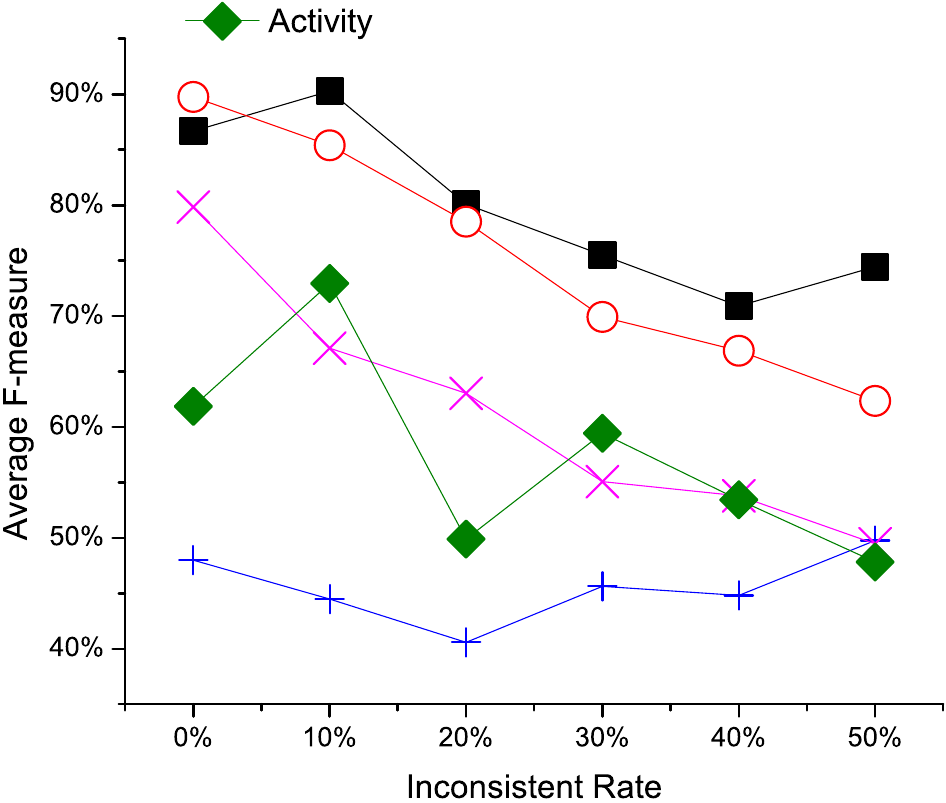}
\label{fig:clr-incons-f}
}
\subfigure{
\includegraphics[width=1.4in,height=1.0in]{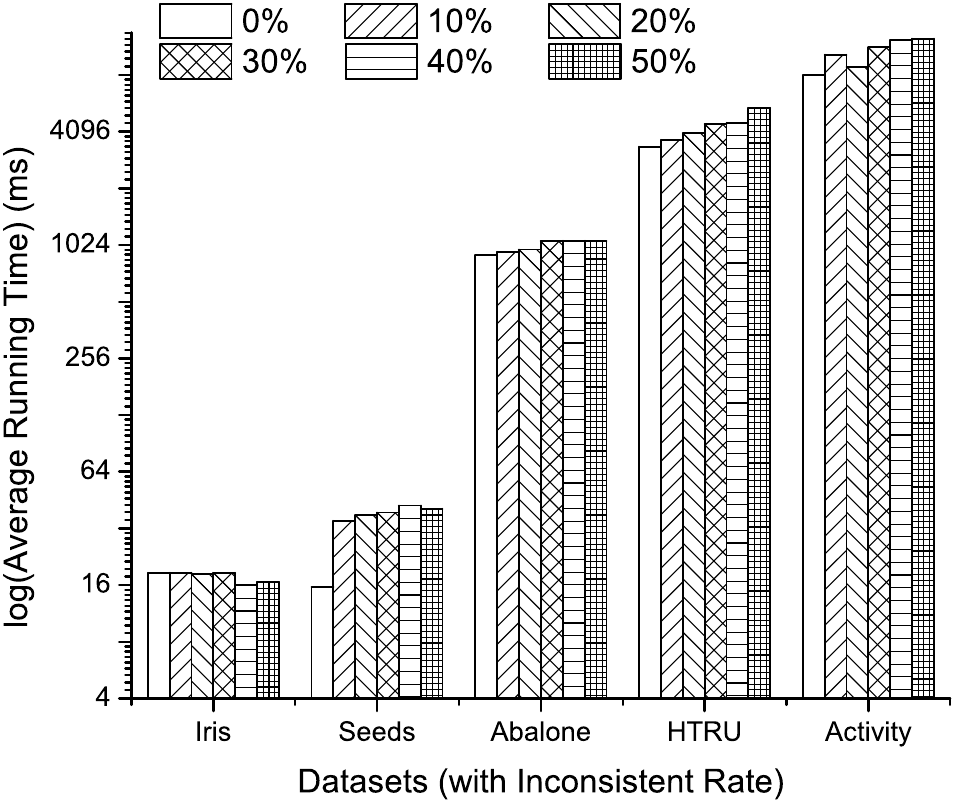}
\label{fig:clr-incons-t}
}
\vspace{-2mm}
\caption{Results on Clustering for CLARANS Algorithm: Varying Inconsistent Rate.}
\vspace{-2mm}
\label{fig:clr-incons}
\end{figure*}

\begin{figure*}[!htb]
\centering
\subfigure{
\includegraphics[width=1.4in,height=1.0in]{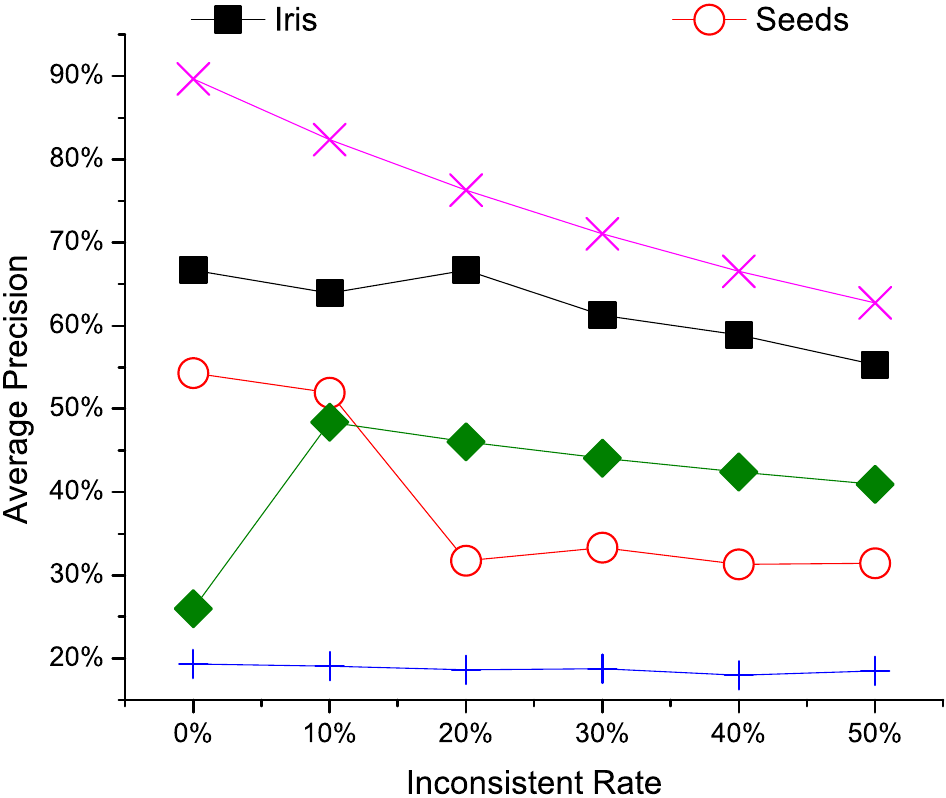}
\label{fig:db-incons-p}
}
\subfigure{
\includegraphics[width=1.4in,height=1.0in]{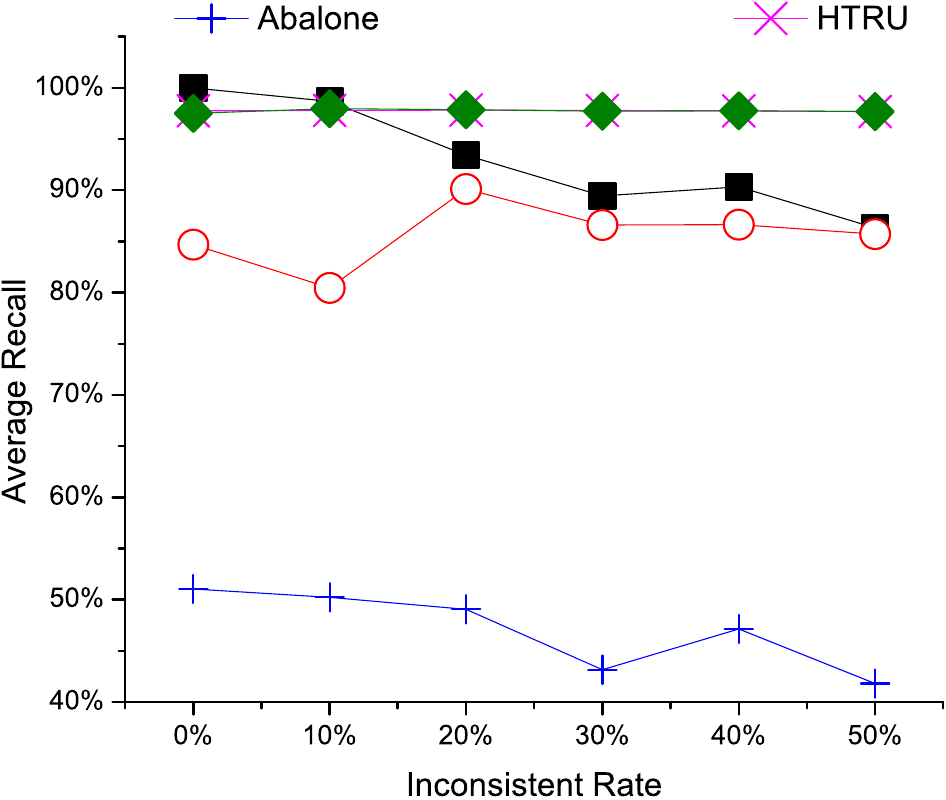}
\label{fig:db-incons-r}
}
\subfigure{
\includegraphics[width=1.4in,height=1.0in]{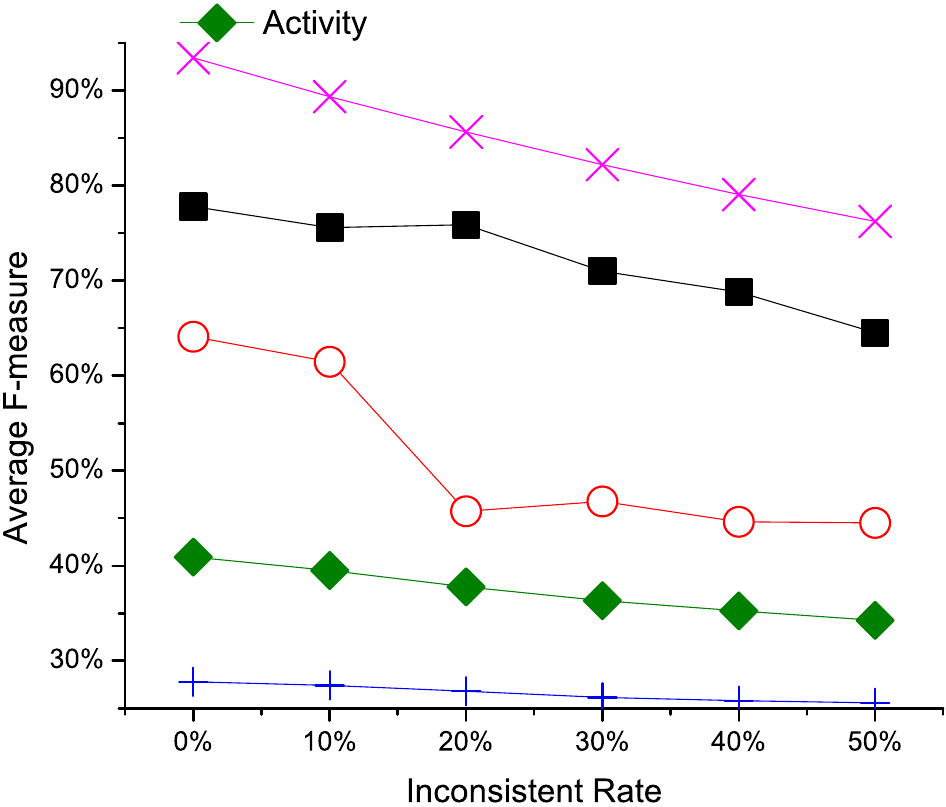}
\label{fig:db-incons-f}
}
\subfigure{
\includegraphics[width=1.4in,height=1.0in]{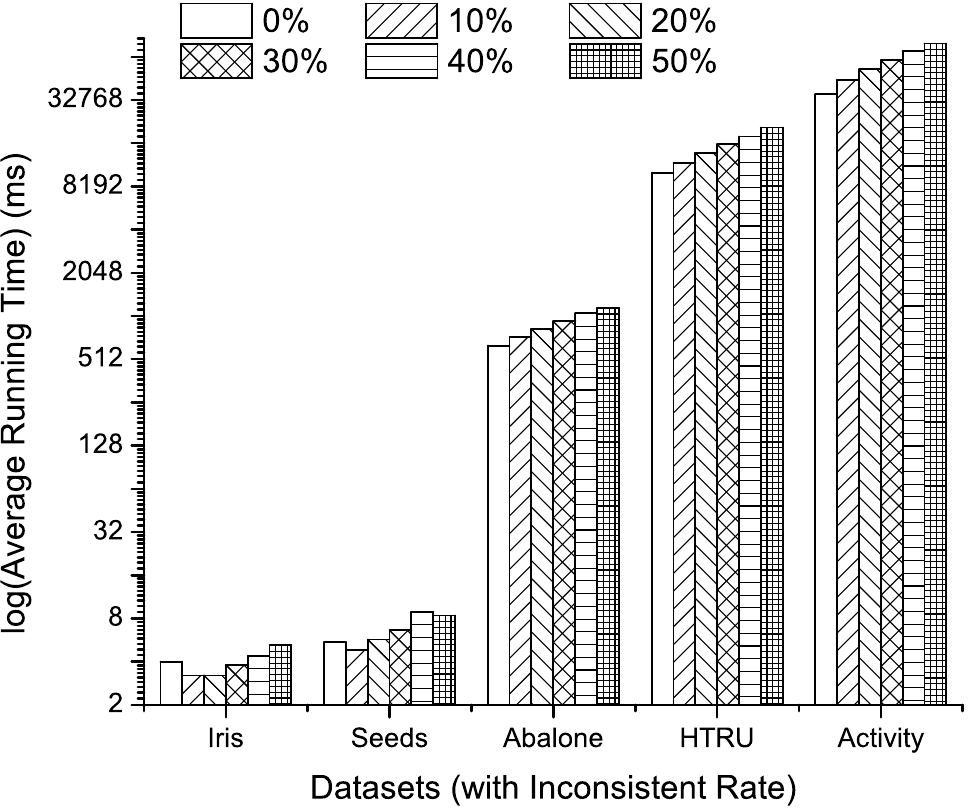}
\label{fig:db-incons-t}
}
\vspace{-2mm}
\caption{Results on Clustering for DBSCAN Algorithm: Varying Inconsistent Rate.}
\vspace{-2mm}
\label{fig:db-incons}
\end{figure*}

\clearpage
\begin{figure*}[!htb]
\centering
\subfigure{
\includegraphics[width=1.4in,height=1.0in]{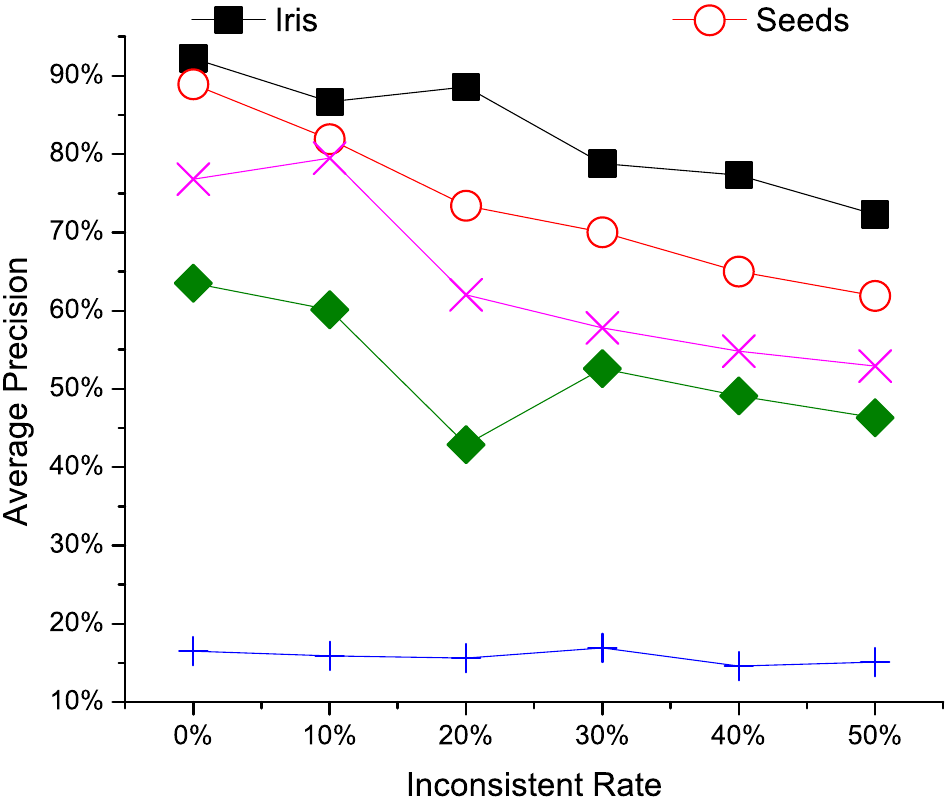}
\label{fig:bir-incons-p}
}
\subfigure{
\includegraphics[width=1.4in,height=1.0in]{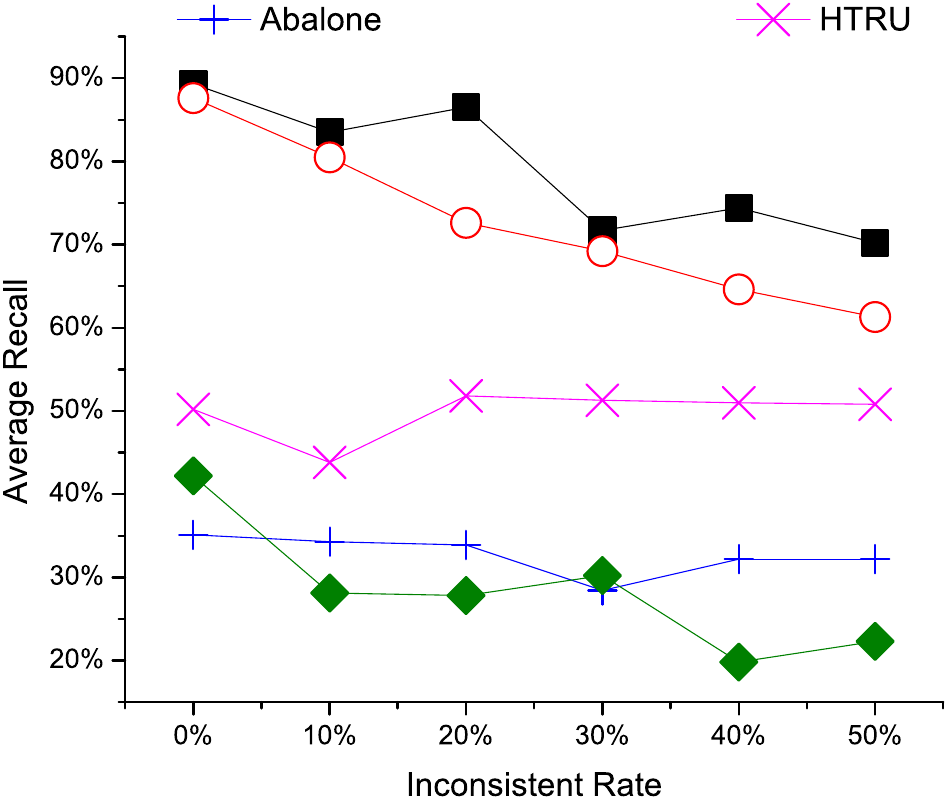}
\label{fig:bir-incons-r}
}
\subfigure{
\includegraphics[width=1.4in,height=1.0in]{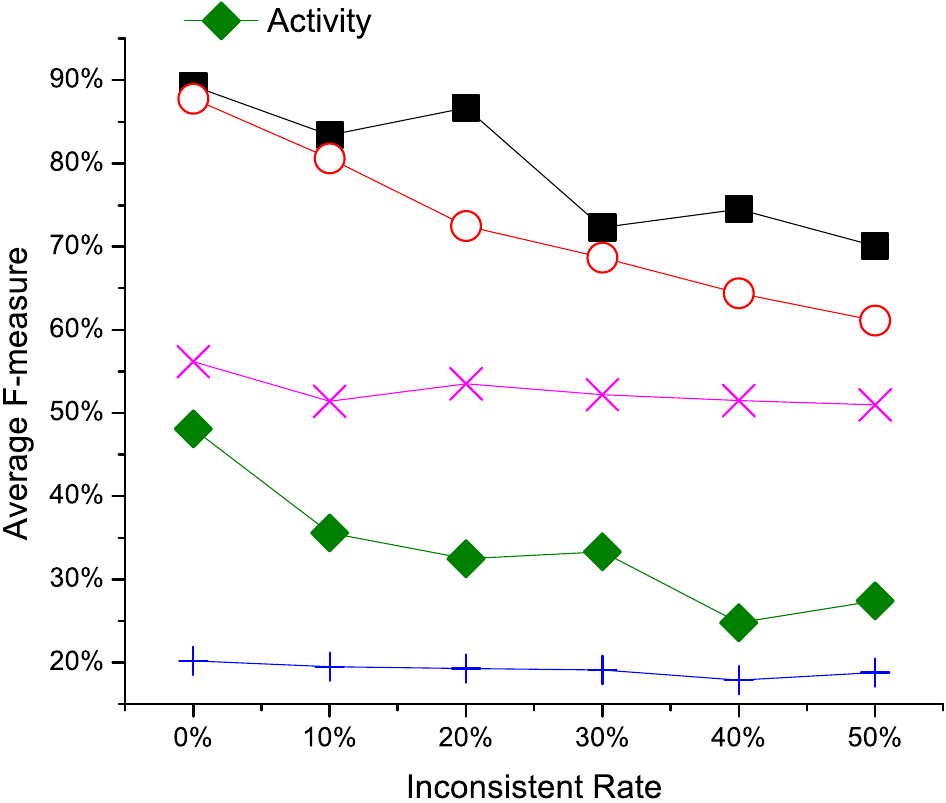}
\label{fig:bir-incons-f}
}
\subfigure{
\includegraphics[width=1.4in,height=1.0in]{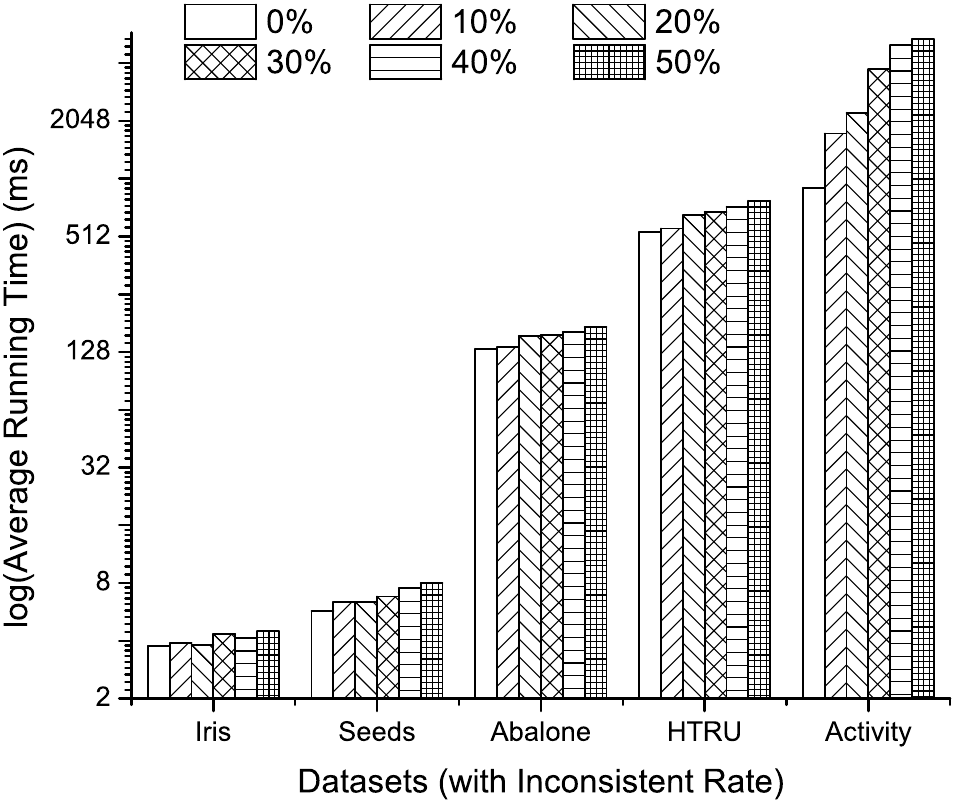}
\label{fig:bir-incons-t}
}
\vspace{-2mm}
\caption{Results on Clustering for BIRCH Algorithm: Varying Inconsistent Rate.}
\vspace{-2mm}
\label{fig:bir-incons}
\end{figure*}

\begin{figure*}[!htb]
\centering
\subfigure{
\includegraphics[width=1.4in,height=1.0in]{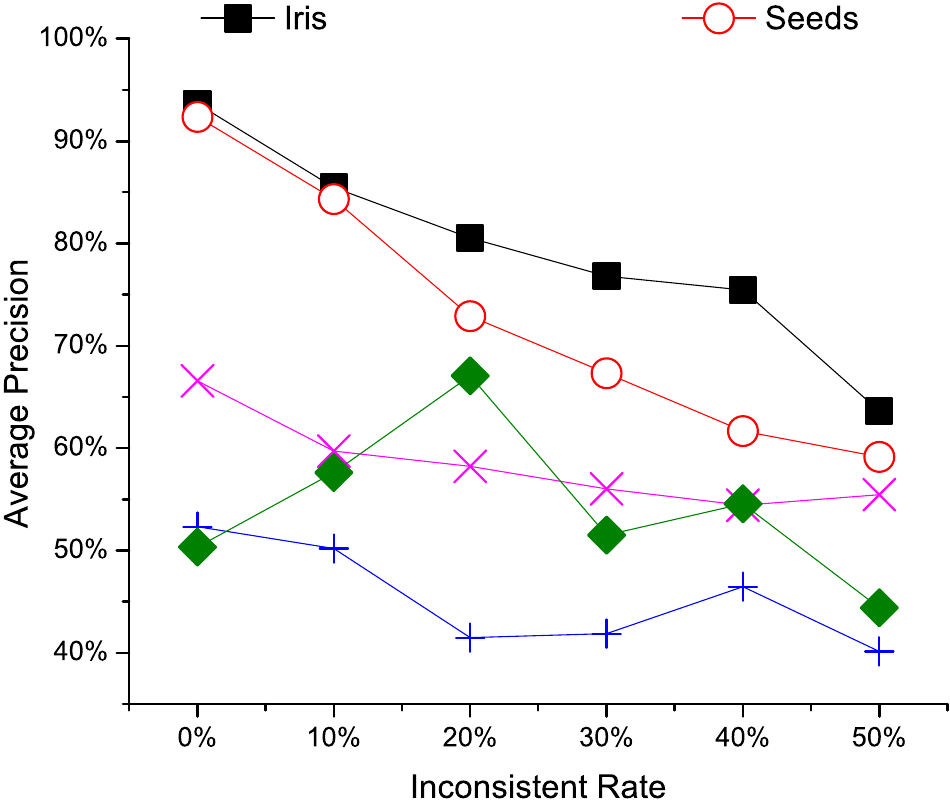}
\label{fig:cure-incons-p}
}
\subfigure{
\includegraphics[width=1.4in,height=1.0in]{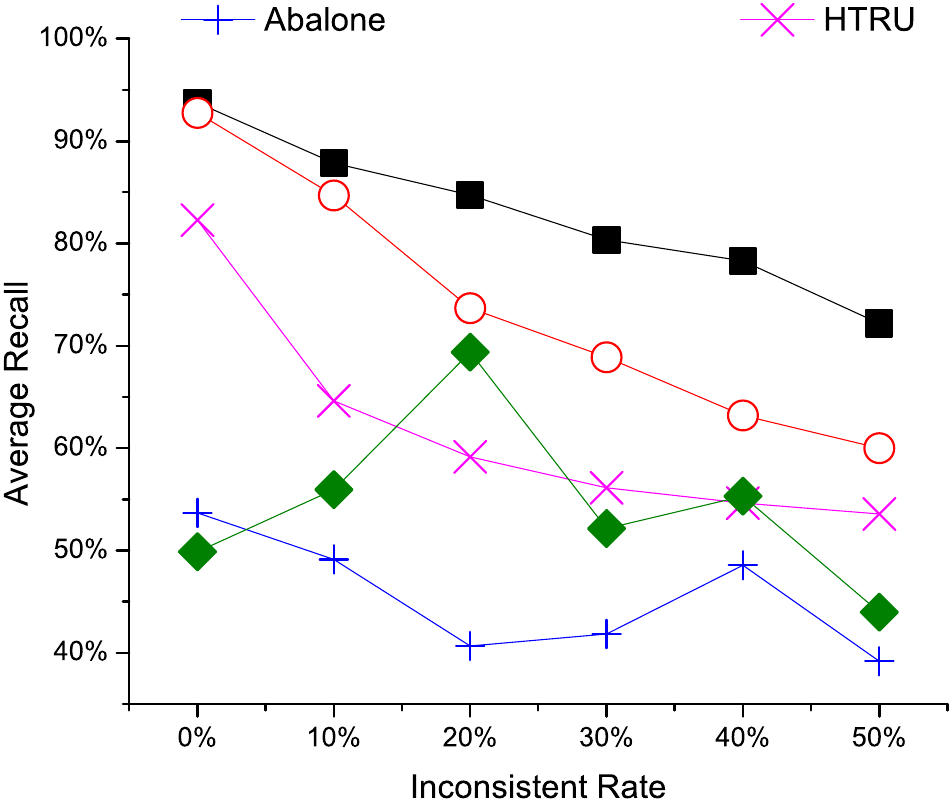}
\label{fig:cure-incons-r}
}
\subfigure{
\includegraphics[width=1.4in,height=1.0in]{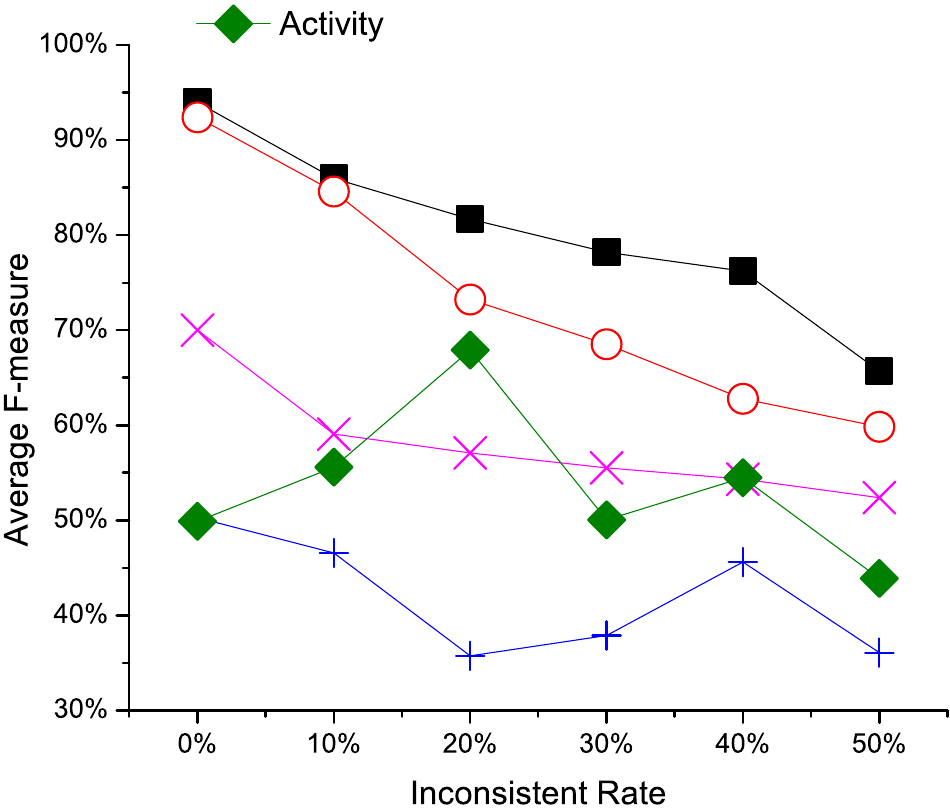}
\label{fig:cure-incons-f}
}
\subfigure{
\includegraphics[width=1.4in,height=1.0in]{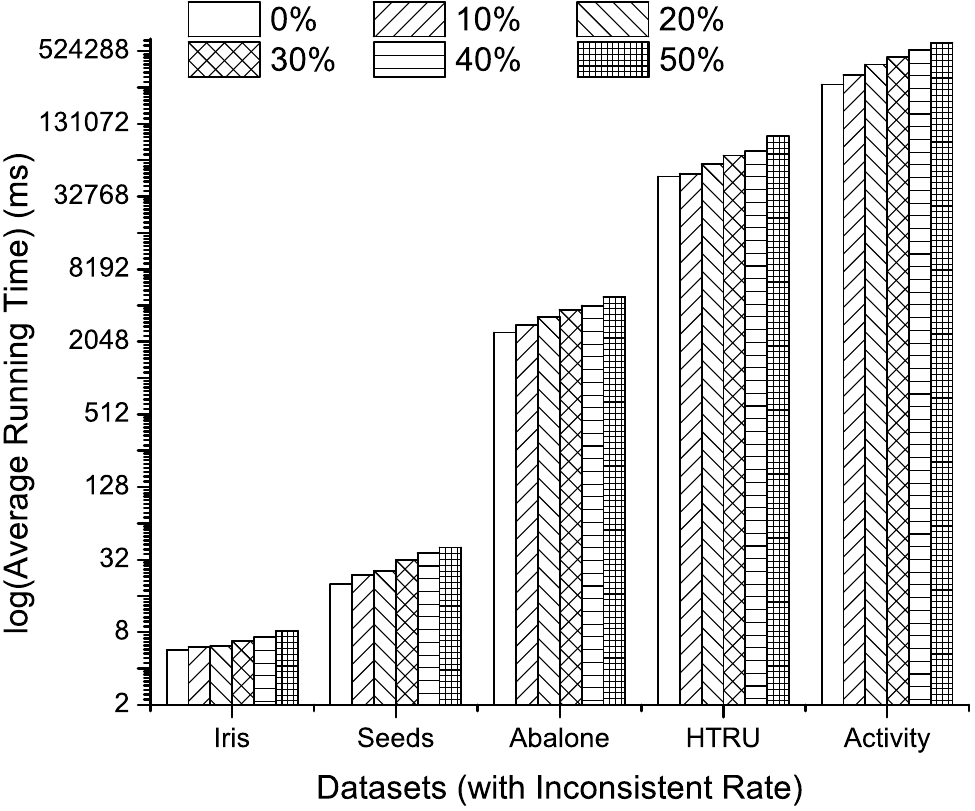}
\label{fig:cure-incons-t}
}
\vspace{-2mm}
\caption{Results on Clustering for CURE Algorithm: Varying Inconsistent Rate.}
\vspace{-2mm}
\label{fig:cure-incons}
\end{figure*}

\begin{figure*}[!htb]
\centering
\subfigure{
\includegraphics[width=1.4in,height=1.0in]{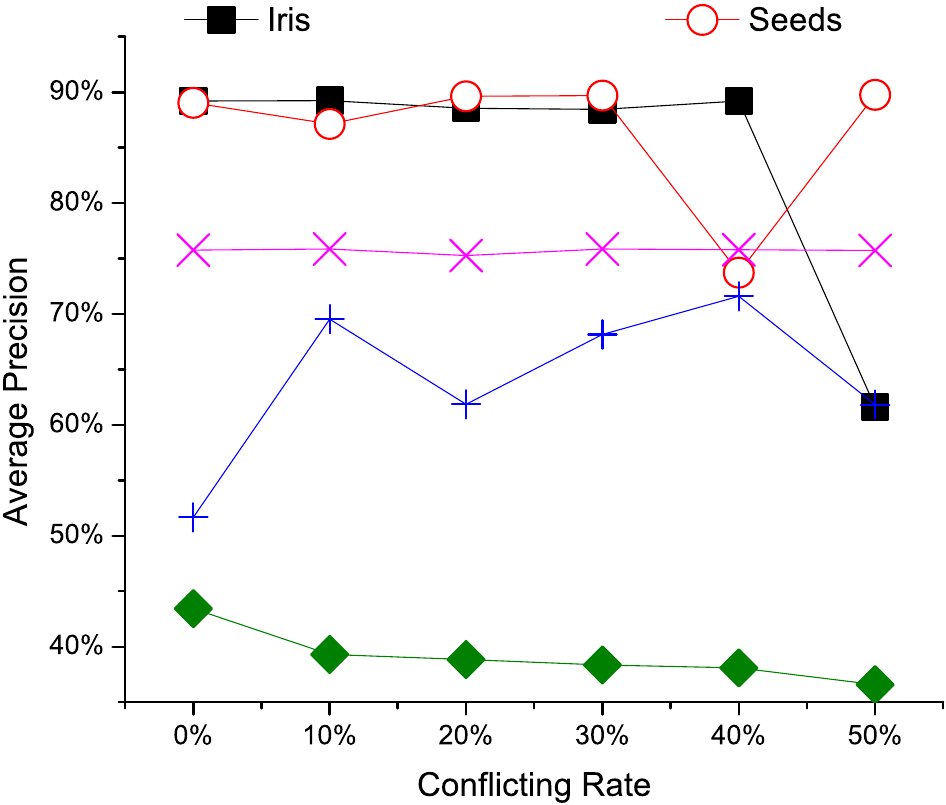}
\label{fig:km-conf-p}
}
\subfigure{
\includegraphics[width=1.4in,height=1.0in]{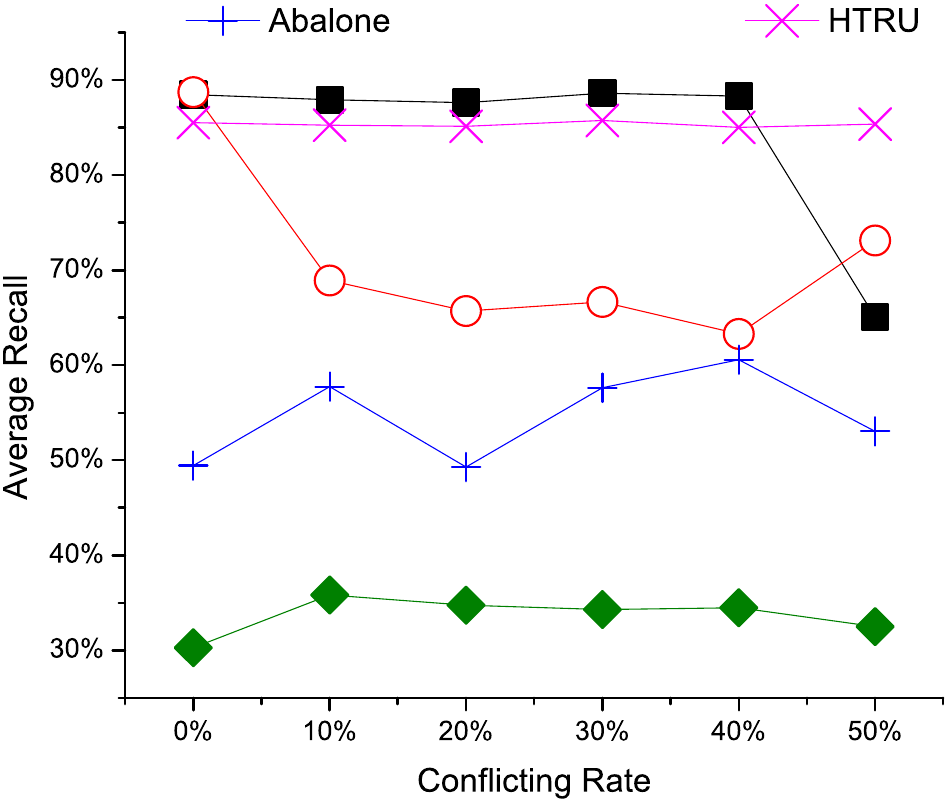}
\label{fig:km-conf-r}
}
\subfigure{
\includegraphics[width=1.4in,height=1.0in]{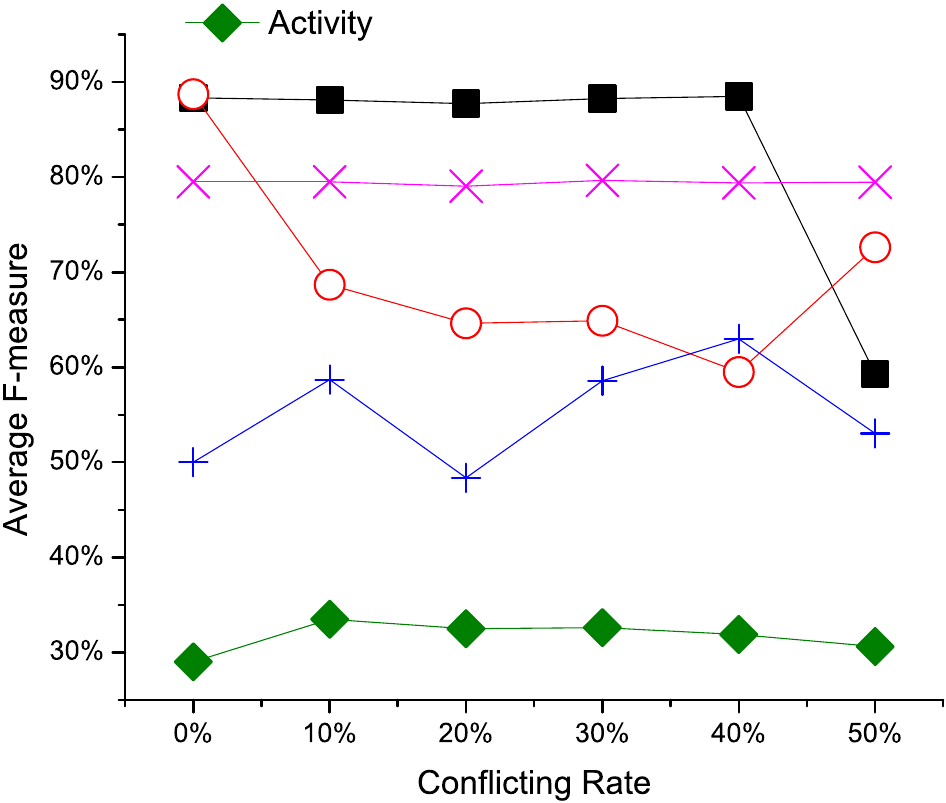}
\label{fig:km-conf-f}
}
\subfigure{
\includegraphics[width=1.4in,height=1.0in]{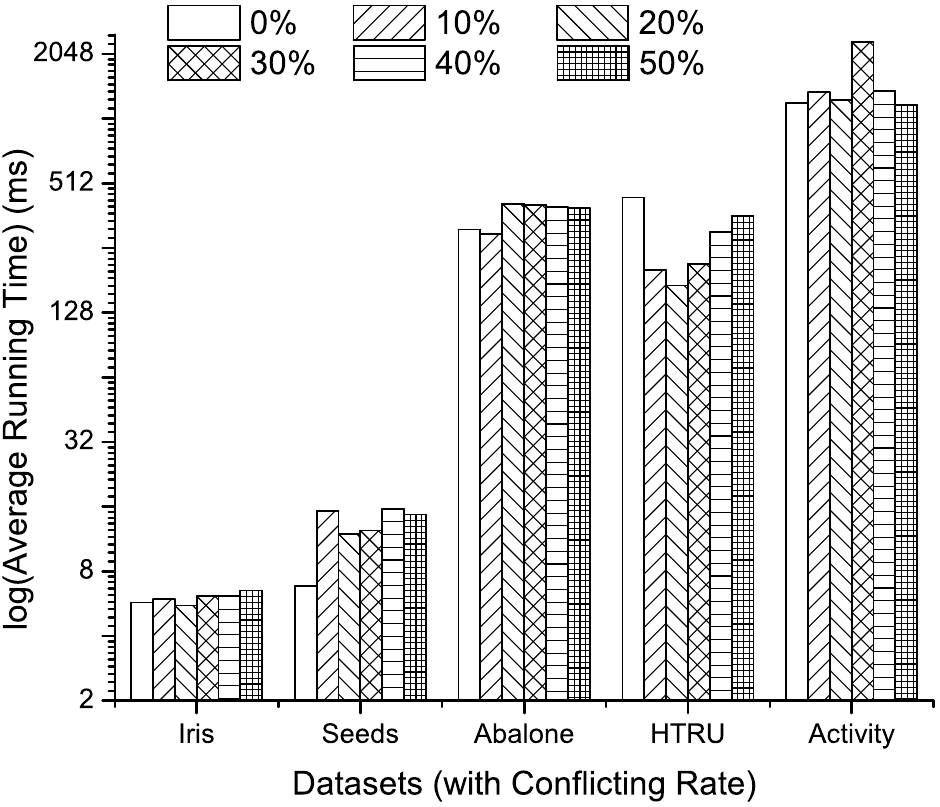}
\label{fig:km-conf-t}
}
\vspace{-2mm}
\caption{Results on Clustering for K-Means Algorithm: Varying Conflicting Rate.}
\vspace{-2mm}
\label{fig:km-conf}
\end{figure*}

\begin{figure*}[!htb]
\centering
\subfigure{
\includegraphics[width=1.4in,height=1.0in]{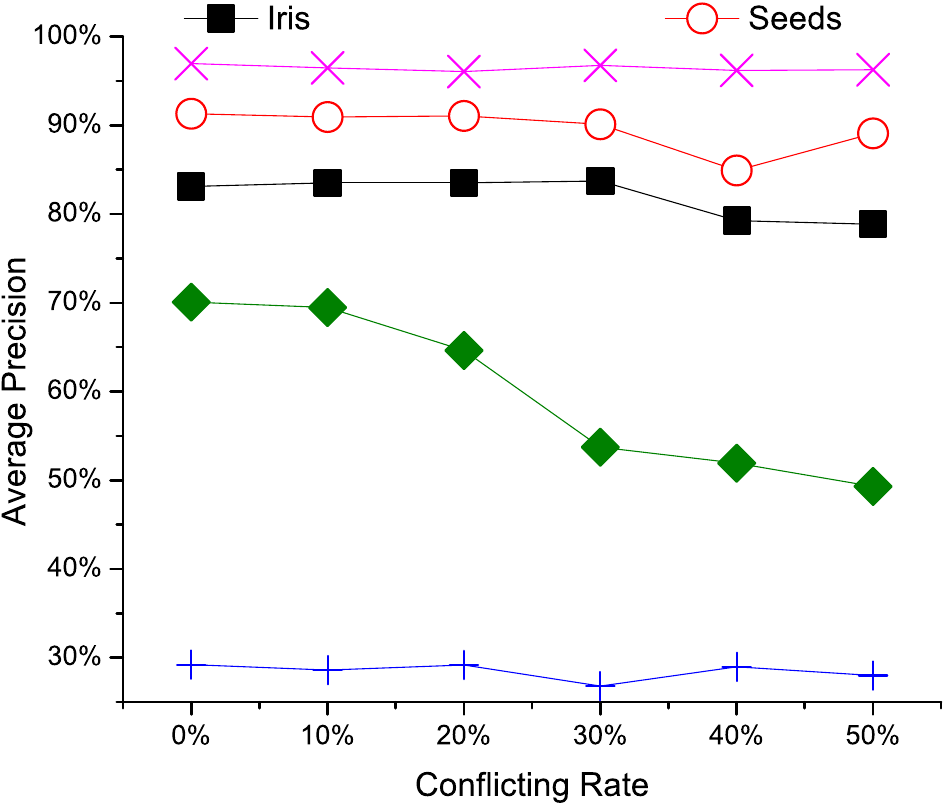}
\label{fig:lvq-conf-p}
}
\subfigure{
\includegraphics[width=1.4in,height=1.0in]{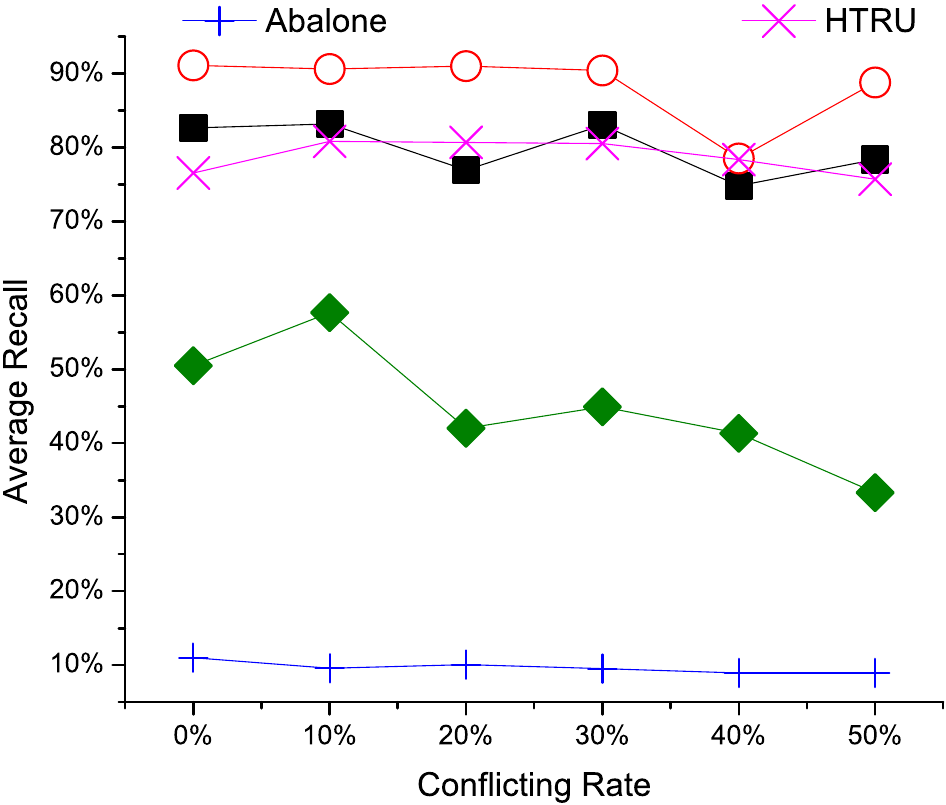}
\label{fig:lvq-conf-r}
}
\subfigure{
\includegraphics[width=1.4in,height=1.0in]{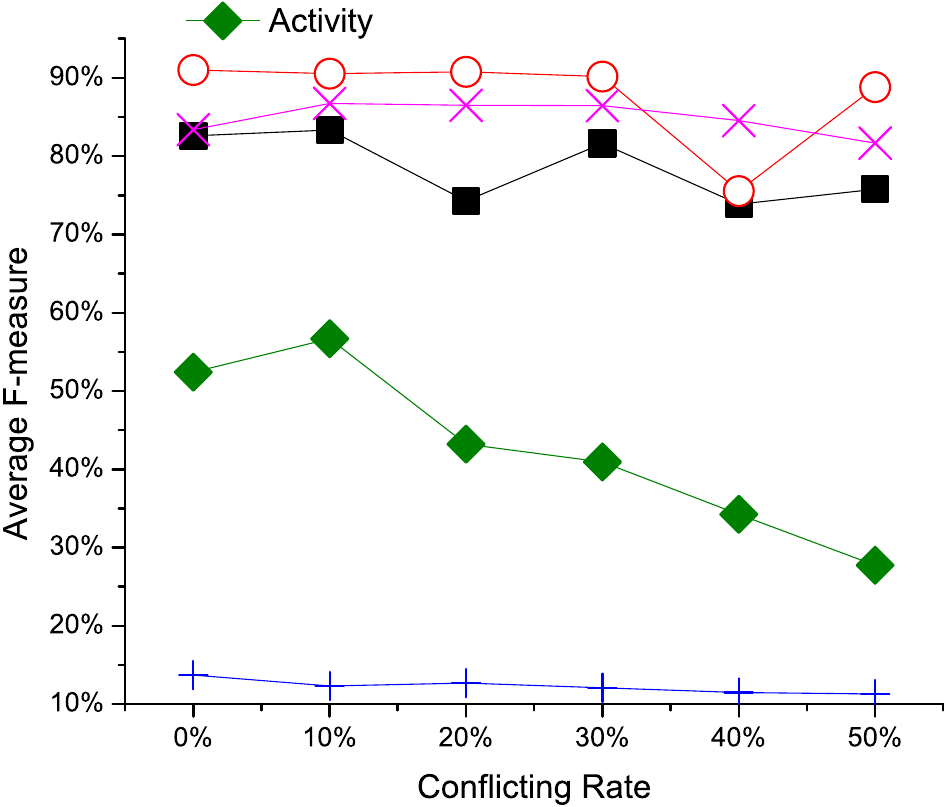}
\label{fig:lvq-conf-f}
}
\subfigure{
\includegraphics[width=1.4in,height=1.0in]{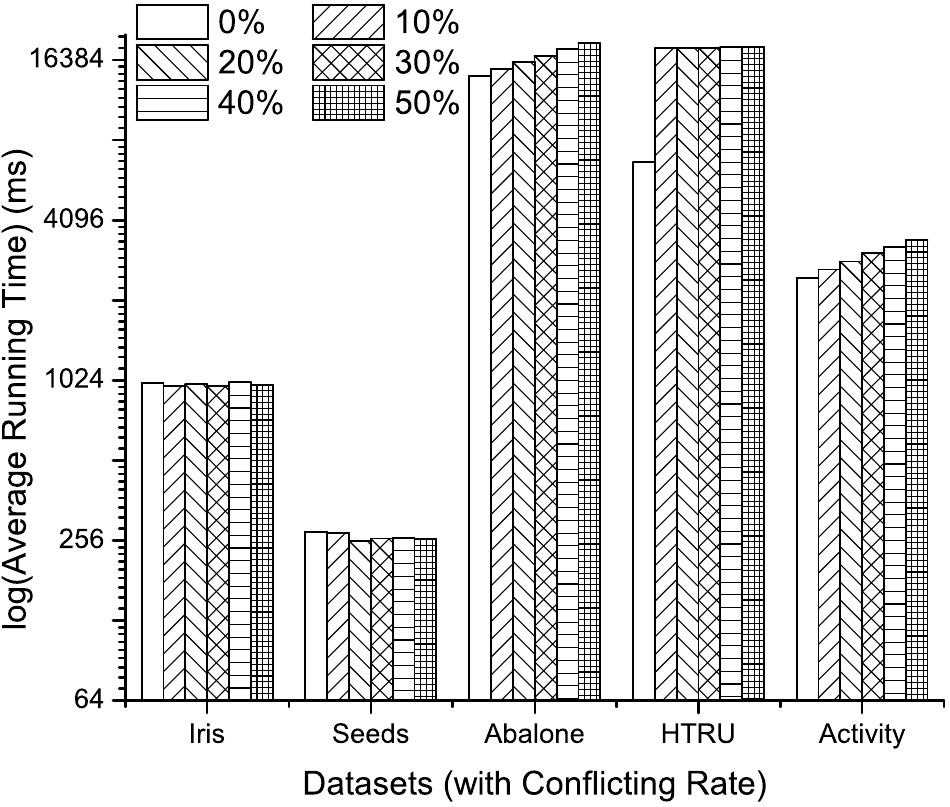}
\label{fig:lvq-conf-t}
}
\vspace{-2mm}
\caption{Results on Clustering for LVQ Algorithm: Varying Conflicting Rate.}
\vspace{-2mm}
\label{fig:lvq-conf}
\end{figure*}

\begin{figure*}[!htb]
\centering
\subfigure{
\includegraphics[width=1.4in,height=1.0in]{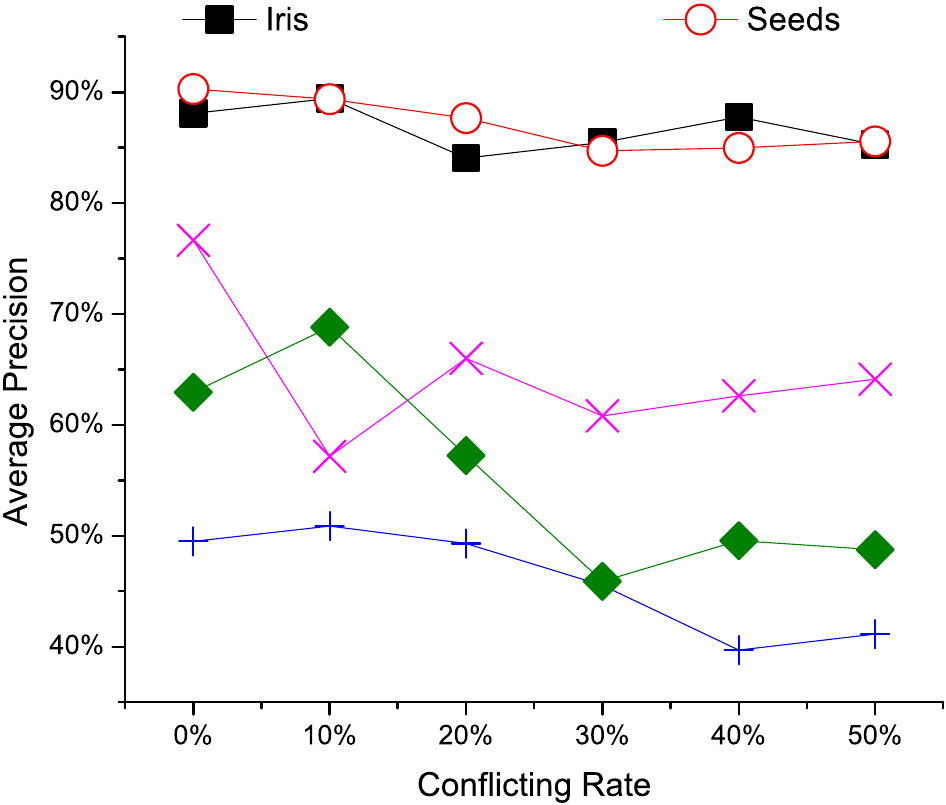}
\label{fig:clr-conf-p}
}
\subfigure{
\includegraphics[width=1.4in,height=1.0in]{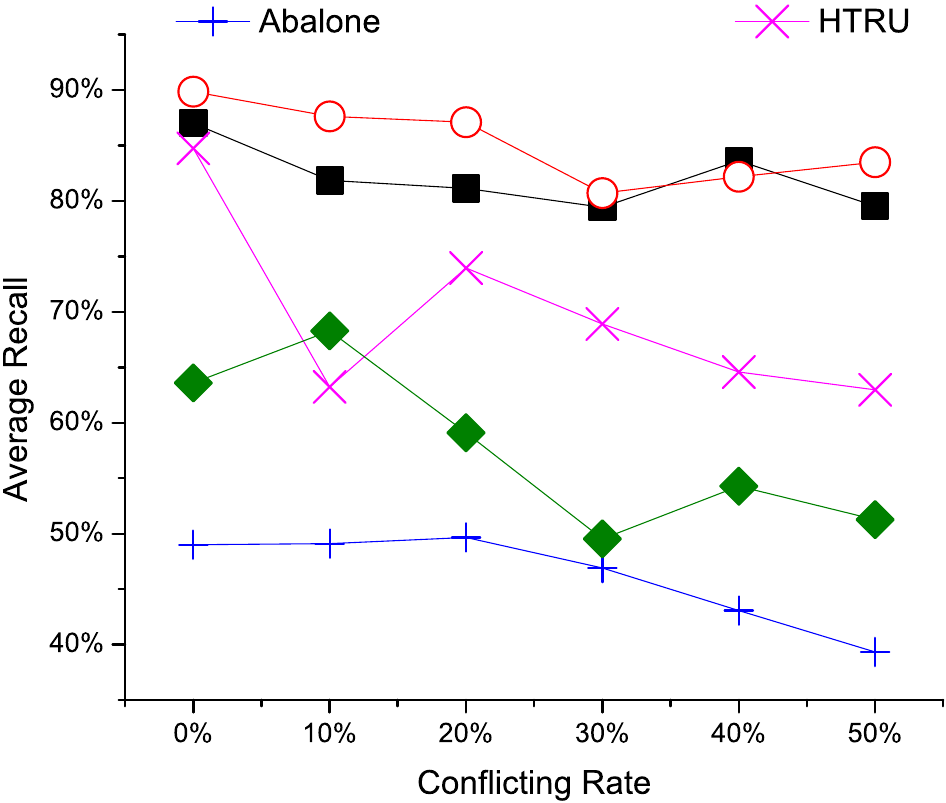}
\label{fig:clr-conf-r}
}
\subfigure{
\includegraphics[width=1.4in,height=1.0in]{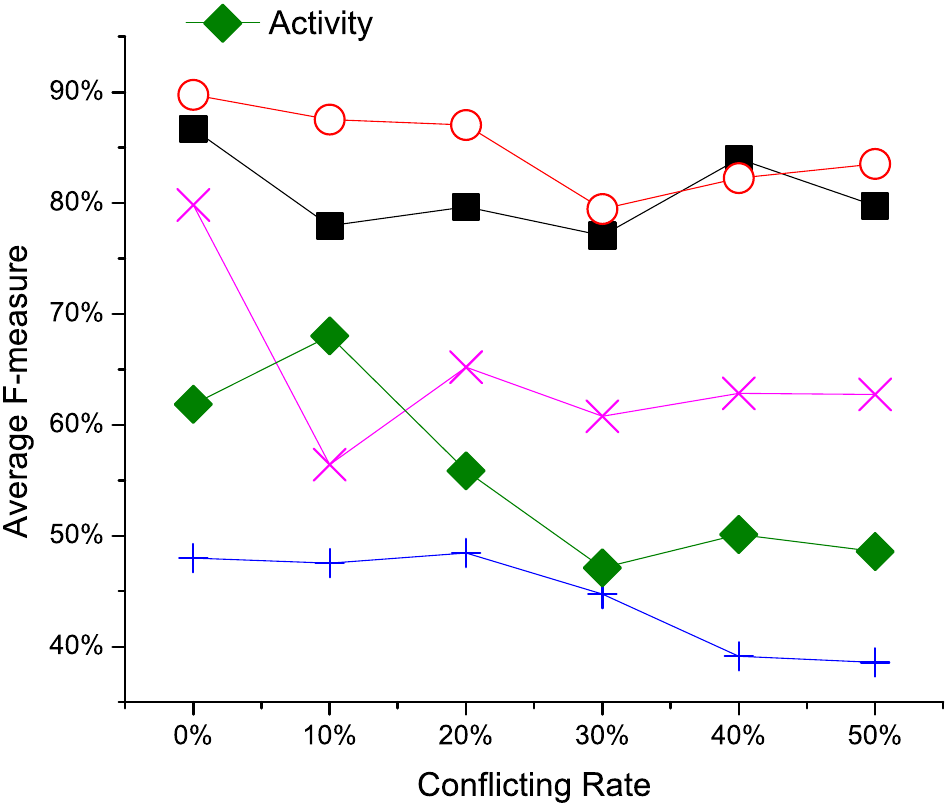}
\label{fig:clr-conf-f}
}
\subfigure{
\includegraphics[width=1.4in,height=1.0in]{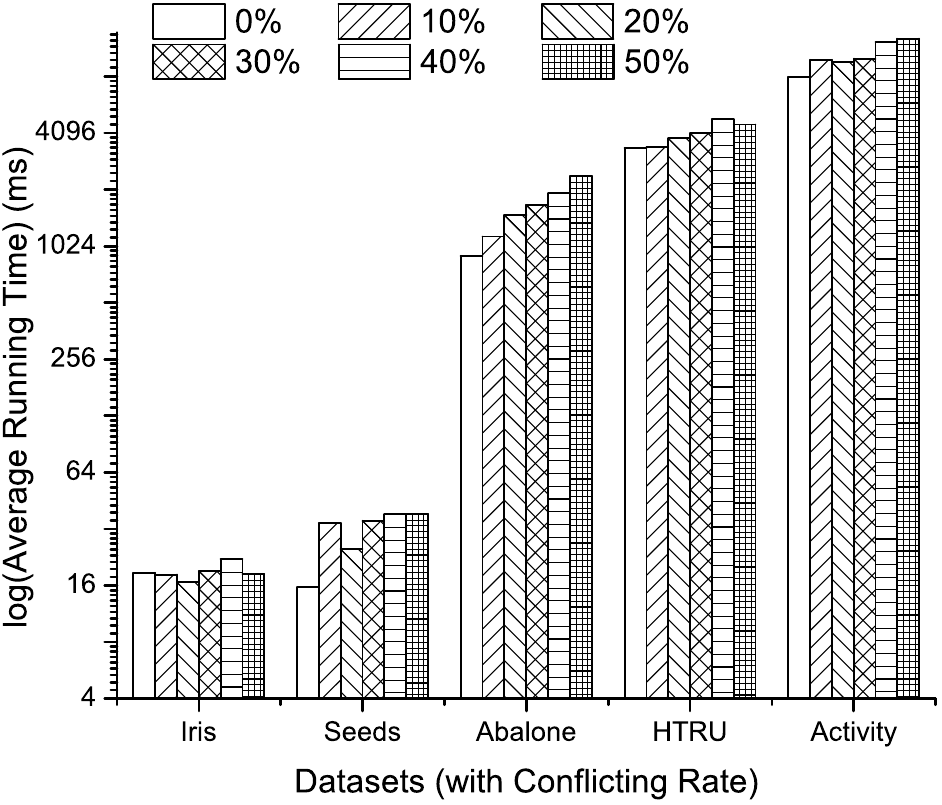}
\label{fig:clr-conf-t}
}
\vspace{-2mm}
\caption{Results on Clustering for CLARANS Algorithm: Varying Conflicting Rate.}
\vspace{-2mm}
\label{fig:clr-conf}
\end{figure*}

\begin{figure*}[!htb]
\centering
\subfigure{
\includegraphics[width=1.4in,height=1.0in]{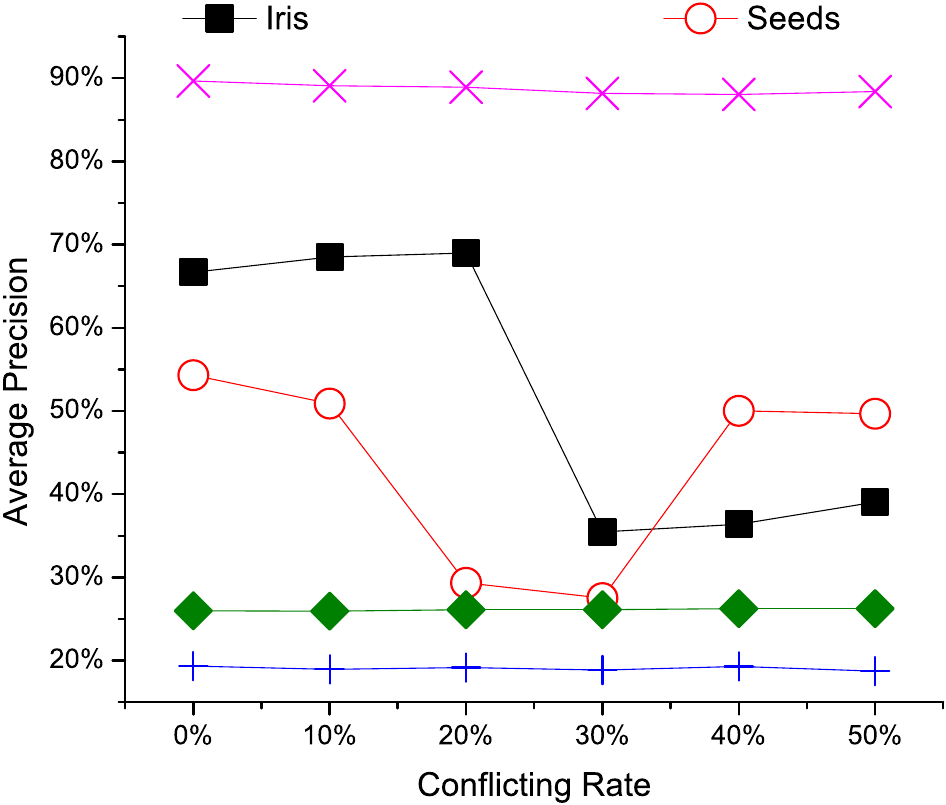}
\label{fig:db-conf-p}
}
\subfigure{
\includegraphics[width=1.4in,height=1.0in]{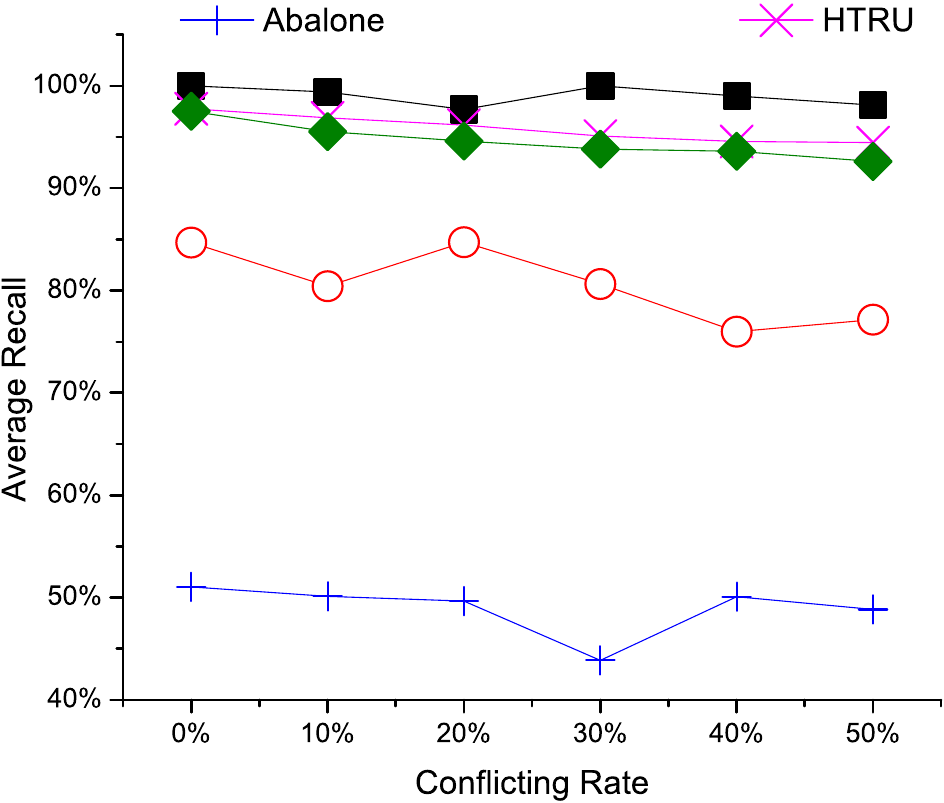}
\label{fig:db-conf-r}
}
\subfigure{
\includegraphics[width=1.4in,height=1.0in]{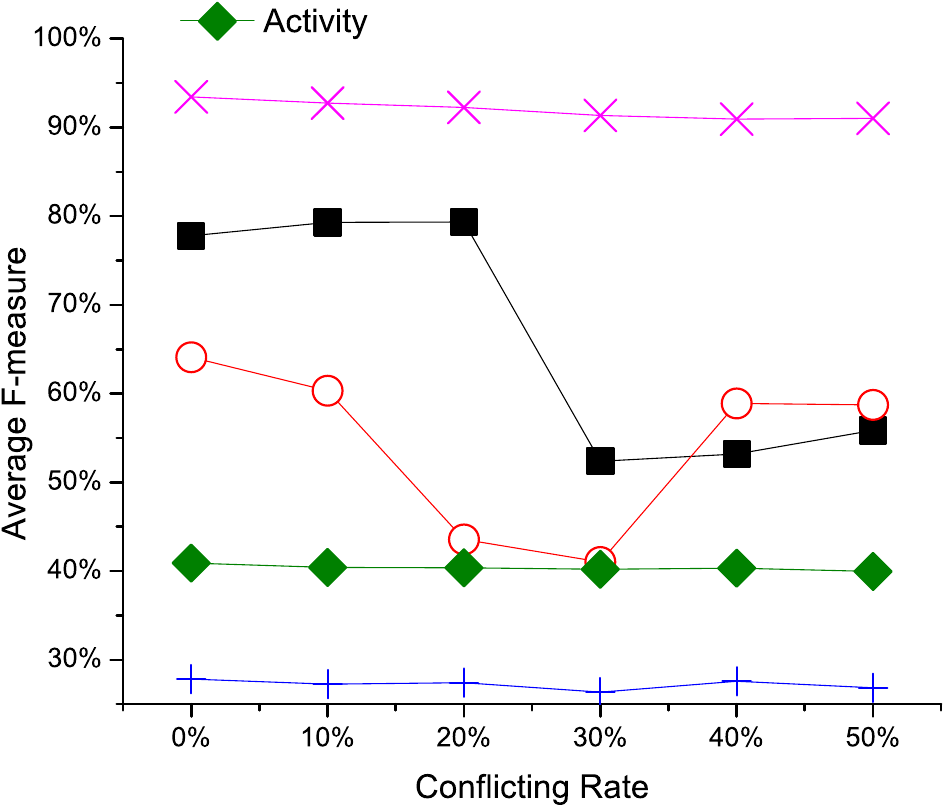}
\label{fig:db-conf-f}
}
\subfigure{
\includegraphics[width=1.4in,height=1.0in]{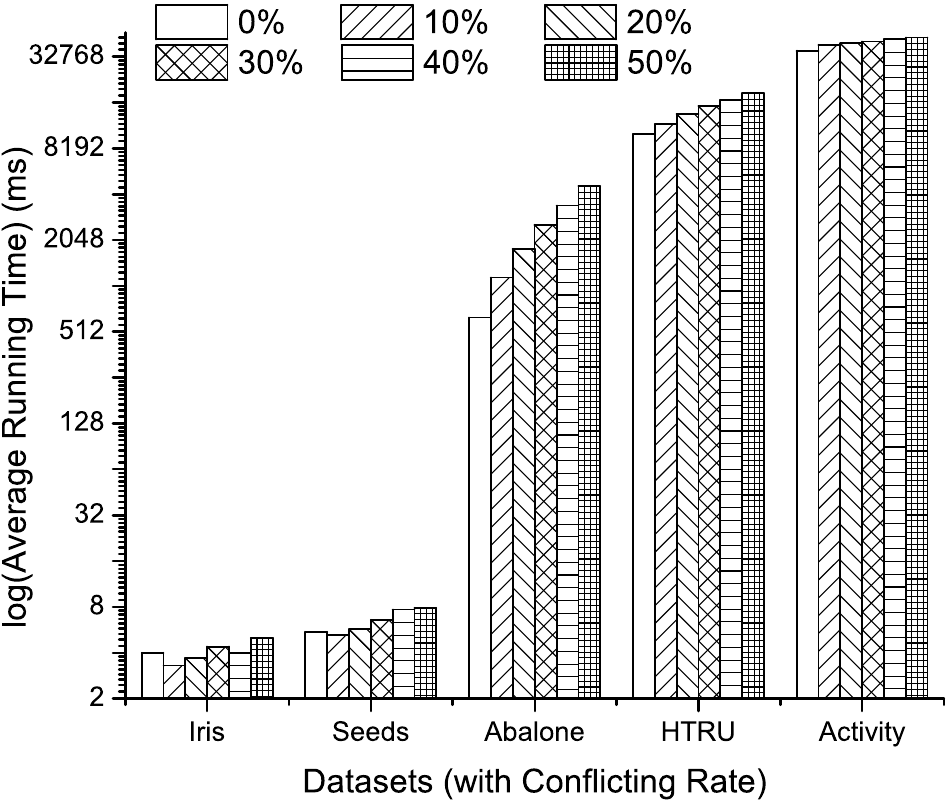}
\label{fig:db-conf-t}
}
\vspace{-2mm}
\caption{Results on Clustering for DBSCAN Algorithm: Varying Conflicting Rate.}
\vspace{-2mm}
\label{fig:db-conf}
\end{figure*}

\begin{figure*}[!htb]
\centering
\subfigure{
\includegraphics[width=1.4in,height=1.0in]{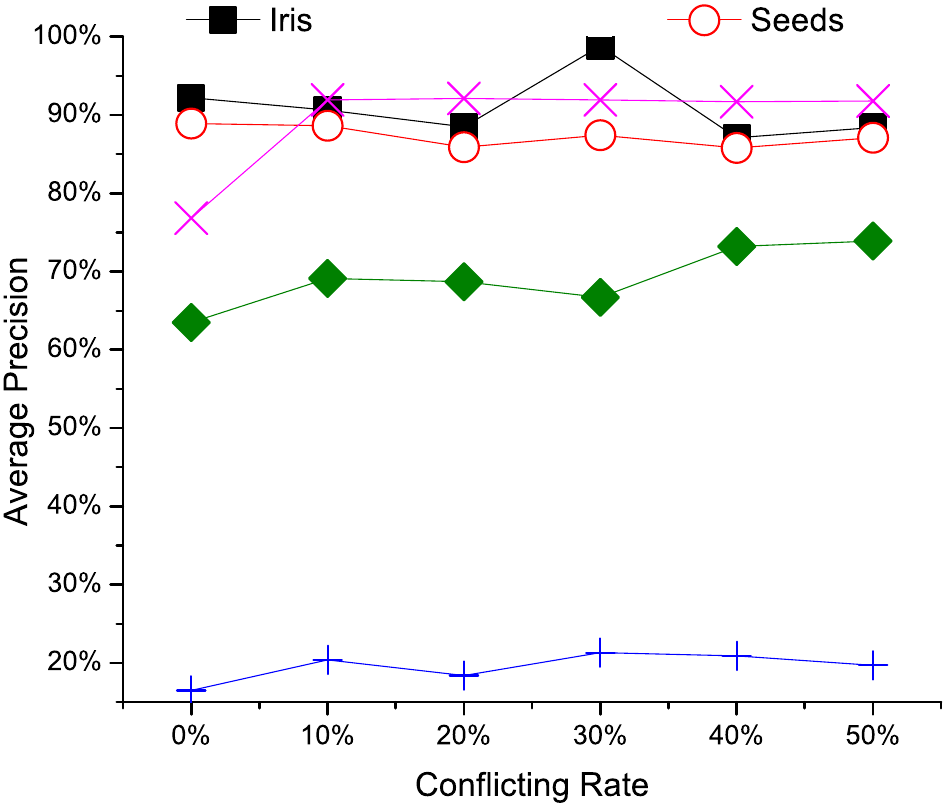}
\label{fig:bir-conf-p}
}
\subfigure{
\includegraphics[width=1.4in,height=1.0in]{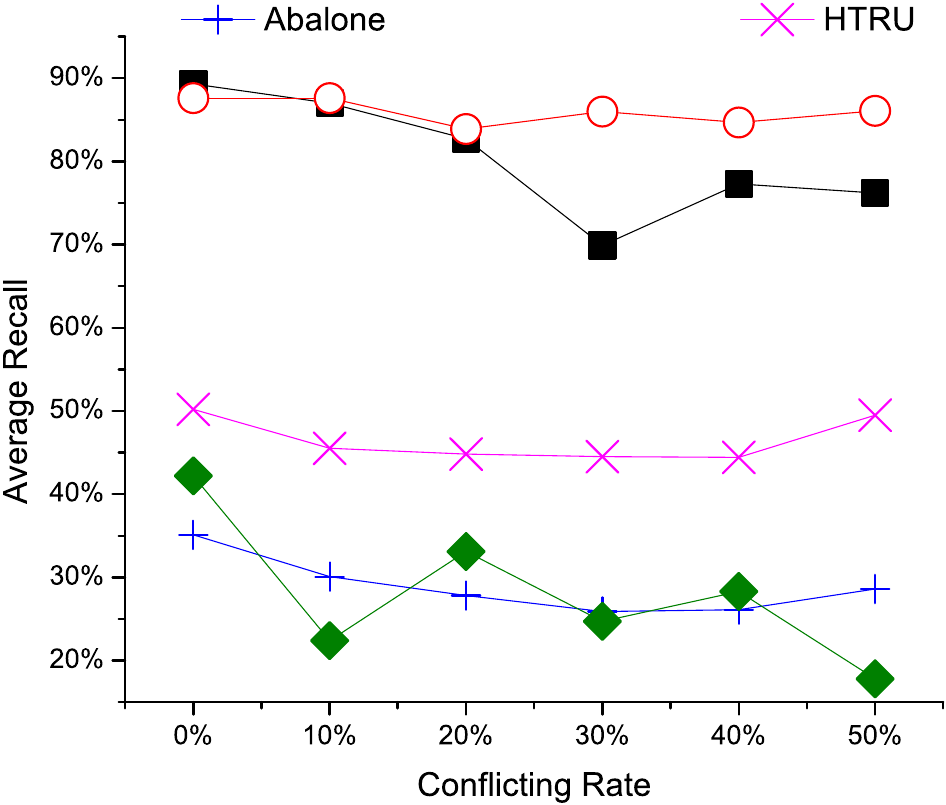}
\label{fig:bir-conf-r}
}
\subfigure{
\includegraphics[width=1.4in,height=1.0in]{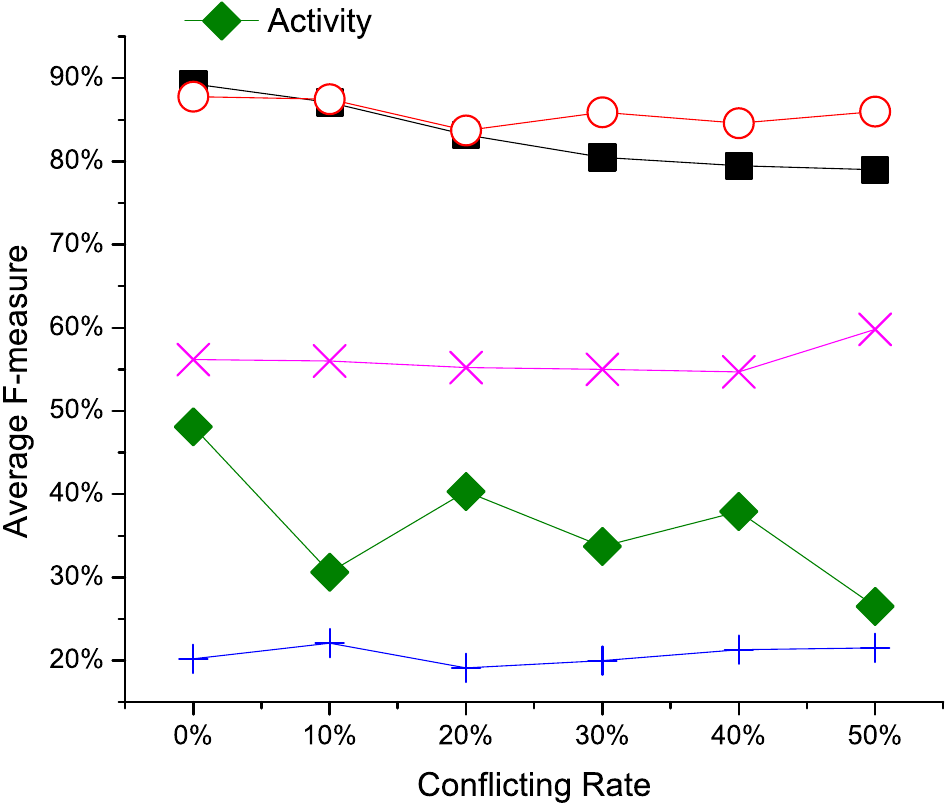}
\label{fig:bir-conf-f}
}
\subfigure{
\includegraphics[width=1.4in,height=1.0in]{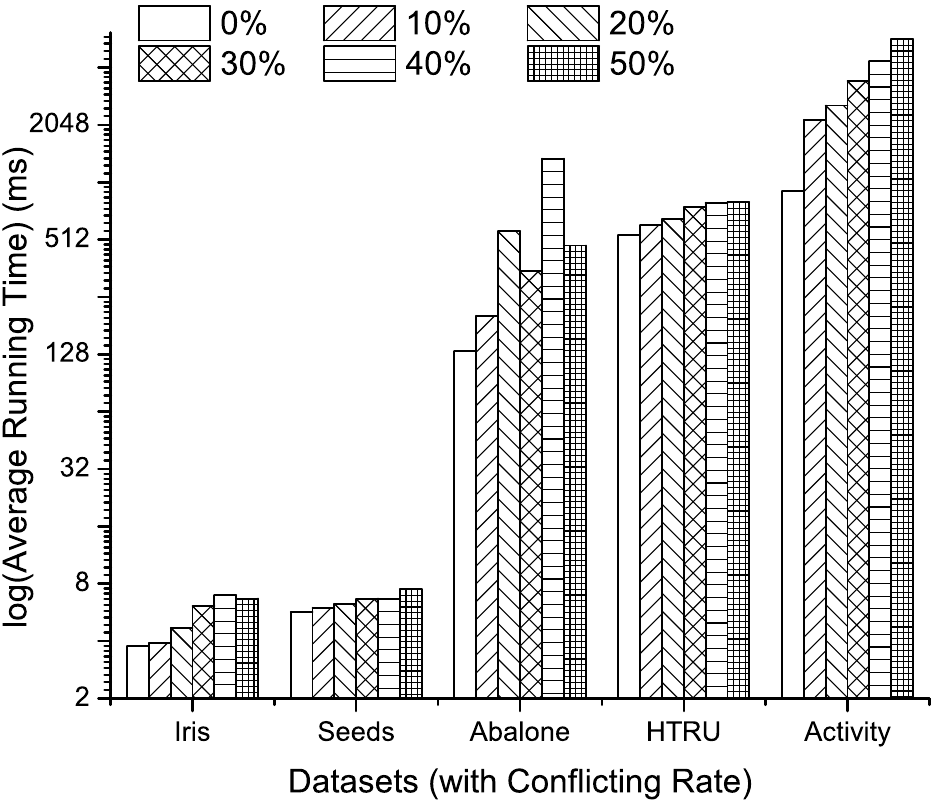}
\label{fig:bir-conf-t}
}
\vspace{-2mm}
\caption{Results on Clustering for BIRCH Algorithm: Varying Conflicting Rate.}
\vspace{-2mm}
\label{fig:bir-conf}
\end{figure*}

\begin{figure*}[!htb]
\centering
\subfigure{
\includegraphics[width=1.4in,height=1.0in]{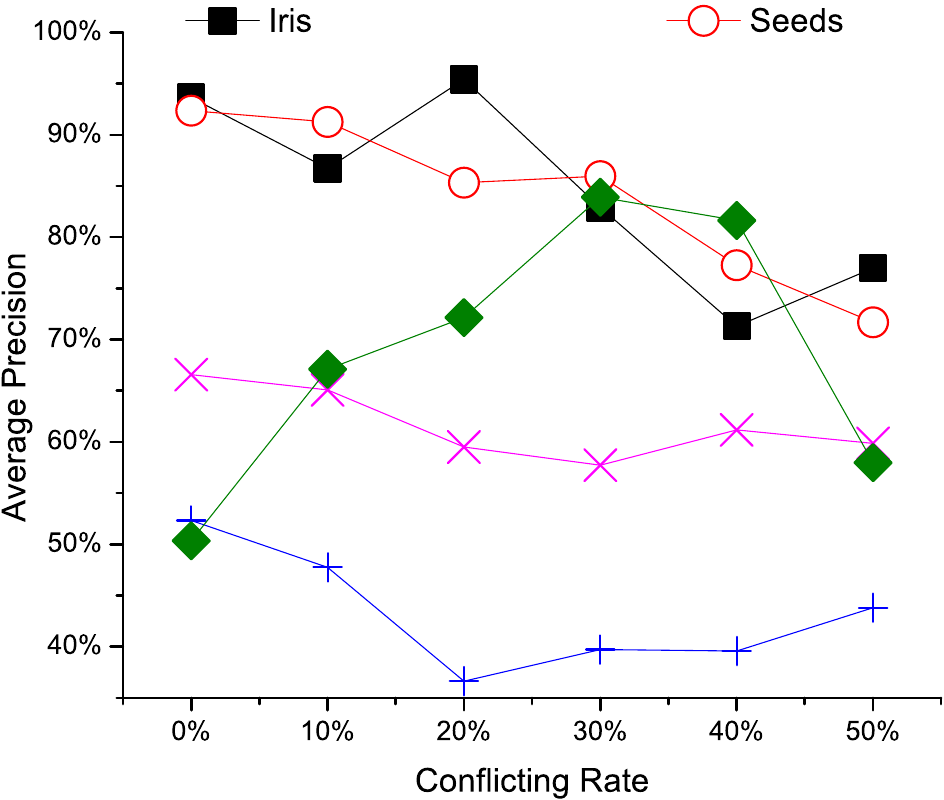}
\label{fig:cure-conf-p}
}
\subfigure{
\includegraphics[width=1.4in,height=1.0in]{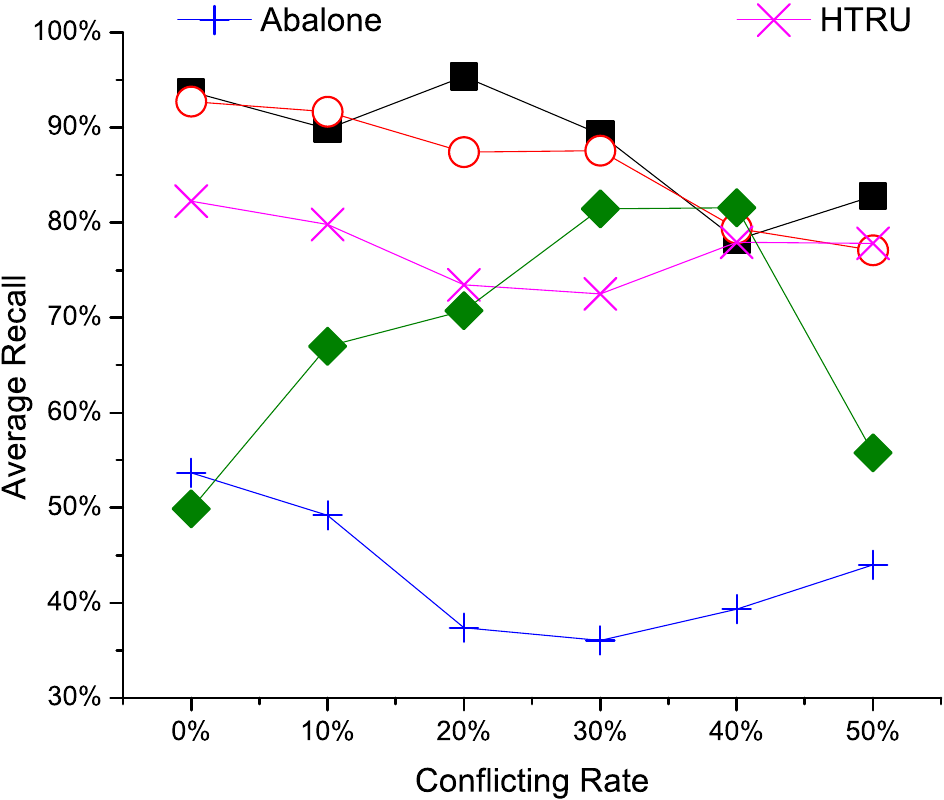}
\label{fig:cure-conf-r}
}
\subfigure{
\includegraphics[width=1.4in,height=1.0in]{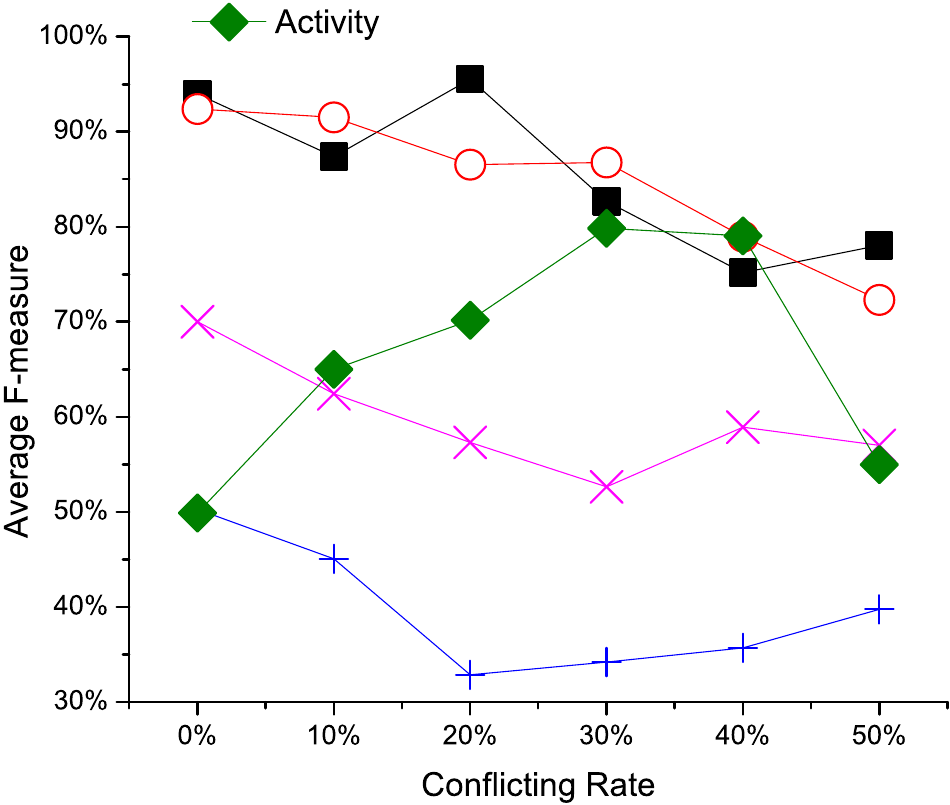}
\label{fig:cure-conf-f}
}
\subfigure{
\includegraphics[width=1.4in,height=1.0in]{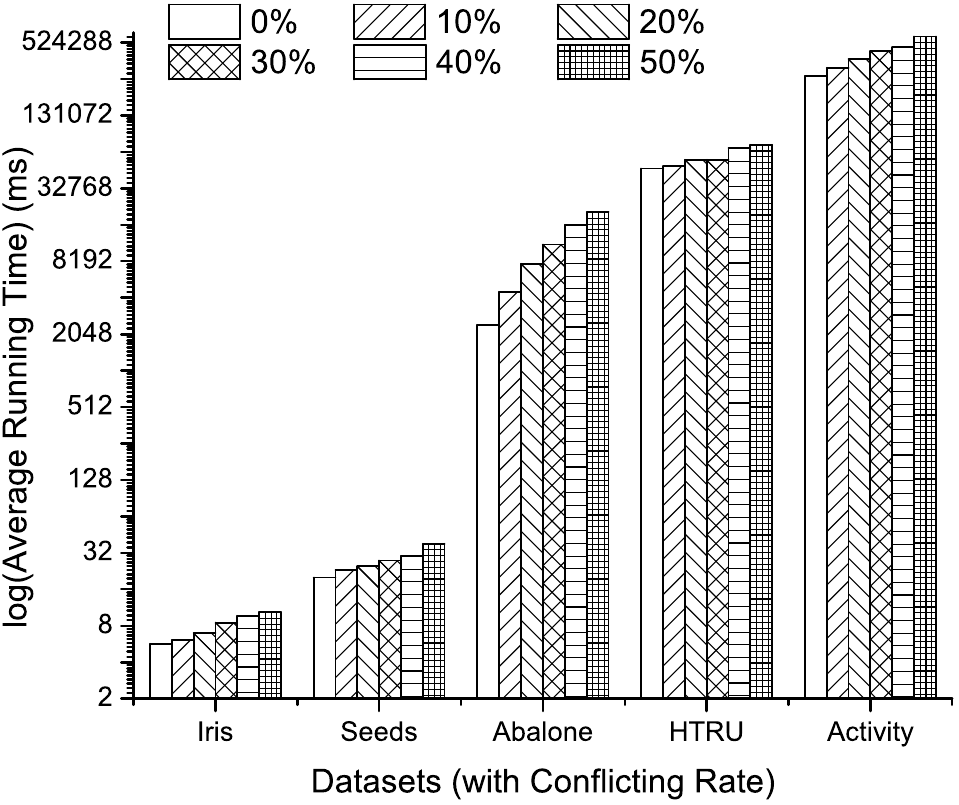}
\label{fig:cure-conf-t}
}
\vspace{-2mm}
\caption{Results on Clustering for CURE Algorithm: Varying Conflicting Rate.}
\vspace{-2mm}
\label{fig:cure-conf}
\end{figure*}